\documentclass[10pt, twoside, openright]{report}

\usepackage{preamble}

\begin{document}

\pagenumbering{gobble}
\begin{center}
\begin{figure}[h]
    \hspace*{-1cm}
    \includegraphics[width=8cm]{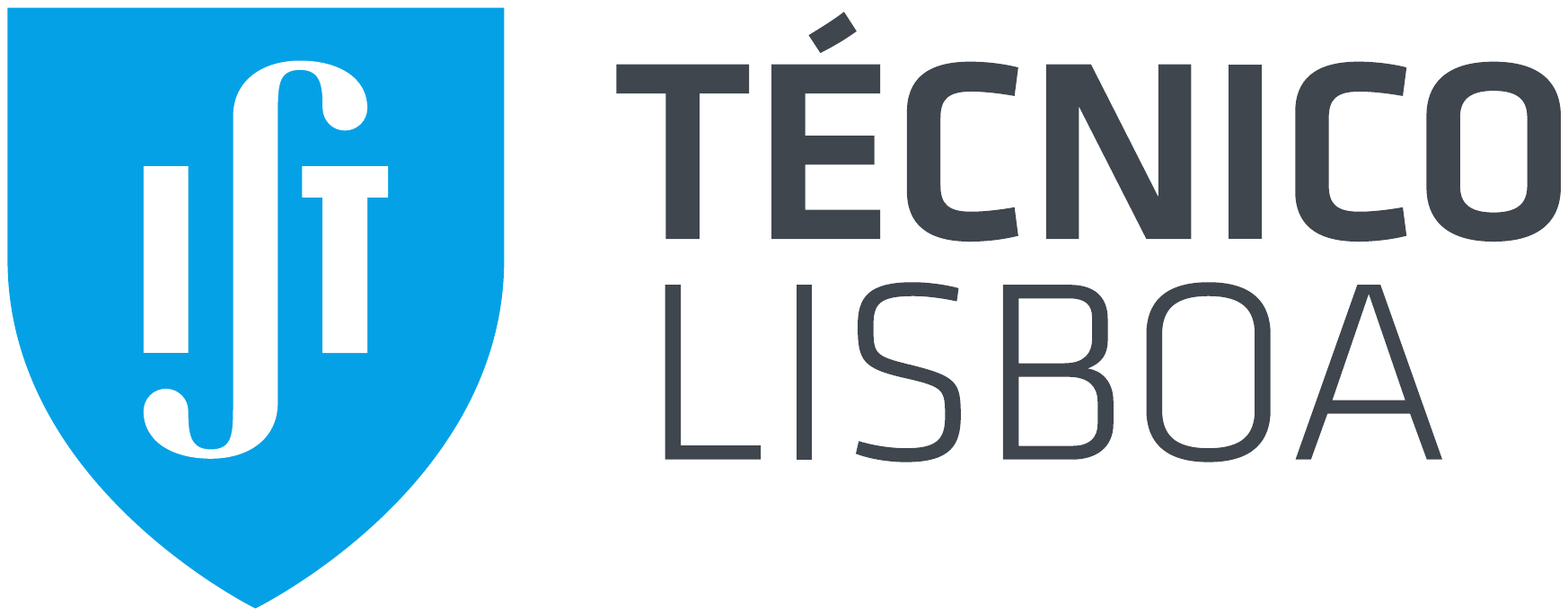}
\end{figure}

\fontsize{16pt}{0}\textbf{A framework for large scale phylogenetic analysis}
\vspace{0.5cm}

\fontsize{14pt}{0}\textbf{Bruno Miguel Leandro Lourenço}
\vspace{2cm}

\fontsize{12pt}{0}\selectfont Thesis to obtain the Master of Science Degree in
\vspace{0.25cm}

\fontsize{16pt}{0}\textbf{Computer Science and Engineering}
\vspace{0.5cm}

\fontsize{12pt}{0}\selectfont Supervisors: Doctor Alexandre Paulo Lourenço Francisco
\vspace{0.25cm}

Doctor Cátia Raquel Jesus Vaz
\vspace{2cm}

\fontsize{14pt}{0}\textbf{Examination Committee}
\vspace{0.5cm}

\fontsize{12pt}{0}\selectfont Provisory
\vspace{0.25cm}

\vspace{5.25cm}
\fontsize{14pt}{0}\textbf{December 2020}
\end{center}
\cleardoublepage
\pagenumbering{roman}
\chapter*{Acknowledgements}

Throughout this thesis I have received a great deal of support and assistance. Hence, I am using this opportunity to express my gratitude to everyone who supported me. 

I would first like to thank my supervisors, Alexandre Francisco and Cátia Vaz, for providing me with the valuable support, availability, and guidance needed to accomplish this thesis.

I would also like to thank a colleague, Miguel Coimbra, for providing me with valuable advices and guidance that I needed to choose the right direction and complete my thesis.

In addition, I would also like to thank my girlfriend for all the help and motivation she gave me to keep going despite all of the challenges, and for always being there for me through all the ups and downs.

Finally, I would like to thank my family and friends for supporting me, and for always providing me fun distractions that create happy memories during this thesis.

\vfill
\noindent This work was partly supported by national funds through FCT -- Fundação para
a Ciência e Tecnologia, under projects PTDC/CCI-BIO/29676/2017 and
UIDB/50021/2020.
\chapter*{Abstract}

With growing exchanges of people and merchandise between countries, epidemics have become an issue of increasing importance and huge amounts of data are being collected every day. Hence, analyses that were usually run in personal computers are no longer feasible. It is now common to run such tasks in \ac{HPC} environments and/or dedicated systems. On the other hand, we are often dealing in these analyses with graphs and trees, and running algorithms to find patterns in such structures. Hence, although graph oriented databases and processing systems can be of much help in this setting, as far as we know there is no solution relying on these technologies to address large scale phylogenetic analysis challenges. This work aims to develop a modular framework that exploits such technologies, namely Neo4j. We address this challenge by proposing and developing a framework which allows representing large phylogenetic networks and trees, as well as ancillary data, that supports queries on such data, and allows the deployment of algorithms for inferring/detecting patterns and pre-computing visualizations, as a Neo4j plugin. This framework is innovative and brings several advantages to the phylogenetic analysis process, like the management of the phylogenetic trees, which will avoid having to compute them again, and the use of multilayer networks, that will make the comparison between them more efficient and scalable. The experimental evaluation results showcase that it can be very efficient in the mostly used operations and that the supported algorithms comply with their time complexity.

\vspace{0.5cm}

\textbf{Key Words:} Phylogeny; data processing; data storage; graphs; database;

\chapter*{Resumo}

Na atualidade, trocas de pessoas e mercadorias entre diferentes países tem vindo a aumentar. Como consequência as epidemias tornaram-se uma preocupação maior, resultando na recolha de grandes quantidades de dados todos os dias. As análises que normalmente eram executadas em computadores pessoais já não são viáveis. Agora é comum executar essas análises em ambientes \ac{HPC} e/ou sistemas dedicados. Por outro lado, nessas análises lidamos frequentemente com gráficos, árvores e com execuções de algoritmos para encontrar padrões nestas estruturas. Embora existam base de dados orientadas a grafos e sistemas de processamento que podem ajudar neste tema, não conhecemos nenhuma solução baseada nestas tecnologias para lidar com os desafios da análise filogenética em larga escala. O objetivo deste projeto é o desenvolvimento de uma plataforma que explore estas tecnologias, nomeadamente o Neo4j. Nós abordamos este desafio com a proposta e o desenvolvimento de uma plataforma que permita a representação de grafos e árvores filogenéticas de maior dimensão, bem como dados auxiliares, que suporta consultar esses dados e que permita a execução de algoritmos, para inferir/detectar padrões e pré-computar visualizações, como plugins do Neo4j. Esta plataforma é inovadora e traz vantagens para a análise filogenética, como por exemplo, o armazenamento dos grafos, que evita ter que computá-los novamente, e o uso de redes multi camadas, que torna a comparação entre eles mais eficiente. A análise dos resultados experimentais mostra que a plataforma pode ser muito eficiente nas operações mais utilizadas e que os algoritmos suportados obedecem à sua complexidade de tempo.

\vspace{0.5cm}

\textbf{Palavras Chave:} Filogenia; processamento de dados; armazenamento de dados; grafos; base de dados;

\tableofcontents
\addcontentsline{toc}{chapter}{\listfigurename}
\listoffigures
\addcontentsline{toc}{chapter}{\listtablename}
\listoftables
\addcontentsline{toc}{chapter}{List of Acronyms}
\chapter*{List of Acronyms}

\begin{acronym}[MPC]
    \acro{HPC}{High-performance computing}
    \acro{RDBMS}{Relational Database Management System}
    \acro{CRUD}{create, read, update, and delete}
    \acro{OLTP}{Online Transaction Processing}
    \acro{ACID}{atomicity, consistency, isolation, durability}
    \acro{RDF}{Resource Description Framework}
    \acro{DPT}{data processing tools}
    \acro{NGS}{Next-Generation Sequencing}
    \acro{DNA}{Deoxyribonucleic acid}
    \acro{MLST}{Multilocus Sequence Typing}
    \acro{MLVA}{Multiple-Locus Variable Number Tandem Repeat Analysis}
    \acro{SNP}{Single Nucleotide Polymorphism}
    \acro{CSV}{comma-separated values}
    \acro{ST}{Sequence Type}
    \acro{eBURST}{eletronic Based Upon Related Sequence Types}
    \acro{goeBURST}{Globally Optimized eBURST}
    \acro{API}{application programming interface}
    \acro{HTTP}{Hypertext Transfer Protocol}
    \acro{AOP}{aspect-oriented programming}
    \acro{APOC}{Awesome Procedures on Cypher }
    \acro{SQL}{Structured Query Language}
    \acro{IoC}{inversion of control}
    \acro{JSON}{JavaScript Object Notation}
    \acro{OGM}{object graph mapper}
    \acro{REST}{representational state transfer}
\end{acronym}

\cleardoublepage
\pagenumbering{arabic}
\chapter{Introduction}

Phylogenetics is the study of the evolutionary history and relationships among individuals or groups of organisms, which aims to produce diagrammatic hypothesis about the history of the evolutionary relationships of a group of organisms known as phylogenetic tree. The relationships are inferred through the analysis of the traits of the individual or group, that is, by applying computational algorithms, methods, and programs to the phylogenetics data.

With the growing exchanges of people and merchandise between countries, epidemics have become an issue of increasing importance. The computational phylogenetics were mostly performed in personal computers and desktops. That is, the data was loaded to the client computer through a text file, the algorithms were executed, and the results of the algorithms executions were then displayed to the user. However, this kind of analysis is not feasible anymore, since huge amounts of data are being collected every day, and there are certain operations that require a considerable amount of memory or time. Instead, it is now common to run such tasks in high performance computing environments and/or dedicated systems. Therefore, there is a need to find a better way to store and maintain the data rather than in personal computers and desktops.

In large scale phylogenetic analysis of microbial population genetics, it is often needed to sequence, and type the information of the organisms, and afterwards to apply a set of phylogenetic inference methods to produce a diagrammatic hypothesis about the history of the evolutionary relationships of a group of organisms. The computation and analysis of microbial population genetics often produce graphs and trees , which have many relationships \cite{artc:trees}. As graph databases naturally apply to these data structures and are optimized to perform queries and operations over them, that is, they are designed specifically to deal with highly connected data, it should be possible to store them in a graph database.

Although graph oriented databases can be of much help in this setting, as far as it is known there is no solution relying on these technologies to address large scale phylogenetic analysis challenges. Thus, a study on which database engine better addresses the needs of this challenge can provide new insights and lead to innovative approaches, comparing graph databases such as Neo4j \cite{artc:neo4j}, Titan Aurelius \cite{web:titan}, JanusGraph \cite{web:janusgraph}, Dgraph \cite{web:dgraph}, Allegrograph \cite{web:allegrograph}, and Apache Rya \cite{artc:rya}. The comparisons made in this context suggest that Neo4j offers the most interesting set of features and capabilities, such as relying on an architecture based on native graph storage and processing engine, having an active community, and allowing to extend itself with plugins to support any graph algorithm. Therefore, Neo4j is the graph database system that is used to address the large scale phylogenetic analysis challenge.

\section{Objectives}

The objective of this project is to develop a modular framework for large scale phylogenetic analysis that exploits a graph oriented database technology which allows to access the phylogenetics data, without needing to load it into the clients computers. This framework should have a data model that allows the representation of large phylogenetic networks and trees, as well as the ancillary data. It should support queries on such data, and allow the deployment of algorithms for inferring/detecting patterns and for pre-computing visualizations.

\section{Document Structure}

This document is composed by several chapters, namely by the Background, Proposed Solution, Implementation, Experimental Evaluation, and Final Remarks chapters. The Background chapter starts by explaining the concept of a graph database and presents several use cases where a graph database is used to store highly connected data. Then, several graph databases are identified and discussed to understand which database suits better. Moreover, it presents a platform that performs phylogenetic analysis, explains how it is accomplished and presents the different types of phylogenetic data. The Proposed Solution chapter enumerates several requirements and use cases for this project. It proposes an architecture for a framework, to cope with the phylogenetic large scale analysis challenge, exploiting several architectural views. Additionally, it describes what technologies shall support this framework, providing several key characteristics of each. The Implementation chapter emphasizes the architectural choices by exemplifying how to extend some parts of the framework. Furthermore, it presents implementation decisions and explains why they were embraced. The Experimental Evaluation chapter explains the types of tests, the operations, the datasets, and the system settings used to evaluate the framework. Afterwards, it provides an analysis of the results obtained from these tests. Finally, the Final Remarks chapter presents the conclusions of this project, and showcases how it can be extended in future work.

This project is publicly available at \url{https://github.com/Brunovski/phyloDB} and it can be deployed using Docker \cite{web:docker}. The repository of the project provides the architecture documentation, the deployment and usage definitions, and examples of how to deploy and use the framework.

\chapter{Background}

This chapter provides an overview of the graph databases, and the large scale phylogenetic analysis of microbial population genetics.

The data that is managed during the large scale phylogenetic analysis of microbial population genetics may have many relationships, and as graph databases are designed specifically to deal with highly connected data, it should be possible to store and manage this data within a graph database. Therefore, the concept of a graph database is introduced, and several use cases of using a graph database system to store and manage similar data are presented. Furthermore, the problems that may result from using a \ac{RDBMS} to handle the same type of data are discussed. To support this discussion, an experiment that compares a \ac{RDBMS} with a graph oriented database is analysed. Due to \ac{RDBMS} not being suitable for large scale phylogenetic analysis, some metrics are defined to better understand which graph database engine is the most suitable. Hence, several graph databases engines are compared using those metrics.

The study of microbial population genetics is composed of several processes, such as the alignment of organisms sequences, the application of a typing methodology and the execution of a set of phylogenetic inference methods. Hence, a platform which performs this kind of analysis is presented. Afterwards, since the usual processes of the study of microbial population genetics constitutes scientific workflows, the definition and examples of scientific workflows are presented. Moreover, the workflow system which is used by the platform is introduced. Each one of the specified processes are detailed to understand how the study of microbial population genetics is accomplished. Finally, throughout the explanation of all these concepts and processes, the several types of data that exist or may appear during an execution of the overall process are also explained.

\section{Graph Databases}\label{graphdatabases}

A graph database management system is an online database management system with \ac{CRUD} methods that expose a graph data model. Graph databases are generally built for use with \ac{OLTP} systems. Accordingly, they are normally optimized for transactional performance, and engineered with transactional integrity and operational availability in mind\cite{book:GraphDatabases, bkchp:GDintroduction}. This type of database addresses the problem of leveraging complex and dynamic relationships in highly connected data. Graph databases offer appealing characteristics, such as performance and flexibility. Regarding the performance of graph databases, it tends to remain relatively constant, even as the dataset grows, because the queries only use the respective portion of the graph. In terms of flexibility, a graph database allows adding new nodes, labels, and relationships, to an existing structure, without disturbing the existing queries and application functionality. A graph database exposes a graph data model, by storing the data in a graph format. A graph is a set of nodes that represent entities and a set of edges that represent the relationships that connect them. Graphs and trees are often produced in the computation and analyses of phylogenetic data. Therefore, it should be possible to use a graph oriented database, which naturally applies to these data structures and are optimized to perform queries and operations over them.

\subsection{Use Cases}

The most common graph database use cases include contexts such as social data, recommendation algorithms, geospatial operations, and network and data centre management. All of these use cases share a common context characteristic, that is they all deal with highly connected data, just like in the context of large scale phylogenetic analysis. In each use case, using a graph data model and the specific characteristics of the graph database, allows to generate competitive insight and significant business value. For instance, in social applications, such as Facebook, using a graph database allows to understand who interacts with whom, and how people are connected. This allows the social network application to generate insight of the aspects that influence individual behaviours.

\subsection{RDBMS Comparison}
The use of a \ac{RDBMS} to store connected, semi-structured datasets is not suitable, as they struggle when attempting to model several depths of relationships \cite{bkchp:connecteddata}. In spite of relationships existing in relational databases, they only exist as a mean of joining tables. In this context it is often needed to disambiguate the semantics of the relationships that connect entities, as well as qualify their weight or strength. Relational relations do nothing of this kind. Another drawback of using a \ac{RDBMS} occurs when the amount of data increases significantly, and the structure of the dataset becomes more complex and less uniform. To cope with this, the relation model has to include large join tables, null-checking logic, and sparsely populated rows. Therefore, as the number of relations increases, it degrades the performance of the system and makes it difficult to evolve the existing database in response to changing business needs \cite{book:GraphDatabases}. 

To exemplify that \ac{RDBMS} are not adequate for this type of context, an experiment was conducted \cite{bkchp:connecteddata} using a relational store and Neo4j \cite{artc:neo4j}. The objective of this experiment was to test the execution time of both database systems when querying for friends of friends in a social network context with different levels of depth of friends, to the maximum level of depth of five. Summarizing the experiment, the idea is that given any two persons chosen at random, find a path that connects them which is at most five relationships long.

The conclusion of the experiment is that for a social network containing about 1,000,000 people, each with approximately 50 friends, the results strongly suggest that graph databases are the best choice for connected data, as shown in Table \ref{tbl:RDBMSvsNeo4j}. 
The most interesting fact of this experiment is that at depth four the relational database exhibits crippling latency, making it practically useless for an online system. Neo4j’s timings have deteriorated a little too, but the latency here is at the periphery of being acceptable for a responsive online system. Finally, at depth five, the relational database simply takes too long to complete the query, however Neo4j returns a result in around two seconds. 

 \begin{table}[t]
 \centering
  \begin{tabular}{ c c c c c} \hline
   Depth & \ac{RDBMS} Execution time (s) & Neo4j Execution Time (s) & Records \\\hline
   2 & 0.016 & 0.01 & 2500 \\
   3 & 30.267 & 0.168 & 110,000 \\
   4 & 1543.505 & 1.359 & 600,000 \\
   5 & Unfinished & 2.132 & 800,000 \\ \hline

  \end{tabular} 
 \caption{Finding extended friends in a relational database versus efficient finding in Neo4j. Adapted from \cite{bkchp:connecteddata}.} 
 \label{tbl:RDBMSvsNeo4j} 
\end{table}

\subsection{Graph Databases}

The \ac{RDBMS} systems might not be then the best choice to address the context of large scale phylogenetic analysis challenge. Although graph oriented databases can be of much help in these contexts, as far as it is known there is no solution relying on these technologies to address this challenge. Therefore, the set of graph database systems that were selected for comparison are Neo4j, Titan Aurelius \cite{web:titan}, JanusGraph \cite{web:janusgraph}, Dgraph \cite{web:dgraph}, Allegrograph \cite{web:allegrograph}, and Apache Rya \cite{artc:rya}.

Neo4j is one of the most popular graph database management systems, and it is used in several companies such as Ebay and Microsoft. It is implemented in Java and accessible from software written in other languages using the Cypher query language \cite{artc:cypher}. Titan Aurelius is a scalable graph database optimized for storing and querying graphs. It is built over backend storage frameworks such as Apache HBase \cite{book:hbase} or Apache Cassandra \cite{artc:cassandra}. JanusGraph, which debuted in 2017, is a database optimized for storing and querying large graphs with billions of edges and vertices distributed across a multi-machine cluster. It is based on the source Java code base of the Titan graph database project and is supported by the likes of Google, IBM and the Linux Foundation. Dgraph, which is written in Go, is a distributed graph database offering horizontal scaling and \ac{ACID} properties. It is built to reduce the disk seeks and minimize network usage footprint in cluster scenarios. It automatically moves data to rebalance cluster shards. AllegroGraph is a commercial graph database which supports many programming languages, and is based in uniform machine architectures. Apache Rya is a scalable system for storing and retrieving the \ac{RDF} data in a cluster of nodes. It introduced a new serialization format for storing the \ac{RDF} data, an indexing method to provide fast access to data, and query processing techniques for speeding up the evaluation of SPARQL queries. \ac{RDF} is based on the idea of making statements about resources in the form of \verb|<subject, predicate, object>| expressions called triples.

To better understand which graph databases are best suited for our context, several of their capabilities and features are compared in Table \ref{tbl:dbcomparisons}. These include the underlying storage and processing engine \cite{bkchp:GDinternals}, writing scalability, storage capacity, semantics extensibility, built-in algorithms, visualization tools, \ac{DPT} integration, and source code and activeness.

\begin{table*}[h]
    \tabcolsep=0pt
    \begin{tabular*}{\textwidth}{@{\extracolsep{\fill}}ccccccccccc@{\extracolsep{\fill}}}
        \hline
        Database & Storage & PE & Scalability & Nodes & Extensible & Algorithms & Visualization & DPT & Source & Active \\ \hline
        Neo4j & N & N & C & 34 Billion & Yes & 31 & Yes & Yes & Open & Yes \\
        Titan Aurelius & NN & N & C & $2^{59}$ & No & 0 & No & Yes & Open & No \\
        JanusGraph & NN & N & C & $2^{59}$ & No & 2 & Yes & Yes & Open & Yes \\
        Dgraph & N & N & C & ? & No & 0 & Yes & No & Open & Yes \\
        AllegroGraph & N & NN & SS & 1 Billion & Yes & 0 & No & Yes & Closed & Yes \\
        Apache Rya & NN & NN & C & ? & No & 0 & No & Yes & Open & Yes \\ \hline
    \end{tabular*}
    \begin{tablenotes}
        \item \textbf{Keys:} \textbf{PE} - Processing Engine, \textbf{N} - Native, \textbf{NN} - Non Native, \textbf{SS} - Single Server, \textbf{C} - Cluster
   \end{tablenotes}
   \caption{Databases comparison summary.}
   \label{tbl:dbcomparisons} 
\end{table*}


\paragraph{Underlying Storage and Processing Engine}
The underlying storage can be a native graph storage, that is optimized and designed for storing and managing graphs. Alternatively, it can serialize the graph data into a relational database, an object-oriented database, or some other general-purpose data store. The processing engine can either be native, if the database exposes a graph data model through \ac{CRUD} operations and uses index-free adjacency, or non-native, if it uses global indexes. A database engine that relies on index-free adjacency is one where each node maintains direct references to its adjacent nodes, and each node acts as a micro-index to other nearby nodes, which is much cheaper than using global indexes. It means that query times are independent of the total size of the graph, and are instead proportional to the amount of the graph data searched. A non-native graph database engine, in contrast, uses global indexes to link nodes together. These indexes add a layer of indirection to each traversal, thereby incurring greater computational cost.

Neo4j and Dgraph are built over a native graph storage and native processing engine system. Titan Aurelius and JanusGraph are based on a native graph processing engine, but they are built over backend storage frameworks such as Apache HBase \cite{book:hbase} or Apache Cassandra \cite{artc:cassandra}, therefore not providing a native graph storage. AllegroGraph has a native graph storage and a non-native graph processing engine. Finally, Apache Rya is not based either on a native graph storage or native graph processing engine.

\paragraph{Writing Scalability}
The writing scalability refers to the possibility of the database to scale, depending on the amount of write operations workload that it is subjected to. The appealing form of scalability is horizontal, as it allows to scale by adding more resources to support the database. AllegroGraph is based on a single server architecture, and all the remaining database engines support cluster architectures that allows them to scale horizontally.

\paragraph{Storage Capacity}
Storage capacity is the space that a database provides to store the data, which in this case is organized as nodes and relationships. AllegroGraph allows to store up to 1 billion nodes, while Neo4j supports the storage of 34 billion nodes \cite{web:neo4junlimitednodes}. Titan Aurelius and JanusGraphIt allows to store up to 1 quintillion relationships and half as many nodes. And finally, Dgraph and Apache Rya do not specify a storage capacity limit.

\paragraph{Semantics Extensibility}
Semantics extensibility is the possibility to extend the database functionalities by writing custom code, which can then be invoked. Neo4j and AllegroGraph provide mechanisms that allow the extension of the database semantics, by implementing custom functions or procedures, whereas the remaining database systems do not.

\paragraph{Built-in Algorithms}
Built-in algorithms are the algorithms which are provided by the database to run over the stored data. Neo4j has community-driven libraries available, that combined with the officially available algorithms, provide 31 implemented algorithms to use over the graphs stored in the system. Finally, JanusGraph provides 2 built-in algorithms, while the remaining database engines do not provide any.

\paragraph{Visualization Tools}
Graph visualization tools allow to visualize the data which are stored on the database. Neo4j, JanusGraph and Dgraph allow to integrate visualization tools to visualize the data which is stored in the database. For instance, Neo4j allows to visualize data with tools that connects directly to the database such as Neovis \cite{web:neovis} and Popoto \cite{web:popoto}. The remaining databases do not support the visualization of the data with visualization tools.

\paragraph{DPT Integration}
Data processing tools allow to collect and manipulate data into a desired form. Dgraph does not provide the possibility of integrating the database with a data processing tool. However, all the other databases provide such functionality. For example, Neo4j allows to integrate with Apache Spark \cite{artc:spark}.

\paragraph{Source Code and Activeness}
The source code of a project or tool can either be open or closed source, depending on whether the code is available for the public or not. Activeness stands for the status of a project, which can be active, if it is still in development, or concluded if not. Thereby, Neo4j, JanusGraph, Dgraph, and Apache Rya are open source and active, AllegroGraph is closed source and also active, and Titan Aurelius is open source and concluded. \\

Considering these graph database system comparisons, it can be recognized that Neo4j offers the most interesting set of capabilities and features, and also has an active community. The most interesting set of capabilities and features will be identified and detailed in Section Proposed Solution.

\section{Large Scale Phylogenetic Analysis}\label{workflow}

In large scale phylogenetic analysis projects, such as INNUENDO \cite{artc:innuendo}, the performance of epidemiological surveillance is necessary, with the data generated also being used to study microbial population genetics. In the particular case of INNUENDO, it provides an effective indicator based surveillance system of priority-listed pathogens, which is fundamental in combating and controlling food-born diseases. This system is able to monitor the geographical location, spread, type and genomic variation of pathogens to rapidly detect the emergence of food-borne outbreaks. It is based on a framework that contains an analytical platform and standard procedures for the use of whole-genome sequencing in the surveillance, outbreak detection, and investigation of food-born pathogens in the context of small countries with limited resources.


The analysis of this type of data is based on the need to pass files through a series of transformations, called a pipeline or a scientific workflow. A scientific workflow is a framework which aims to compose and execute a series of computational or data manipulation steps over some data. An example of a workflow system is shown in Figure \ref{fig:workflow}.


The early forms of workflow systems are based on scripts and the \textit{make} utility. However, for phylogenetic analysis it is desired that these frameworks should be able to accommodate production pipelines consisting of both serial and parallel steps, complex dependencies, varied software and data file types, fixed and user-defined parameters and deliverables \cite{artc:workflow}. Thereby, there are many workflow frameworks of different types, that were built specifically to solve some mentioned problems. The INNUENDO project relies on a workflow framework named FlowCraft \cite{web:flowcraft}. FlowCraft is a python engine that automatically builds pipelines by assembling previously made and ready to use components, which are modular pieces of software or scripts. Hence, FlowCraft is used to assemble a set of transformations to perform the study of microbial population genetics. The alignment of genetic sequences, the application of a typing methodology, and finally the execution of a set of phylogenetic inference methods, comprise the typical sequence of steps in microbial population genetics studies. The alignment task deals with small parts of the organism genome, specifically with locus and alleles. Subsequently, by applying a typing methodology several data types arise such as isolates, allelic profiles, and ancillary data. Finally, the last task applies inference and visualization algorithms, which generate distances between allelic profiles, and visualization coordinates for each allelic profile. Each one of these concepts are explained in the following subsections.

\subsection{Phylogenetic Data}

The input of Flowcraft is the organism information that comes from the laboratories. Hence, given such information that comes as biological samples, a \ac{NGS} process is applied to obtain the genetic sequences. Then, alignment tools or assembly tools are executed to assembly the genomes \cite{artc:NGS, artc:NGS2}. This process can be defined as the reconstruction of the organism genome system with random small parts of it to determine the nucleic acid sequence, that is, the order of nucleotides in the \ac{DNA}. It allows to map genomes of new organisms, finish genomes of previously known organisms, or to compare genomes across multiple samples. It can be performed by two different processes, the matching and the assembly. In the matching process, the alignment is accomplished by building the current genome based on a previously built genome of the same organism. In the assembly process, all the needed combinations of the alleles are made to obtain the correct form of the genome. These processes can be compared as building a puzzle, where the matching has the solution for the puzzle available, while the assembly does not.

\paragraph{Locus and Alleles}
The sequences assembled in the alignment process may occupy a given position of a locus and define distinct alleles of that locus. A locus is a specific location in the chromosome, and every unique sequence, either \ac{DNA} or peptide depending on the locus, is defined as a new allele. An allele can also be defined as a viable \ac{DNA} coding sequence for the transmission of traits, and it is represented with a number identifying the allele and string containing the sequence. An example of this data is presented in Figure \ref{fig:alleles}, where the allele with identifier 1 belongs to locus \verb|aroe|, and the sequence of the allele is the following string with the nucleic acid sequence, which is a succession of letters that indicate the order of nucleotides. It is expected that, the alleles are represented through files which follow the FASTA format. FASTA is a text-based format for representing either nucleotide sequences or amino acid sequences, in which nucleotides or amino acids are represented using single letter codes. This format is represented in Figure \ref{fig:alleles}.

\begin{figure}[!ht]
\centering
\begin{minipage}{.65\textwidth}
\begin{verbatim}
>aroE_1
GAAGCGAGTGACTTGGCAGAAACAGTGGCCAATATTCGTCGCTACCAGATGTTTGGCATC
GCGCGCTTGATTGGTGCGGTTAATACGGTTGTCAATGAGAATGGCAATTTAATTGGATAT
CTAGACAAGTTACAGGAGCAGACAGGCTTTAAAGTGGATTTGTGT
>aroE_2
GAACCGAGTGACTTGGCAGAAACAGTGGCCAATATTCGTCGCTACCAGATGTTTGGCATC
GCGCGCTTGATTGGTGCGGTTAATACGGTTGTCAATGAGAATGGCAATTTAATTGGATAT
CTAGACAAGTTACAGGAGCAGACAGGTTTTAAAGTGGATTTGTGT
\end{verbatim}
\end{minipage}
\caption{Example of alleles.}
\label{fig:alleles}
\end{figure}

\subsection{Microbial Typing}

After the alignment phase, a typing methodology is applied to identify or fingerprint each organism based on the genes that are presented in almost all organisms, which are named as conserved genes. Bacterial identification and characterization at subspecies level is commonly known as microbial typing. This process provides the means to execute phylogenetic inference methods, which then produces a hypothesis about the history of the evolutionary relationships about a group of organisms. There are several typing methodologies, such as the \ac{MLST} \cite{artc:MLST, artc:MLST2}, \ac{MLVA} \cite{artc:mlva}, and \ac{SNP} \cite{artc:snp}. These are available for a multitude of bacterial species, and are being used globally in epidemiological microbial typing and bacterial population studies.

\paragraph{Isolate}
The main goal of the typing methods is the characterization of organisms existing in a given sample. However, some microorganisms from the sample collected need to be isolated to be characterized. Thus, each organism isolated from the microbial population becomes an isolate.

\paragraph{Ancillary Data}
An isolate can be associated with typing information and ancillary details. Ancillary details include information about the place where the microorganism was isolated, the environment, the host, and other possible contextual details. These details are usually represented through \ac{CSV} formatted files, where the tab character is used to separate values. An example of this data is shown in Figure \ref{fig:ancillary}, which shows that the ancillary detail line with id 1 is related to the allelic profile with id 1, and is composed by some information that was extracted from the isolate with the identifier \verb|AU2523|. This information include the specie, that is \verb|A.denitrificans|, and the isolation localization, which was in North America, more precisely in USA.

\begin{figure}[!ht]
    \centering
    \begin{tabular}{c c c c c c c}
       id & ST & isolate & species & country & continent \\
       1 & 1 & AU2523 & A.denitrificans & USA & North America \\
       2 & 2 & AU8059 & A.denitrificans & Unknown & \\
       3 & 3 & AU8060 & A.denitrificans & France & Europe\\
       4 & 4 & AU8080 & A.insolitus & USA & North America\\
       5 & 5 & ACH26 & A.insolitus & USA & North America\\
    \end{tabular}
    \caption{Example of ancillary details.} 
    \label{fig:ancillary} 
\end{figure}

\paragraph{Allelic Profile}
One of the most popular methodologies is the \ac{MLST}, which is an unambiguous procedure for characterizing isolates of bacterial species using the sequences of internal fragments. This methodology types several species of microorganisms, and when applied, the set of alleles identified at the \textit{loci} are considered to define a \ac{ST}, a key identifier for this methodology, that can also be defined as an allelic profile. The chosen \textit{loci} are usually different for each species, although some species may share some or even all \textit{loci} in their \ac{MLST} schemas. The number of chosen \textit{loci} can vary and be greater or smaller than the seven \textit{loci} more commonly adopted. The generated sequences are compared to an allele database and for each gene, the different sequences are assigned as distinct alleles and, for each isolate, the alleles at each of the \textit{loci} define the allelic profile, also known as \ac{ST} in this methodology. These allelic profiles are represented through files which follow a tabular format. Usually, it is a delimited text file that uses a tab character to separate values. The \ac{MLVA} methodology follows a similar concept, in the sense that by applying the methodology it is obtained sequences which are identified by some identifier, and it has several columns representing each element that characterizes the sequence. The \ac{SNP} methodology differs from the latter, since the data which is originated by it follows a sequential format, where each sequence is normally constituted by an identifier and a set of numbers. A value of 0 in any locus represents the allele that was mostly found on that locus, while a 1 represents any other possible allele.

Examples of several allelic profiles, created by applying the \ac{MLST} and \ac{SNP} methodologies, are presented in Figure \ref{fig:profiles} and in Figure \ref{fig:snp} respectively. In Figure \ref{fig:profiles} it can be interpreted that the profile with \ac{ST} identifier 1 is identified by the \textit{loci} \verb|nusA|, \verb|rpoB|, \verb|eno|, \verb|gltB|, \verb|lepA|, \verb|nuoL|, and \verb|aroe|. The number below of each locus is the allele identifier, which ends up mapping to the different alleles, that were described before. For example the locus \verb|aroe|, with the allele identifier with value 1 maps for the first entry of the Figure \ref{fig:alleles}. A \ac{ST} having the locus \verb|aroe| with the allele identifier with value 2 maps to the second entry. \ac{MLVA} allelic profiles would have a similar representation. 

\begin{figure}[!ht]
 \centering
 \begin{tabular}{ l l l l l l l l}
   ST & nusA & rpoB & eno & gltB & lepA & nuoL & aroe \\
   1 & 1 & 26 & 2 & 2 & 59 & 8 & 1 \\
   2 & 1 & 26 & 2 & 4 & 59 & 2 & 1 \\
   3 & 1 & 26 & 2 & 2 & 62 & 8 & 2 \\
   4 & 1 & 26 & 7 & 2 & 59 & 3 & 2 \\
   5 & 1 & 27 & 1 & 1 & 62 & 9 & 1 \\
 \end{tabular} 
 \caption{Example of \ac{MLST} profiles.}
 \label{fig:profiles}
\end{figure}

In Figure \ref{fig:snp} it can be interpreted that the profile with identifier 1 is characterized by the following sequence, which contains a set of 0's and 1's that represent if this profile is characterized by the alleles that were the mostly found on that locus.

\begin{figure}[!ht]
\centering
\begin{minipage}{.65\textwidth}
\begin{verbatim}
1 0100000111101010001000101010101001010100011101011000101010
2 1111010010101100100101010101001000001010010001010101010100
\end{verbatim}
\end{minipage}
\caption{Example of \ac{SNP} profiles.}
\label{fig:snp}
\end{figure}

\subsection{Algorithms}

Succeeding the typing process follows the execution of a phylogenetic inference method to the results. A phylogenetic inference method is the application of computational algorithms, methods, and programs to phylogenetic data that allows to produce a diagrammatic hypothesis about the history of the evolutionary relationships of a group of organisms. There are several types of phylogenetic inference methods, such as distance matrix methods, maximum parsimony, maximum likelihood, and Bayesian inference. However, in INNUENDO the inference methods which are applied are based on distance matrix methods \cite{artc:phyloviz, artc:phyloviz2.0, artc:10.1093/bib/bbaa147}.

\paragraph{Distances}
Distance matrix methods rely on the genetic distance between the sequences being classified. The distances are often defined as the fraction of mismatches at aligned positions, with gaps either ignored or counted as mismatches. An example of a distance matrix, resultant from applying the hamming distance between each profile represented in Figure \ref{fig:profiles}, is shown in Figure \ref{fig:matrix}.

\begin{figure}[!ht]
$$D = 
\begin{bmatrix}
0 & 2 & 2 & 3 & 5 \\
2 & 0 & 4 & 4 & 5 \\
2 & 4 & 0 & 3 & 5 \\
3 & 4 & 3 & 0 & 6 \\
5 & 5 & 5 & 6 & 0 \\
\end{bmatrix}
$$
\caption{Example of a distance matrix.}
\label{fig:matrix}
\end{figure}

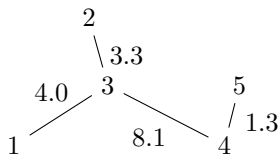
\begin{figure}[!ht]
 \centering
  \begin{tikzpicture}
	\node (a) at (0, 3) {1};
    		\node (b) at (1, 4.7) {2};
    		\node (c) at (1.25, 3.8) {3};
    		\node (d) at (2.8, 3) {4};
    		\node (e) at (3.0, 3.8) {5};
        	
    		\node (1) at (0.5, 3.7) {4.0};	
    		\node (2) at (1.5, 4.2) {3.3};	
    		\node (3) at (1.8, 3.1) {8.1};	
    		\node (7) at (3.3, 3.3) {1.3};
    		
    		\draw (a)-\\(c);
    		\draw (b)-\\(c);
    		\draw (c)-\\(d);
    		\draw (d)-\\(e);
		\end{tikzpicture}
  \captionof{figure}{Example of an evolutionary hypothesis.}
  \label{fig:rooted}
\end{figure}

\paragraph{Inference Algorithms}
An inference algorithm is then executed to compute the diagrammatic hypothesis based on the distance matrix previously calculated. The diagrammatic hypothesis calculated comes in the form of a graph or a tree, where each node represents the allelic profile, and the relationships between them are quantified by the distances. Figure \ref{fig:rooted} is an example of such evolutionary hypothesis. The semantics of these relationships depends on the algorithm used. The \ac{goeBURST} algorithm \cite{artc:goeburst} is an example of an implementation of these algorithms. It is a globally optimized implementation of the \ac{eBURST} algorithm \cite{artc:eburst} that identifies alternative patterns of descent for several bacterial species. This algorithm can be stated as ﬁnding the maximum weight forest or the minimum spanning tree, depending on weight definition. 

The \ac{goeBURST} algorithm uses the Kruskal algorithm, described in Algorithm \ref{alg:kruskal}, to achieve its goals. It relies on this algorithm since it can obtain the optimal tree, with respect to the deﬁned partial order on the set of relationships between organisms. However, in case of a tie when comparing the weight of the relations, it should consider the number of relations at distance one, two and three, the occurrence frequency of the relation, and lastly the sequence type identifier.

When executing many inference algorithms over the same allelic profiles, the concept of multilayer networks, which consists of using the same nodes to represent different graphs, becomes appealing. That is, due to the several algorithm outputs being expressed over the same context, it becomes easier to compare them. For example, if the \ac{goeBURST} algorithm and any other algorithm are run, then the nodes of both resulting graphs would be the same, but for each algorithm it would produce different relationships.

\begin{algorithm}[!ht]
 \textbf{Input:} A graph ${G}$ whose edges have distinct weights. \\
 \textbf{Output:} ${F}$ is the minimum spanning forest of ${G}$.\\
 
	\textbf{Initialization:} Create a graph ${F}$ (a set of trees), where each vertex in the graph is a separate tree. Create a set ${S}$ containing all the edges in the graph. 
	\hfill \\

\textbf{Loop:} While ${S}$ is not empty and ${F}$ is not yet spanning do:
 
	\begin{enumerate}
		\item Remove an edge with minimum weight from ${S}$. 
		\item If the removed edge connects two different trees then add it to the forest F, combining two trees into a single tree.
	\end{enumerate}
	
 \textbf{Finalize:} Return the hierarchy ${F}$.
	\caption{ Kruskal algorithm.}
	\label{alg:kruskal}
\end{algorithm}

\paragraph{Visualization Algorithms}
After executing an inference algorithm, a visualization algorithm, such as Radial Static Layout \cite{book:radial}, which is described in Algorithm \ref{alg:radial}, or GrapeTree \cite{artc:grapetree}, is executed to compute the optimal coordinates for each node of the received graph or tree. Afterwards, the coordinates are provided to a render framework which then presents each profile and relationship to a user interface. The Radial Static Layout algorithm is a method of displaying tree structures in a way that expands outwards, radially. It is one of many ways to visually display a tree and can be used when the size of the node is not considered. 

\begin{algorithm}[!ht]
    \textbf{Input:} ${T = (V, E, \delta)}$. \\
    \textbf{Output:} Coordinates ${x}$, ${y}$: ${V}$ ${\rightarrow}$ ${R^+}$ for the nodes.\\
    \textbf{Data:} Queue ${Q}$, leafcount: ${V}$ ${\rightarrow}$ ${N^+}$ \{ from a previous postorder traversal \}. \\
 
	\textbf{Initialization:} ${r \leftarrow}$ root(${T}$); rightborder(${r}$) ${\leftarrow}$ 0; wedgesize(${r}$) ${\leftarrow}$ 2${\pi}$; ${x}$(${r}$) ${\leftarrow}$ ${y}$(${r}$) ${\leftarrow}$ 0. \\

\textbf{Loop:} While !${Q}$.empty() do:
 
	\begin{enumerate}
		\item ${v}$ ${r \leftarrow}$ ${Q}$.delete.first().
		\item ${\eta}$ ${\leftarrow}$ rightborder(${v}$).
		\item Loop: For Each child ${w}$ of ${v}$ do: 	    
		    \begin{enumerate}
    		    \item ${Q}$.insert(${w}$)
    		    \item rightborder(${w}$) ${\leftarrow}$ ${\eta}$.
    		    \item wedgesize(w) ${\leftarrow}$ 2${\pi\times}$leafcount(${w}$)) ${\div}$ leafcount(${r}$)
    		    \item ${\alpha}$ ${\leftarrow}$ rightborder(${w}$) + wedgesize(${w}$) ${\div}$ 2)
    		    \item ${\alpha}$ ${\leftarrow}$ rightborder(${w}$) + wedgesize(${w}$) ${\div}$ 2)
    		    \item ${x}$(${w}$) ${\leftarrow}$ ${x}$(${v}$) + cos(${\alpha}$) ${\times}$ ${\delta}$((${v, w}$)); ${y}$(${w}$) ${\leftarrow}$ ${y}$(${v}$) + sin(${\alpha}$) ${\times}$ ${\delta}$((${v, w}$))
    		    \item ${\eta}$ ${\leftarrow}$ ${\eta}$ + wedgesize(w) 
        \end{enumerate}
    \end{enumerate}
	\caption{Radial Static Layout algorithm.}
	\label{alg:radial}
\end{algorithm}

\section{Discussion}

This chapter provided an overview of the graph databases, the phylogenetic analysis process and the types of data related to it. It started by presenting the definition of a graph database, and several use cases where a graph database can be useful. Additionally, it discusses the problems which may arise of utilizing a \ac{RDBMS}, to manage data that can be represented by graphs, by analysing an experiment. To understand which graph database could be suitable, several graph database engines were compared. Then, the large scale phylogenetic analysis process is presented emphasizing the phylogenetic data and algorithms that are used by it. There are several types of data such as, the alleles, the sequence allelic profiles, the isolates, the ancillary data, the relations between each profile, and the optimal coordinates for the visualization rendering. Some of these data can be represented by dataset files of different formats, such as \ac{MLST}, \ac{MLVA} or \ac{SNP} data. Since graph databases have been applied in contexts where there is highly connected data, it should be possible to use a graph database to store this types of data.

\chapter{Proposed Solution}\label{cpt:proposedSolution}

This chapter defines the functional and non functional requirements, the uses cases, the architecture, and the technologies that shall compose the proposed solution.

The proposed solution relies on a framework that allows to store and manage the data resulting from the phylogenetic analysis in a graph database, execute inference and visualization algorithms over such data, and more. There are several uses cases for a framework like this, for instance the possibility to query data and to load datasets. To provide functionalities like those, the framework is composed by different components. These components are described with several architectural views, such as a data model, client server, layered and decomposition. These views are complemented with the respective reasoning. At last, these components must rely on some technology to be executed. Hence, the technologies that shall support the different components of the framework are specified.

\section{Functional Requirements}

Functional requirements state what the system must do, and how it must behave or react to. The functional requirements that were identified for the framework of large scale phylogenetic analysis are as follows:

\begin{itemize}
    \item Store in a database the data resulting from the phylogenetic analysis considering multilayer networks;
    \item Development of a framework which allows to perform queries over the data stored in the database;
    \item Loading of several datasets with different formats, such as \ac{CSV} and FASTA, into the database;
    \item Execution of inference and visualization algorithms, like goeBURST and Radial Static Layout respectively, over the data stored in the database;
    \item Support of authentication and authorization.
\end{itemize}

\section{Non Functional Requirements}

Non-functional requirements or quality attributes requirements are qualifications of the functional requirements or of the overall product. The non-functional requirements that were identified for the framework of large scale phylogenetic analysis are as follows:

\begin{itemize}
    \item The framework should be complemented with documentation that provides different types of views.
    \item The framework code should be modular, in a way that it allows to easily reuse and extend the code.
\end{itemize}

\section{Use Cases}

\begin{figure}[!ht]
  \centering
  \includegraphics[scale=0.28]{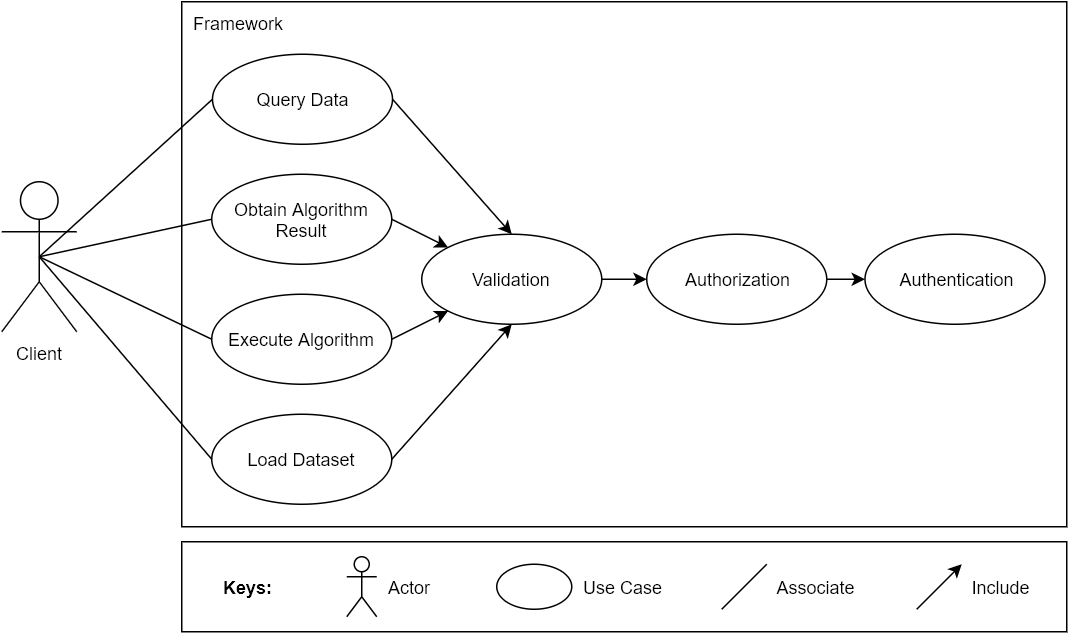}
  \caption{Use cases.}
  \label{fig:usecase}
\end{figure}

A use case is a written description of how users will perform tasks on a system. It outlines, from the user perspective, the behaviour of the system as it responds to a request. The use cases identified shall allow to reason about who are the users of the framework, their objectives, the actions they are able to perform, and how the framework shall respond to each action. The framework should allow its users, which are other applications, to query data, load datasets, execute algorithms, and obtain results. Each of these operations start by verifying if the request is authenticated, the permissions of the user, and the request. Each use case is represented in Figure \ref{fig:usecase}

\paragraph{Query Data} The framework should allow its users to query any type of data that is stored on the database. For example, it should be possible to find all isolates for which it already exists typing information under a specific typing method schema. Thus, the user should perform the several requests, that identifies the resources to obtain. Once the request is received by the framework it should return the data gathered from the database. The several types of data may include profiles, isolates, and others.

\paragraph{Load Dataset}
The framework should allow its users to load several datasets into the database. For example, it should be possible to load a file containing the profiles of a taxonomic unit. Thus, the user should perform a request sending the dataset to be loaded. Once the request is received by the framework, it should parse the data into database entities, store them, and return a code identifying if the operation was successful. The datasets may come from several file formats, such as FASTA, \ac{CSV}, and others.

\paragraph{Execute Algorithm}
The framework should allow its users to run an algorithm, either for inference or visualization, over the data that is stored in the database. For example, it should be possible to run the goeBURST algorithm to calculate the minimum spanning tree, given a set of profiles, and to run the Radial Static Layout algorithm to calculate the respective visualization coordinates. Thus, the user should perform a request identifying which algorithm it wants to run. Once the request is received by the framework, it should schedule the execution of the algorithm and return a code identifying if the operation was scheduled.

\paragraph{Obtain Algorithm Result}
The framework should allow its users to obtain the resulting graph from an algorithm execution, after it has been completed. Thus, the user should perform a request identifying which algorithm execution result to retrieve. Once the request is received by the framework, it should return the result stored in the database.

\section{Architecture}

The software architecture of a system is the set of structures needed to reason about the system, which comprise software elements, relations among them, and properties of both. There are several views to document an architecture. Their purpose is to provide insight of how the system is structured as a set of implementation units, in contrast to others that aim to understand how the system is structured as a set of elements that have run-time behaviour and interactions \cite{book:architecture}. Thus, the architecture of the proposed framework should be represented by several views. However, only the most relevant and generic shall be presented in this document, noting that all the others are available in the project repository.

\subsection{Data Model}\label{datamodel}

\begin{figure}[!ht]
 \centering
 \includegraphics[scale=0.49]{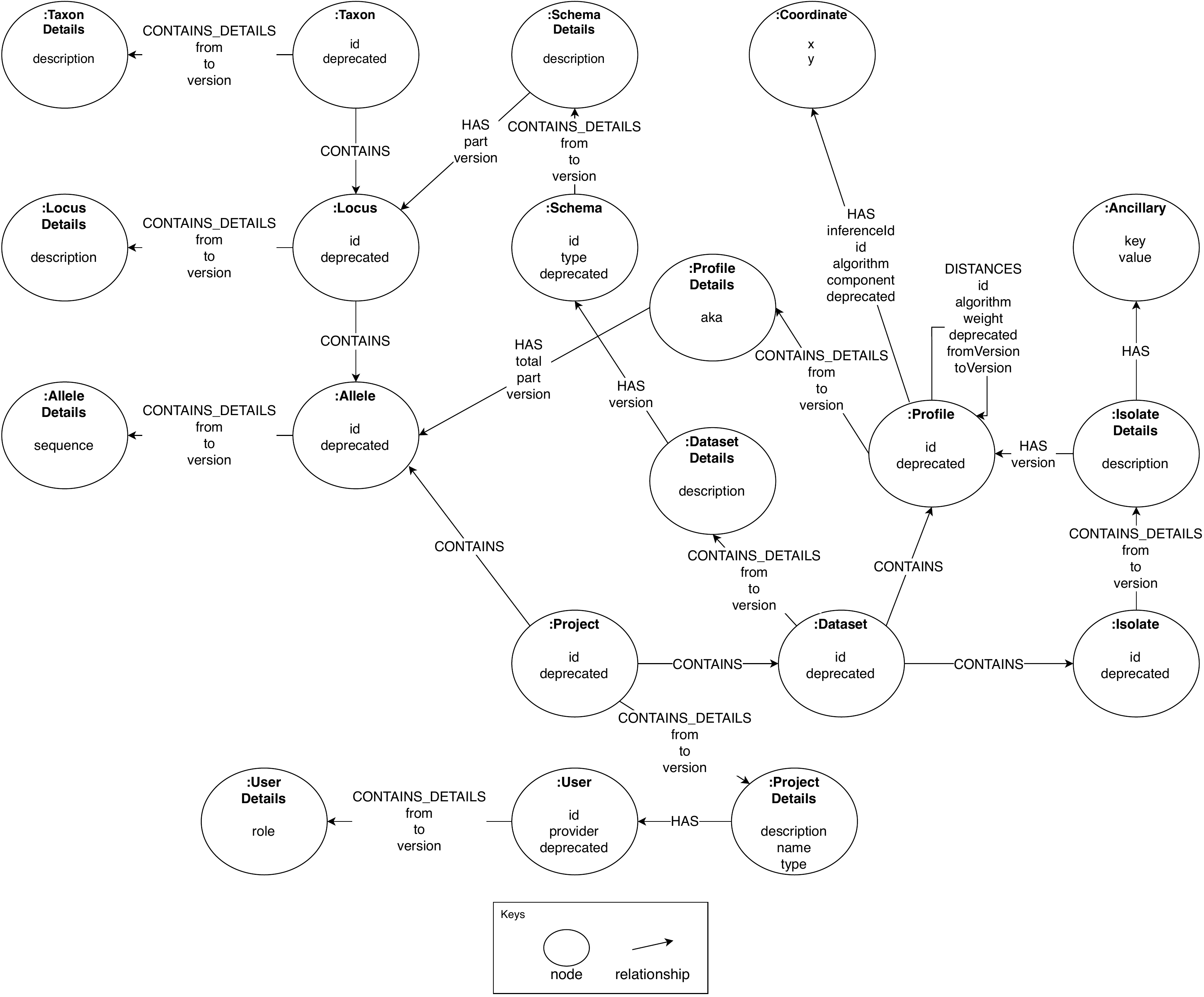}
 \caption{Graph data model architectural view.}
 \label{fig:datamodel}
\end{figure}

The data model style \cite{bkchp:datamodel} describes the information structure in terms of data entities and their relationships. It is used to perform impact analysis of changes, to enforce data quality by avoiding redundancy and inconsistency, and to guide the implementation of modules that access the data. It should be designed as a graph, since the domain of large scale phylogenetic analysis is composed of highly connected data. Therefore, the data model is described as a connected graph of nodes and relationships, where the nodes represent domain entities, and the relationships represent how the different nodes relate. By using a graph data model extensibility should be achieved by itself. That is, it should be possible to extend the system just by adding a set of nodes and relationships. However, it must be ensured that the domain is not violated, since the graph data model does not impose a strict schema. The data model of this project is shown in Figure \ref{fig:datamodel}.

This data model incorporates versioning and soft deletes concerns to mitigate some problems that occur nowadays, such as the impossibility to delete a wrongly inserted profile after executing an analysis that generates a graph containing it. Such profiles can not be removed, because the generated graphs would then become invalid. In this case, by considering a versioning and soft delete strategy, these removals should be possible, since the graphs would be linked to the statuses of the profiles and not to the profiles themselves.

The versioning strategy to achieve this behaviour is to separate each object from its state, link them through a relationship with the respective version number, and capture changes by having different state nodes \cite{web:neo4jversioning}. In the data model, the name of the status nodes shall end with \verb|Details|, and the version relationships shall be named as \verb|CONTAINS_DETAILS|. For example, the allele node is connected to the respective version by the \verb|CONTAINS_DETAILS| relationship between the \verb|Allele| and the \verb|Allele Details| nodes. If the allele is modified, then a new \verb|Allele Details| node is connected to the \verb|Allele| node, while all profiles that were related to that allele version are still valid.

\begin{figure}[!ht]
 \centering
 \includegraphics[scale=0.65]{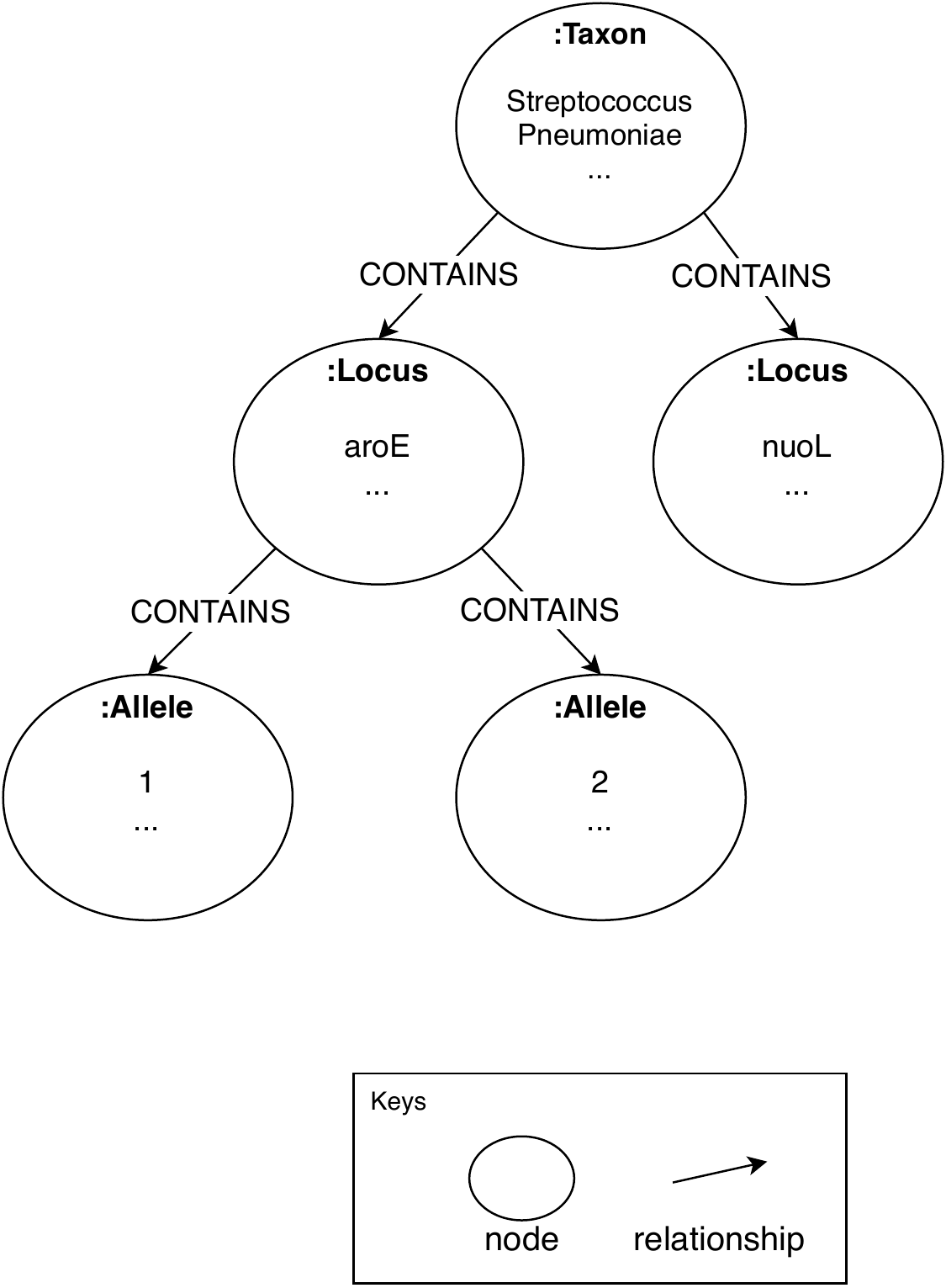}
 \caption{Example representation of phylogenetic data.}
 \label{fig:phylogenetic}
\end{figure}

Phylogenetic data are composed of taxonomic units, \textit{loci}, and alleles. Taxonomic units consist of several \textit{loci}, thus this is represented by a relationship named \verb|CONTAINS| between taxonomic units and \textit{loci} nodes. For instance, to represent the taxonomic unit \verb|Streptococcus pneumoniae| which contains several \textit{loci}, such as \verb|aroE| and others, the \verb|Streptococcus pneumoniae| node would have a \verb|CONTAINS| relationship to \verb|aroE| node. \textit{Loci} may hold specific locations for a set of alleles, as shown in Figure \ref{fig:alleles}. Therefore, to represent this relationship in the data model, a relationship named \verb|CONTAINS| is used between the nodes of each locus and the associated nodes of the alleles. For instance, to express the Figure \ref{fig:alleles} in the data model, the locus \verb|aroE| would have relationships \verb|CONTAINS| to two allele nodes. In Figure \ref{fig:phylogenetic} a simplified example of this case is presented.

The typing methods execution relies on and generates different types of data \cite{artc:10.1007/978-3-642-28062-7_3, artc:relations}, such as typing schemas, allelic profiles, isolates and ancillary data, as explained before. 

\begin{figure}[!ht]
 \centering
 \includegraphics[scale=0.65]{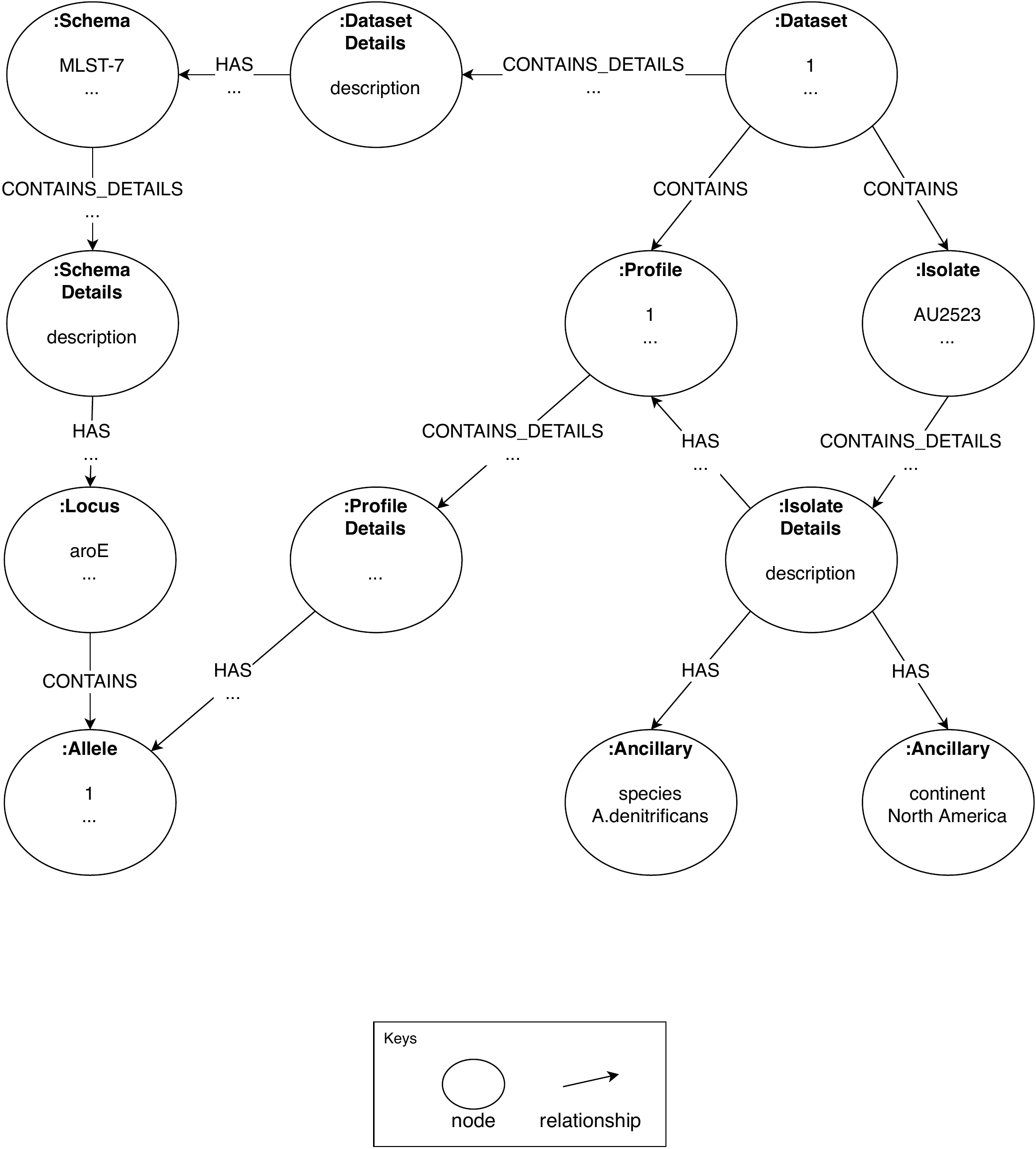}
 \caption{Example representation of typing data.}
 \label{fig:typing}
\end{figure}

Typing schemas can use several \textit{loci} to characterize different allelic profiles, and in the data model this is expressed by the relationships \verb|HAS| between the details of a schema node and the respective \textit{loci} nodes. For instance, to describe the \ac{MLST} schema used in Figure \ref{fig:profiles}, the details of the schema node should contain relationships \verb|HAS| to the \textit{loci} \verb|nusA|, \verb|rpoB|, \verb|eno|, \verb|gltB|, \verb|lepA|, \verb|nuoL|, and \verb|aroe|. Figure \ref{fig:typing} presents an example dataset that follows this \ac{MLST} schema. However, only the relationships to the \verb|aroe| locus are shown to keep the figure simple, but in the complete scenario it would have a relationship to each of the mentioned locus.

Allelic profiles belong to a specific dataset as they are a result of applying a typing methodology. Thus, this is described by the relationship \verb|CONTAINS| between dataset and profile nodes. These profiles follow the same schema, hence they should be related to the typing method used. To impose this concern, the \verb|HAS| relationship is used between the dataset and schema nodes, which means that all profiles from that dataset follow the related schema. For example, the details node of a dataset containing the profiles shown in Figure \ref{fig:profiles}, would have a relationship named \verb|HAS| to a \ac{MLST} schema node. However, having the dataset connected to the schema, only allows to perceive what \textit{loci} were used in the typing operation.

Therefore, to know what is the allele that characterizes a profile for each locus used in the schema, the details node of a profile must be connected to the respective allele nodes. Hence, this is represented in the data model by using a relationship called \verb|HAS| between the profile details and the alleles nodes. For instance, in Figure \ref{fig:profiles} the profile with \ac{ST} value 1 is represented by the allele with identifier 1 of locus \verb|nusA|, consequently that profile should have a detail with a relationship called \verb|HAS| to that allele. This example is also presented in Figure \ref{fig:typing}.

Isolates may have related ancillary data, as it is represented in Figure \ref{fig:ancillary}. Thus, in the data model this is expressed as a relationship named \verb|HAS| between the detail of an isolate and ancillary data nodes. Since an isolate may be associated to a profile, there is also a relation between the two, which is called \verb|HAS|. For example, to represent the associations between profiles, isolates, and ancillary data, shown in Figure \ref{fig:ancillary}, the detail of the isolate with name \verb|U2523| would have a relationship called \verb|HAS| to the profile with id 1. The detail of this isolate would also have several relationships \verb|HAS| to each ancillary data associated to it. Figure \ref{fig:typing} also represents a simplified variation of this example.

Graphs and trees are generated by performing analysis over the datasets generated by the typing process. These analyses are based on the execution of inference and visualization algorithms. 

\begin{figure}[!ht]
 \centering
 \includegraphics[scale=0.65]{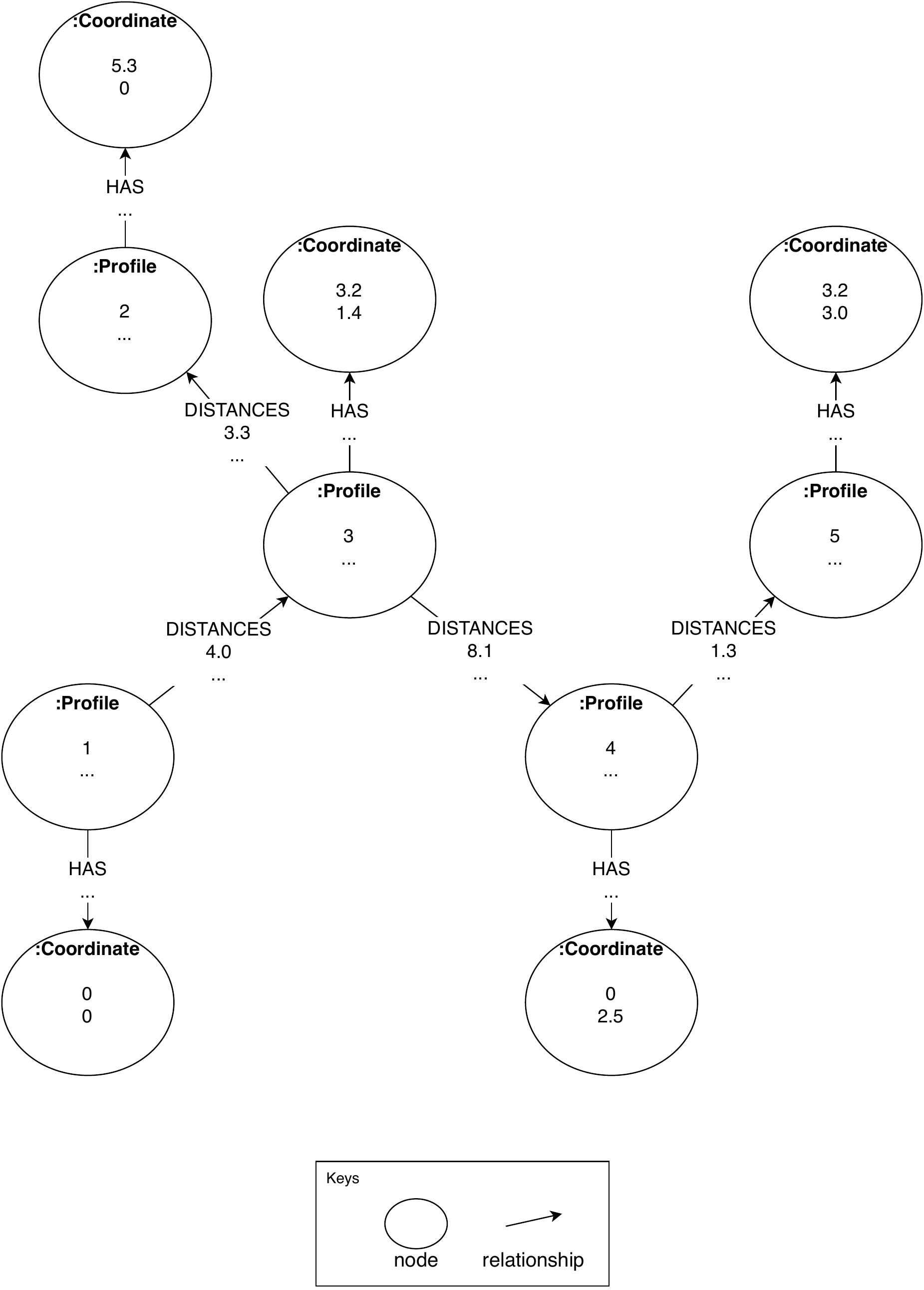}
 \caption{Example representation of algorithms data.}
 \label{fig:algorithms}
\end{figure}

The inference algorithms rely on the genetic distances between profiles. These distances are calculated by computing a distance matrix, as shown in Figure \ref{fig:matrix}. Based on these distances, the algorithm is then executed and relationships \verb|DISTANCES| are created between the different profiles composing the graph. For example, to describe the result of an inference algorithm shown in Figure \ref{fig:rooted}, the several profiles that were used in the execution of the algorithm would have relationships \verb|DISTANCES| between them, quantified by the respective weights. This example is also presented in Figure \ref{fig:algorithms}, in a simplified way. This strategy allows to  consider multilayer networks since the same nodes shall be used to represent different graphs.

Visualization algorithms execute over the graphs resulting from the inference algorithms, and create visualization coordinates for each node of the graphs. Hence, a relationship \verb|HAS| between a profile and coordinate nodes exists to represent the coordinate of some profile, for a given inference and visualization algorithm. Figure \ref{fig:algorithms} also represents an example of the coordinates for the nodes of a given graph.

Users and projects are concepts introduced to handle the authorization concern. The datasets of the profiles may belong only to some users, hence this access restriction should be expressed in the data model. Thus, a user may participate in several projects, and a project may contain a dataset of profiles and several algorithm executions. By restricting the access to a project, the ancillary details, and the results of the algorithms are also restricted. The data model embodies these concerns by using the relationships \verb|HAS| between project and user nodes, and \verb|CONTAINS| between project and dataset nodes.

\subsection{Client Server}\label{clientserver}

\begin{figure}[!ht]
 \centering
 \includegraphics[scale=0.80]{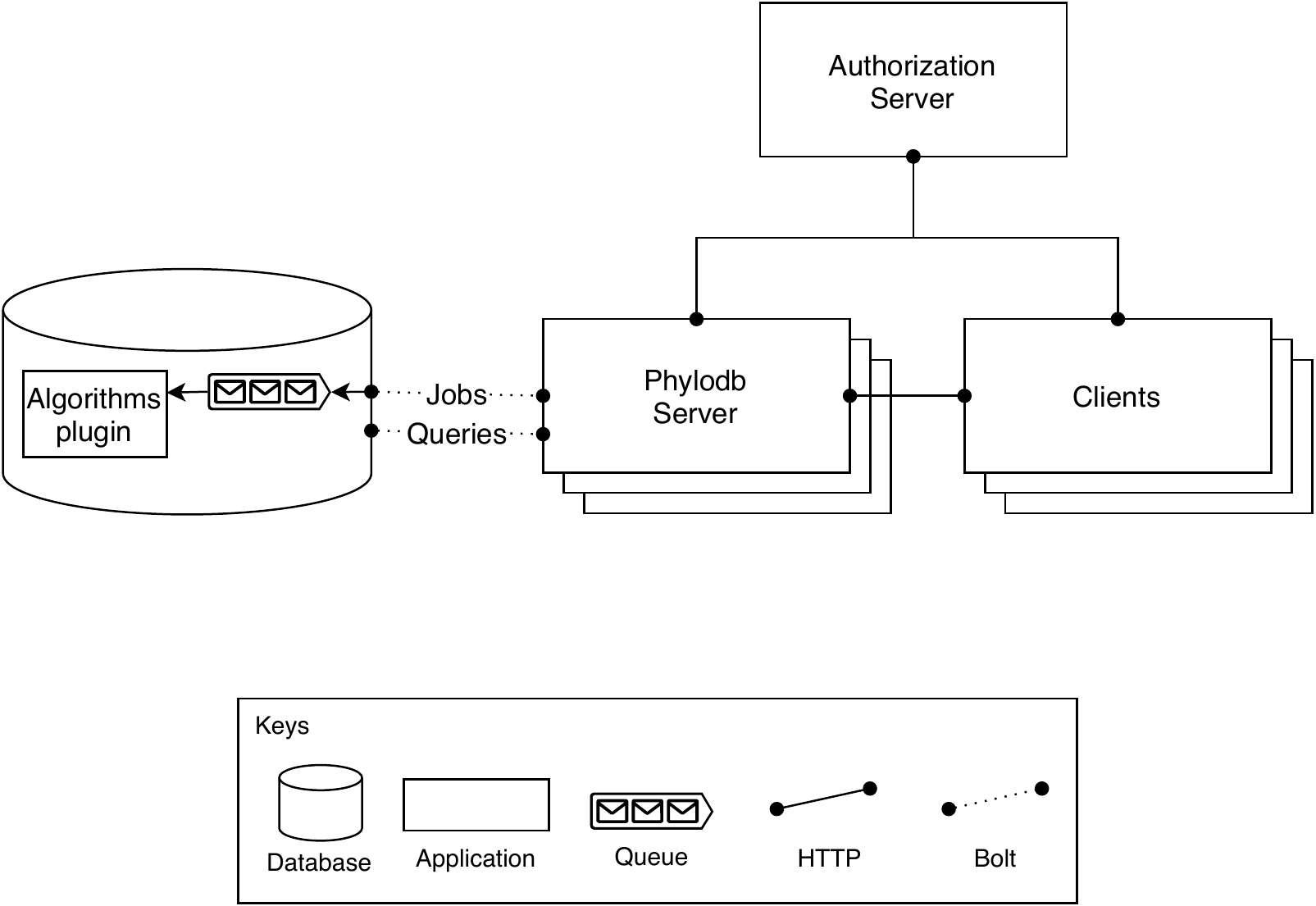}
 \caption{Client-Server architectural view.}
 \label{fig:clientserverview}
\end{figure}

The client-server style is used for analysing the modifiability, reusability, scalability and availability of the solution, by factoring out common services, to better understand how the several units of computation interact between themselves. The client is a component that invokes services of a server component, and the server is a component that provides services to the client components. The client-server view for this framework is presented in Figure \ref{fig:clientserverview}.

The \verb|Phylodb Server| component provides a web \ac{API}, to perform several operations over the data stored in the database, namely access data, load datasets, execute algorithms, and obtain results. It should be possible to scale horizontally to handle more operations by adding more instances of the \verb|Phylodb Server| component. The communication between the users and the \verb|Phylodb Server| component should be done over the \ac{HTTP} protocol. 

The \verb|Phylodb Server| interacts with the database component in two different ways. That is, it can normally query data, but it can also queue executions of the algorithms that are deployed in the database and reside in the \verb|Algorithms Plugin| component. These algorithms also read the needed inputs and write the computed results. The communication between the database and the \verb|Phylodb Server| component shall be done over the Bolt protocol \cite{web:bolt}. 

The \verb|Authorization Server| component manages the user information, and provides operations to perform the authentication of a user. It needs to receive the authentication requests over the \ac{HTTP} protocol.

\subsection{API}\label{api}

\begin{figure}[!ht]
 \centering
 \includegraphics[scale=0.87]{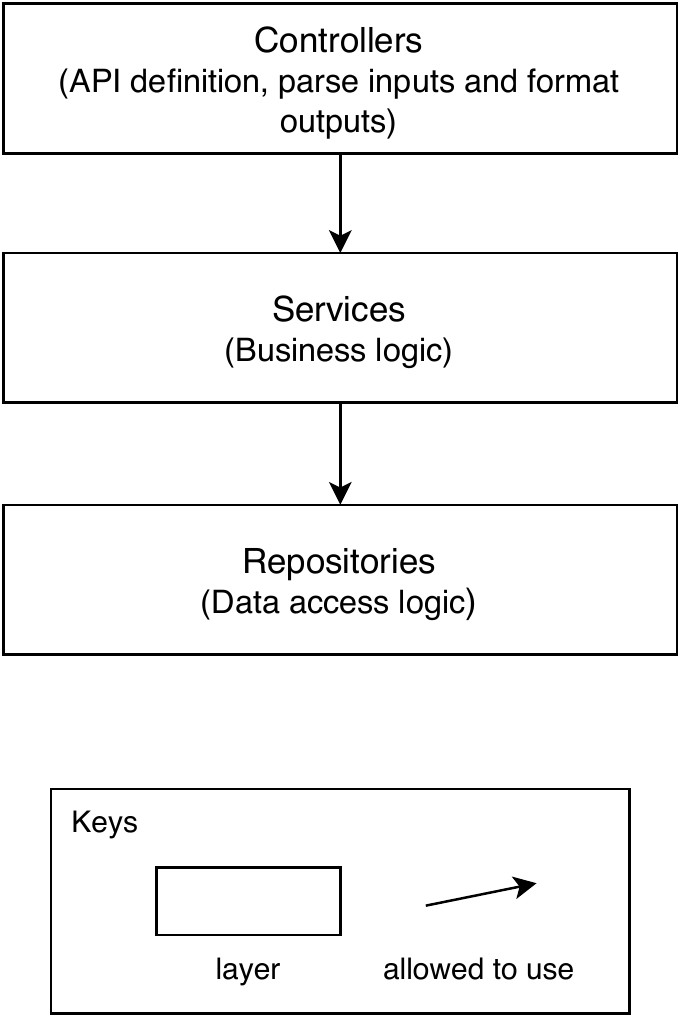}
 \caption{Layered architectural view.}
 \label{fig:apilayeredview}
\end{figure}

The \ac{API} structure should be based on three layers, namely \verb|Controllers|, \verb|Services| and \verb|Repositories|, as represented in Figure \ref{fig:apilayeredview}. This figure represents a layered view that puts together layers, which are groupings of modules that offer a cohesive set of services, in a unidirectional allowed-to-use relation with each other. 

When a request is received by the \ac{API}, it is passed through the \verb|Controllers| layer. This layer contains the controllers that parse the received input, execute the respective service, and retrieve the response containing the respective status code and the formatted content. The \verb|Services| layer contains the services that perform the business logic and use the needed repositories. The \verb|Repositories| layer holds the repositories that shall provide operations to interact with the database. Apart from these layers, there is a validation logic that verifies the request authenticity and the user permissions before the request is processed by them.

\begin{figure}[!ht]
 \centering
 \includegraphics[scale=0.85]{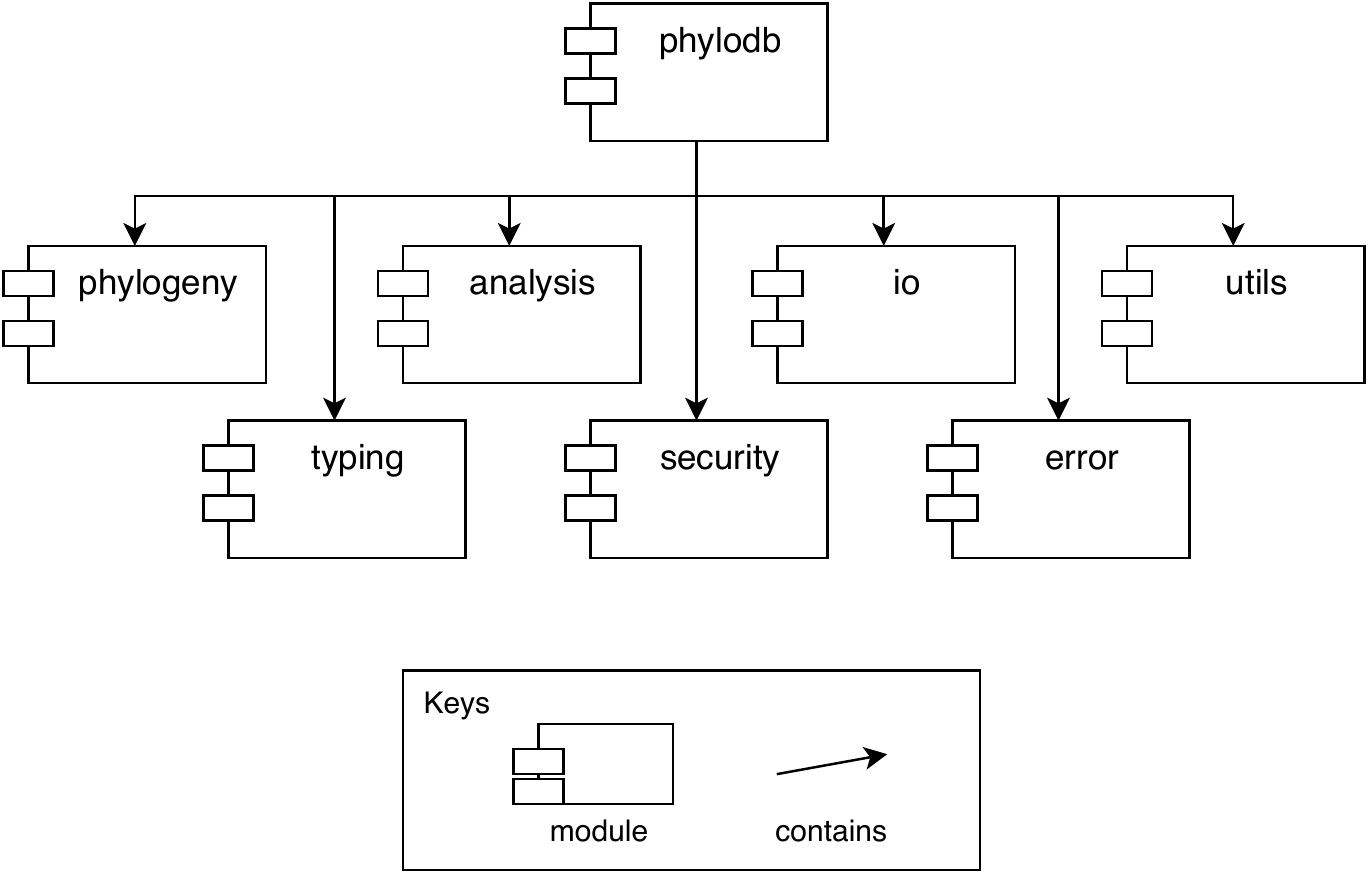}
 \caption{Decomposition architectural view.}
 \label{fig:apidecompositionview}
\end{figure}

Another approach for decomposing the structure of the \ac{API}, is to separate it by modules. This approach can be visualized by relying in a decomposition view. This view is used for decomposing a system into units of implementation, which allows to describe the organization of the code as modules, to better understand how the system responsibilities are partitioned across them, and to reason about the location of changes. The modules that shall compose the structure of the \ac{API} are the \verb|phylogeny|, \verb|typing|, \verb|analysis|, \verb|security|, \verb|io|, \verb|error|, and \verb|utils|. These modules are represented in Figure \ref{fig:apidecompositionview}. 

The \verb|phylogeny|, \verb|typing|, and \verb|analysis| modules contain the main operations to handle the different types of data. The \verb|phylogeny| module should combine the logic of managing taxonomic units, \textit{loci} and alleles. The \verb|typing| module shall contain the operations to administer datasets, profiles, isolates, and schemas. The \verb|analysis| intends to contain the operations to manage the graphs and coordinates, resulting from the execution of inference and visualization algorithms. Each of these modules follows the layered structure previously explained, that is, each module contains the respective controller, service and repository.

The \verb|security|, \verb|io|, \verb|error|, and \verb|utils| modules represent the remaining concerns, which are important to the overall functioning of the \ac{API}. The \verb|security| module aims to hold the logic of managing users and projects, and contains the authentication and authorization concerns. The \verb|io| module shall aggregate the input parsing and output formatting logic. The \verb|error| module should provide operations to perform error handling. Finally, the purpose of the \verb|utils| module shall be to provide common operations which do not belong to a specific module and can be used by several modules.

\begin{figure}[!ht]
 \centering
 \includegraphics[scale=0.85]{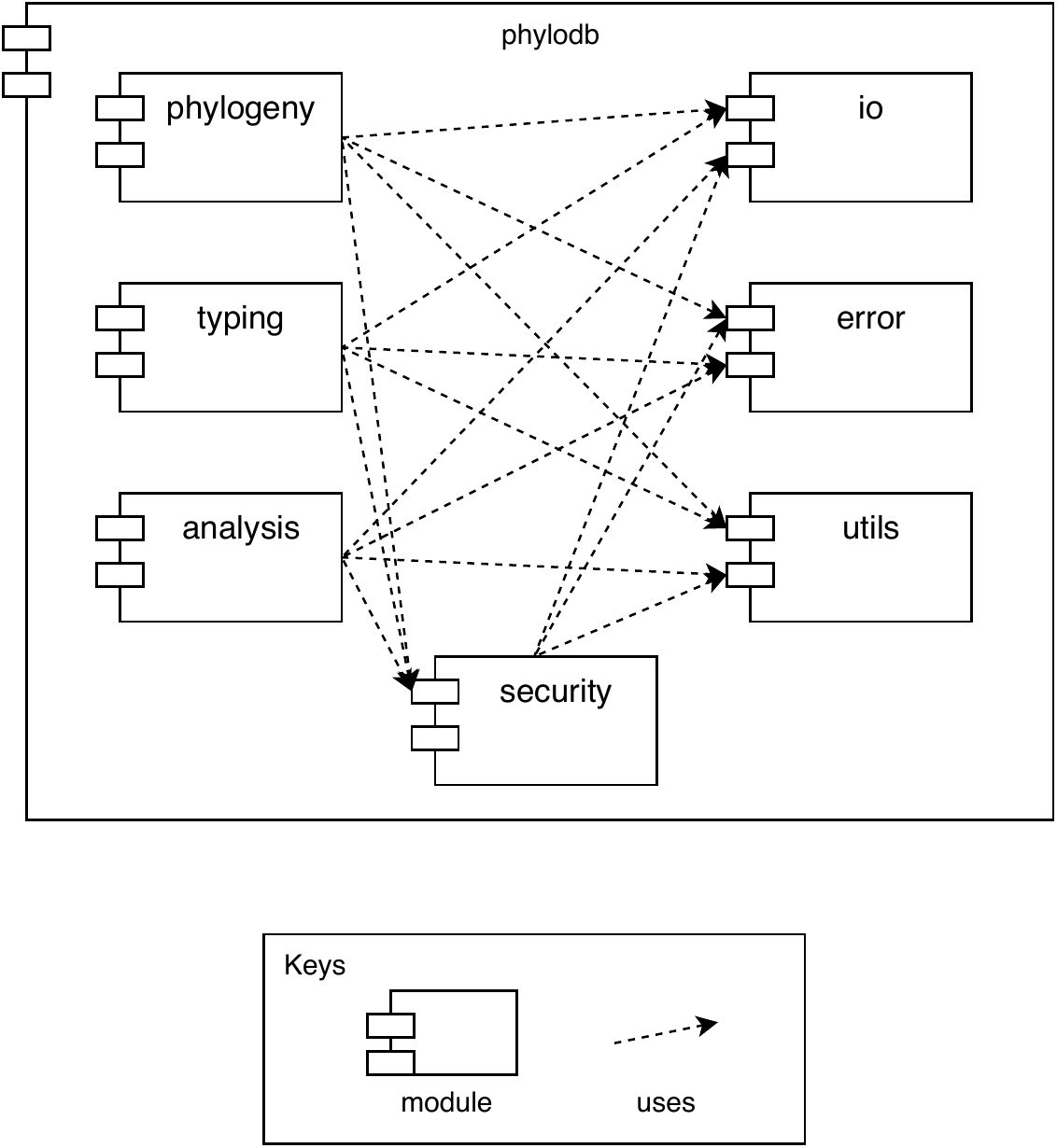}
 \caption{Uses architectural view.}
 \label{fig:apiusesview}
\end{figure}

Since the structure is decomposed in several modules, it is also possible to understand how they interact with each other by analysing an uses view. This view shows how modules depend on each other and which modules can be affected by changes in other modules. This view is represented in Figure \ref{fig:apiusesview}.

It shows that the modules \verb|phylogeny|, \verb|typing|, and \verb|analysis| should rely on the \verb|security| module, which allows them to define the roles and permissions of the users that can perform their respective operations. They shall also depend on the \verb|io| module to parse and format their respective inputs and outputs. And they intend to use the common operations provided by the \verb|utils| module. The \verb|security| module shall depend on \verb|io| and \verb|utils| modules for the same reasons. Finally, the \verb|io| module and \verb|utils| modules shall not rely on other modules. It can be concluded that if changes are made in the \verb|phylogeny|, \verb|typing|, or \verb|analysis| modules, they do not influence any other module. However, if a change is conducted in any of the other modules then the \verb|phylogeny|, \verb|typing|, \verb|analysis|, and \verb|security| modules might need to be reviewed.

\subsection{Plugin}\label{datamodel}

\begin{figure}[!ht]
 \centering
 \includegraphics[scale=0.87]{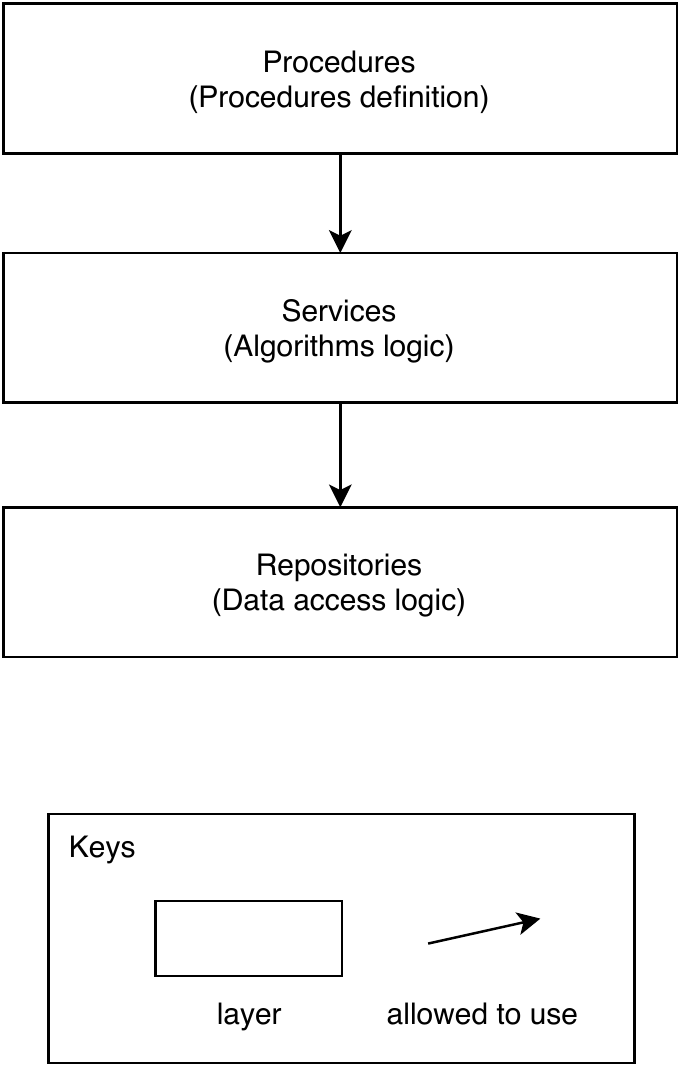}
 \caption{Layered architectural view.}
 \label{fig:pluginlayeredview}
\end{figure}

The structure of the plugin can also be based on three layers, namely \verb|Procedures|, \verb|Services| and \verb|Repositories| as demonstrated in the layered view represented by Figure \ref{fig:pluginlayeredview}. This view implies that the modules within the \verb|Procedures| layer are allowed to use the modules belonging to the \verb|Services| layer, and so on, but the opposite is not allowed. The \verb|Procedures| layer intends to hold the definitions of the operations that allow to execute the supported algorithms, hence a call to an algorithm is directed to them. Their implementation shall parse the received input and execute the respective service provided by the \verb|Services| layer. The \verb|Services| layer shall read the input data for the algorithm from the database, compute the respective algorithm and store the obtained result back to the database. The reading and writing of data is accomplished by using the methods provided by the \verb|Repositories| layer. 

\begin{figure}[!ht]
 \centering
 \includegraphics[scale=0.87]{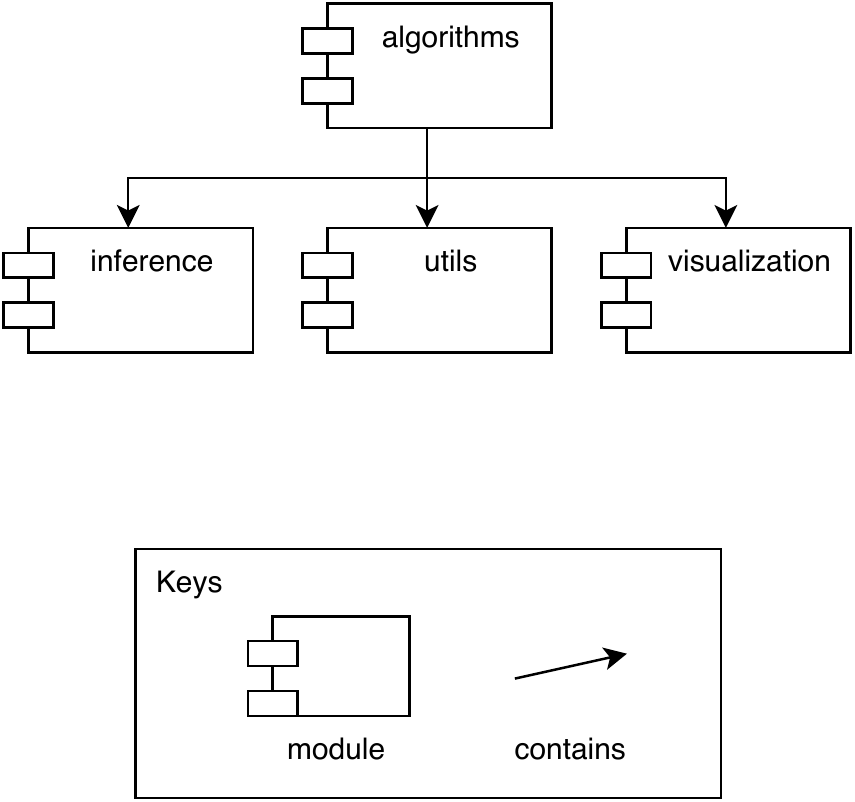}
 \caption{Decomposition architectural view.}
 \label{fig:plugindecompositionview}
\end{figure}

Such structure is constituted by several modules and it can be further decomposed into the \verb|inference|, \verb|visualization|, and \verb|utils| modules. These are shown in the decomposition view presented in Figure \ref{fig:plugindecompositionview}. The \verb|inference| module should contain the inference algorithms and hold the operations to retrieve and store data, related to an inference, from the database. The \verb|visualization| module is similar, in the way that it aims to support the visualization algorithms and hold the operations to retrieve and store data, related to a visualization, from the database. Finally, the \verb|utils| module shall provide common operations which do not belong to a specific module and can be used by different modules. Each module, except for the \verb|utils| module, follow the layered structure previously explained, that is, each module contains the respective procedure, service and repository.

The interaction between these modules can be understood by analysing the uses view represented by Figure \ref{fig:apiusesview}. In this view, it can be recognized that the \verb|inference| and \verb|visualization| modules rely only on themselves and on the common operations provided by the \verb|utils| module. Hence, a change can be conducted without the need to review the other module, since they do not interact with each other. Additionally, the \verb|utils| module shall not rely on any other module.

\begin{figure}[!ht]
 \centering
 \includegraphics[scale=0.87]{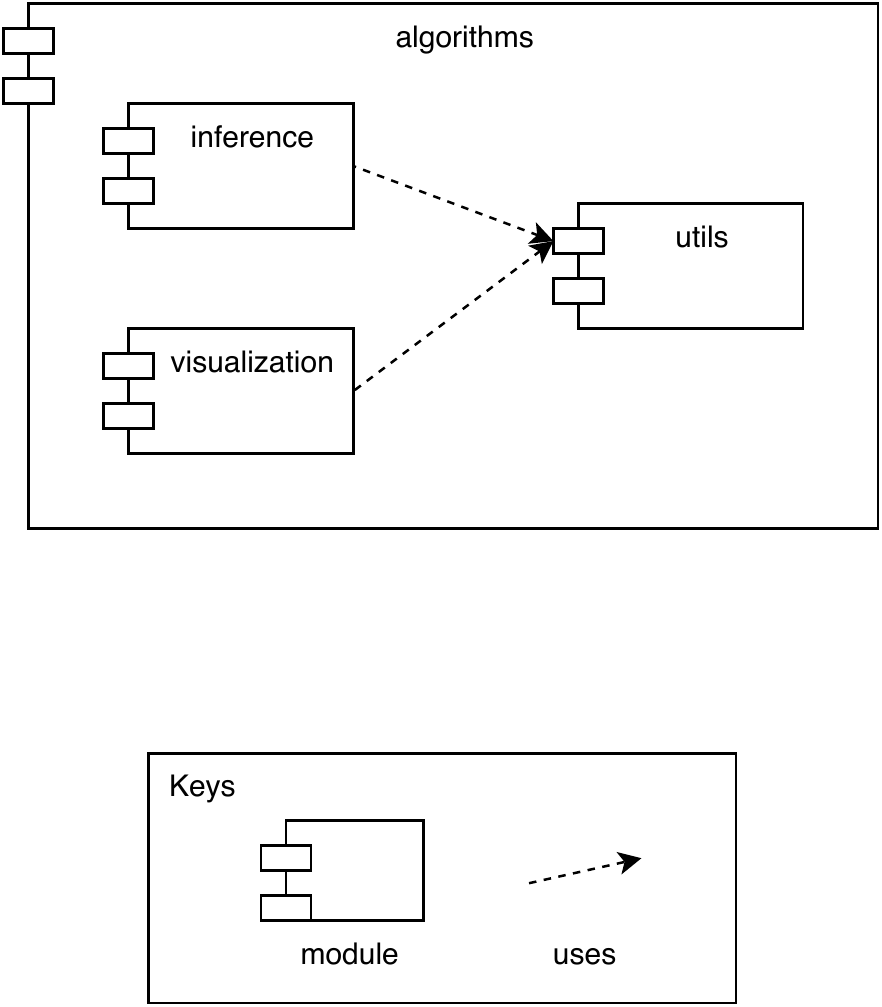}
 \caption{Uses architectural view.}
 \label{fig:pluginusesview}
\end{figure}

\section{Technology}\label{technology}

The technologies that shall support the framework are the Spring Framework \cite{artc:spring}, Google Identity Provider \cite{web:googleidp}, and Neo4j. These technologies will be further explained ahead, and it will be specified what component of the framework they shall support.

\subsection{Spring Framework}\label{springframework}

The Spring Framework is an application framework and inversion of control container for the Java platform. The appealing features of Spring are that it allows to develop applications of different types and provides appealing \ac{AOP} features to deal with cross-cutting concerns \cite{web:spring}. Furthermore, a library which connects the Spring system to the Neo4j database already exists. This library benefit from the \ac{AOP} framework of Spring to make the access to the database simpler. For instance, instead of managing the transactions manually, spring offers the annotation \verb|@Transactional|, which specifies that the method annotated with it must be run in a transactional environment. Finally, the extensions for the query language of Neo4j and most of the already implemented phylogenetic inference algorithms are implemented in the Java language. Considering these facts, the Spring Framework shall support the application server component of the framework.

\subsection{Google Identity Provider}\label{googleauth}

An identity provider is a system that vouches for the identity of a user. The identity provider role is to authenticate a user and provide an authentication token to the service provider. The OAuth 2.0 \ac{API}s from Google, which conform to the OpenID Connect specification \cite{web:oidc}, will be used to authenticate the users. Hence, the authorization server of the framework will rely on Google Identity Provider, which shall authenticate the client applications by any of the existing flows \cite{artc:rfcoauth2.0}. Once the authentication process is completed, an identity token should be granted to the client application. It must then be sent within the request to the application server, so it can authenticate the request. 

\begin{figure}[!b]
 \centering
 \includegraphics[scale=0.25]{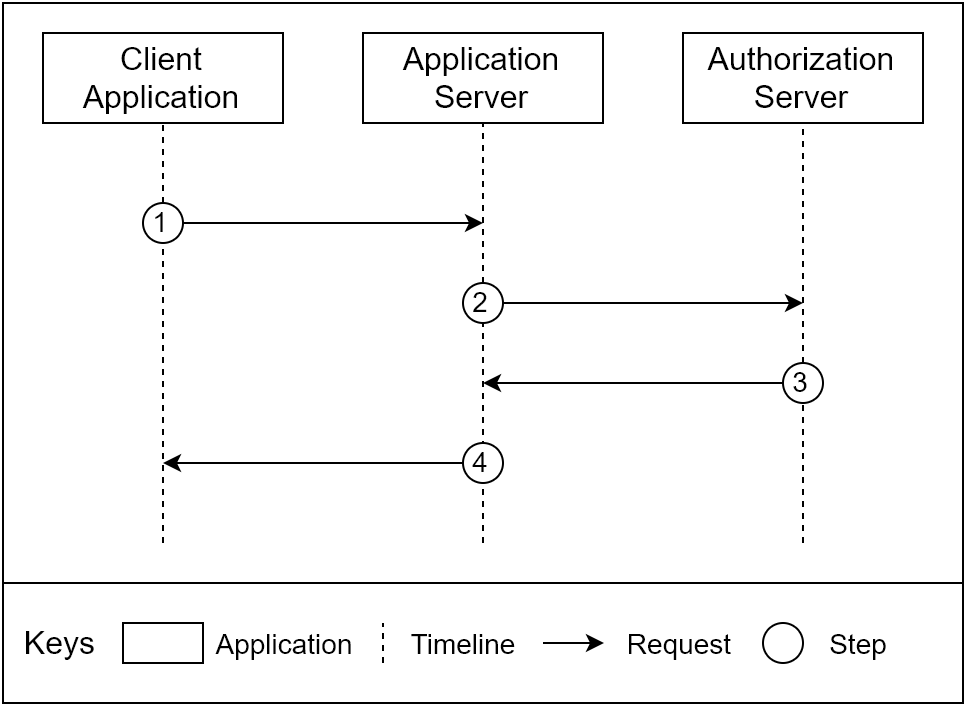}
 \caption{Authentication process using an identity provider.}
 \label{fig:authentication}
\end{figure}

An example of the authentication process is shown in Figure \ref{fig:authentication}, where there are three main components, namely the client application, the application server and the authorization server. It is assumed that the client application already obtained the identity token, thus step 1 starts by sending a request to the application server to perform a given operation. Once the application server receives it, the step 2 begins by sending a request to the authorization server to validate the token. Afterwards, in step 3 the authorization server validates the token and sends a response to the application server. Finally, in step 4, the application proceeds with the operation, and sends a response to the client.

\subsection{Neo4j}\label{neo4jdatabase}

Considering the graph database system comparisons made previously, it can be recognized that Neo4j offers the most interesting set of capabilities and features, and also has an active community. The most interesting features being the fact that it uses Cypher as its query language and is able to extend the query language semantics by implementing custom procedures. As well as the fact that it allows scheduling query executions over the database, and provides a library of graph algorithms, utility functions, and trigger mechanisms. It also allows to perform data sharding.

\paragraph{Cypher}
An effort was started to standardize Cypher as the query language for graph processing. It allows to store and retrieve data from the graph database, and the syntax of Cypher provides a visual and logical way to match patterns of nodes and relationships in the graph. It is a declarative, \ac{SQL}-inspired language for describing visual patterns in graphs and allows to state what to select, insert, update, or delete from the graphs without a description of exactly how to do it. Through Cypher, it is possible to construct expressive and efficient queries to handle the needed create, read, update, and delete functionalities.

\paragraph{Apoc}
The library of graph algorithms, which Neo4j yields \cite{manl:neo4jgraphmanual}, is the \ac{APOC}. It provides algorithms with different focus such as, community detection, centrality, path finding, similarity, and link prediction. These algorithms are presented respectively in Table \ref{tbl:communitydetectionalgorithms}, \ref{tbl:centralityalgorithms}, \ref{tbl:pathfindingalgorithms}, \ref{tbl:similarityalgorithms}, and \ref{tbl:linkpredictionalgorithms}, in the appendices. Besides algorithms, the library contains a collection of functions and procedures, that cover collective operations such as sorting, graph operations, text searches, conversions, geospatial operations, data integration, and reporting.

\paragraph{Triggers}
The \ac{APOC} library provides some background operations, such as trigger mechanisms. These trigger mechanisms allow to register Cypher statements that are executed when data in database are changed. For instance, it is possible to run a trigger before or after events such as when data are created, updated, or deleted. The trigger mechanism is an appealing feature to achieve a kind of incremental computation when executing the algorithms.

\paragraph{Scheduler}
The \ac{APOC} library also provides other type of background operations, which are the background jobs. They allow procedures to run in the background or asynchronously, and execute procedures periodically. The background jobs mechanism is based in a Java scheduler, which relies on a queue, to post the tasks to execute, and on a thread pool, where the threads will execute the tasks that are on the queue. This feature is appealing since some algorithms may take several minutes to finish executing.

\paragraph{Custom Procedures}
A user defined procedure or custom procedure, is a mechanism that enables to extend the Neo4j semantics by writing custom code, which can be invoked directly from Cypher. Procedures can take arguments, perform operations on the database, and return results. These procedures are implemented through Java plugin projects, that after compiled, must be placed in the Neo4j database instance, where it is needed to run the procedure.

\paragraph{Sharding}
Data sharding allows to divide data into horizontal partitions that are organized into various instances or servers, most often to spread load across multiple access points. These shards can be accessed individually or aggregated to see all of the data, when required. This is particularly interesting when the quantity of data is becoming large enough that it makes sense to divide the data into smaller graphs to run on smaller sized hardware and be accessed by necessary parties. 

\section{Discussion}

This chapter defines the proposed solution, which outlines the development of a framework that conforms to the phylogenetic analysis. This framework shall allow to manage the data resulting from the phylogenetic analysis in a graph database, execute inference and visualization algorithms, retrieve their results, and load datasets of different formats. Hence, the framework shall be composed of a web \ac{API}, a graph database with a plugin to support the algorithms and an authorization server. There are different types of architectural views to describe such components, namely data model, client server, layered, decomposition and uses. Each of these architectural views is complemented with the respective explanation. 

The components of the framework shall rely on the Spring Framework, the Google Identity Provider, and Neo4j technologies. The Spring Framework shall support the web \ac{API}, since it has some appealing features, such as providing aspect oriented programming features to deal with cross-cutting concerns. The Google Identity Provider shall handle the authentication of the framework clients. It allows to perform authentication conforming to the OpenID Connect specification. Finally, Neo4j will support the database, as it offers the most interesting set of capabilities and features, and also has an active community. Some of these features include the custom procedures, the \ac{APOC} library and the scheduler mechanism.
\chapter{Implementation}

This chapter describes and reasons about the implementation details of the framework presented in Chapter \ref{cpt:proposedSolution}. It starts by describing what are the components of the solution. Then, it explains the details of the \ac{API} component, such as the interface, the parsing and formatting of files, the error handling, and the security concerns. Finally, it describes details of the plugin component, such as the algorithms. Besides these details, it explains how the respective architecture for each component was materialized, and presents several exemplifications of the different concerns that are described throughout this chapter.

The solution was implemented considering an agile methodology and is publicly available at \url{https://github.com/Brunovski/phyloDB} along with its issues, milestones, and documentation. For testing purposes, the solution is hosted in a server as a Docker image.

\section{Structure}

This implementation relies on three components, namely the \ac{API}, the plugin, and the database. The \ac{API} receives requests and processes them into queries that are executed over the database. The plugin holds the inference and visualization algorithms and allows the \ac{API} to execute them over the data stored in the database. The database contains the phylogenetic data, which is stored as a set of nodes and edges. A database schema is not implied by Neo4j, hence the database schema is defined by the queries performed through the operations provided by the \ac{API}. 

\section{API}\label{api}

The \ac{API} provided by the framework was implemented considering the \ac{REST} architecture, which is based on statelessness. That is, the server should not store any state about the client session on the server-side. Each request from the client to the server must contain all the information necessary to interpret the request. Therefore, the session state must be kept entirely on the client. This architecture has many advantages, such as the possibility to scale the \ac{API} to several instances by deploying it to multiple servers, since any server can handle any request, as they do not possess any session related dependencies.

This \ac{API} relies on the Spring Framework, which allows to use the \ac{IoC} features that it provides. This principle has several advantages, such as allowing to have a greater modularity and facilitating the change between different implementations. The \ac{IoC} pattern used in the \ac{API} is the dependency injection, which supports setting the dependencies of the objects. That is, the act of injecting objects into other objects is done by an assembler rather than by the objects themselves. 

the \verb|Component| annotation and its implementations were used to achieve the dependency injection. This annotation is used to denote classes as components, which then Spring uses to auto detect them for dependency injection. Besides the \verb|Component| annotation Spring offers more specific annotations, such as the \verb|Controller| annotation, that is used to define controllers, the \verb|Service| annotation, which allows to depict classes that hold business logic, and the \verb|Repository| annotation, that is utilized to describe data access classes. These annotations are useful in the implementation of the framework, since they follow the same paradigm as the proposed architecture, which is specifically presented by Figure \ref{fig:apilayeredview}.

\subsection{Interface}

The interaction between the \ac{API} and users is accomplished through \ac{HTTP} requests. Thus, the \ac{HTTP} protocol allows the \ac{API} to retrieve different types of responses depending on several factors, such as the parameters used in the request, the type and result of the operation that is being executed.

The \ac{API} relies on models to parse the input which comes within the body of the requests. These models are defined by the \verb|InputModel| interface. It defines the \verb|toDomainEntity| method, which intends to parse the \ac{JSON} object contained in the body of the request. That is, the input models are used to parse the input data from the requests into domain entities. For instance, the \verb|ProjectInputModel|, which extends from \verb|InputModel| and implements the respective method to retrieve a \verb|Project|, would be used to parse the data contained in the \verb|--data-raw| section of the following request. This request is an example of a curl request that is used to create a project.

\begin{verbatim}
    curl --location --request POST 'http://localhost:8080/projects?provider=google'
        --header 'Content-Type: application/json'
        --header 'Authorization: Bearer {Access Token}'
        --data-raw '{
            "name": "Example Project",
            "visibility": "private",
            "description": "Example project",
            "users": [{"email": "example1@gmail.com", "provider": "google"}, ...]
        }'
\end{verbatim}

Likewise, the \ac{API} also relies on a different type of models to format the domain entities into output objects. These models are defined by the \verb|OutputModel| interface. It outlines the \verb|toResponseEntity| method, that aims to format a domain entity into an output object, which is then retrieved in a response to the user. For example, the \verb|ProjectOutputModel| extends from \verb|OutputModel| and implements the respective method to retrieve a response containing the \verb|200 OK| status code and an object with some of the properties of the respective project. Thus, the database schema is not exposed since these output models contain only the data that is meant to be retrieved.

The status code of the responses, which contains these output models, are defined conforming to the \ac{HTTP} protocol. That is, the status codes depend on the type and result of the respective operations. For example, the \verb|200 OK| status code is used for operations that shall retrieve representations of domain entities, the \verb|201 CREATED| status code is used for operations that create a domain entity and retrieve the generated identifier, and the \verb|204 NO CONTENT| status code is used for operations which do not retrieve anything. While the \verb|400 BAD REQUEST| status code is used to indicate that the request is not valid due to some input, and the \verb|401 UNAUTHORIZED| is utilized for indicating that the request is not valid due to lack of authentication or permissions. Another example is the \verb|500 INTERNAL SERVER ERROR| which is used to signal that an internal error occurred while processing a request.

The mime type for the responses containing these output models is always the \verb|application/json|, since these output models only allow to format objects or lists of objects. However, this \ac{API} also supports the parsing and formatting of more complex data types, namely files, that can have several formats. Thus, the file parsing and formatting will be further detailed ahead. However, the mime type used to retrieve a list of domain entities formatted in a file format is the \verb|text/plain|.

\subsection{Formatters}

Imports and exports of large quantities of data are based on files. There are different formats of files depending on the type of data that they hold. Thus, the interface \verb|Formatter| defines methods to support the imports and exports operations, namely the \verb|parse| and \verb|format| methods. The \verb|parse| method iterates over the lines of a file and calls another method to translate them into domain entities. The latter must be implemented by the inheriting classes. The \verb|format| method iterates over a list of domain entities to transform them into a \verb|String| representation formatted in a specific file format.

There are many file formats used in the phylogenetic analysis. These formats are used to represent alleles, profiles, isolates, and inferences. The alleles can be formatted in the FASTA format. The profiles can be represented in the \ac{MLST}, \ac{MLVA} or \ac{SNP} formats. The \ac{MLST} and \ac{MLVA} formats are very similar, that is, the only difference is the number of columns used to represent the profiles. The isolates follow a tabular format. And the inferences can be represented in Newick or Nexus.

\begin{figure}[!ht]
 \centering
 \includegraphics[scale=0.85]{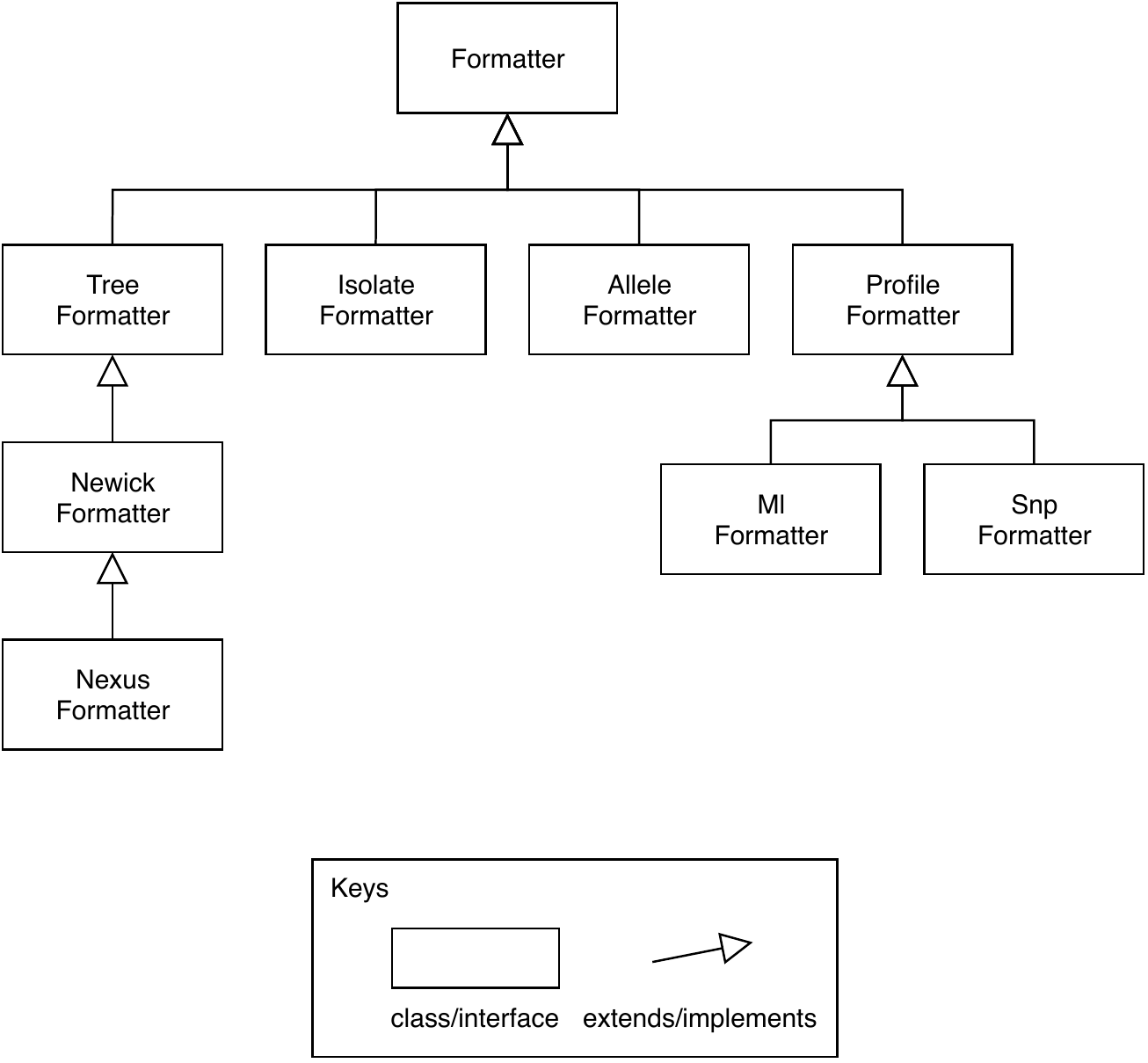}
 \caption{Formatter generalization architectural view.}
 \label{fig:apigeneralazationformatters}
\end{figure}

A class that extends from the \verb|Formatter| interface exists for each of these formats. Each of these classes implements the inherited methods according to the format that they represent. For instance, the \verb|AlleleFormatter|, which extends from \verb|Formatter|, implements the method used by the \verb|parse| method to allow the parsing of alleles from a FASTA file, and the \verb|format| to represent alleles in a FASTA formatted \verb|String|. This structure of classes is presented in Figure \ref{fig:apigeneralazationformatters}.

These formatters can rely on certain settings, that are defined in the \verb|application.properties| configuration file, namely the missing symbols to be considered in some formats and the maximum length of the lines to be formatted in the FASTA format. These settings are stored in the configuration file, so they can be modified without having to recompile the project.

This structure of formatter classes can be easily extended. For instance, to support a new file format it is only needed to create a class that extends from the \verb|Formatter| interface and implement the method used by the \verb|parse| method, and implement the \verb|format| method, according to the format that it shall represent.

\subsection{Errors}

Errors can occur when processing requests. The cause of these errors can either be a user mistake or an internal processing issue. When errors occur, an error response is retrieved to the user with the respective status code. These responses are built based on error models. The \verb|ErrorOutputModel| receives the type of error that occurred and produces the respective error response. The type of errors is defined by the enum \verb|Problem|. For example, an \verb|ErrorOutputModel| is built using the \verb|Problem.UNAUTHORIZED| problem to retrieve an error response which signals that the request did not have a valid authentication token.

User errors can happen due to lack of authentication or permissions, and wrong inputs. However, internal errors may only occur when an unexpected exception is thrown. When internal errors and some of the user errors occur, they are handled by a \verb|ControllerAdvice| component. This component allows to declare exception handling methods using the \verb|ExceptionHandler| annotation. These methods are globally applied to all controllers and allow to receive exceptions as arguments. That is, this component acts as a try catch which encloses all the controllers, where each method catches the exceptions corresponding with the ones declared in their arguments. For example, a method that receives an \verb|HttpMediaTypeNotAcceptableException| exception as argument is executed when an exception of the same type occurs.

Besides retrieving error responses when internal errors occur, log entries are also created. These logs are useful to understand the internal errors that occur. The \verb|org.slf4j.Logger| logger is used to perform these logs. It is configured by defining the respective settings in the configuration file. For instance, this file should contain a property that holds the name of the file to where the logs should be written to.

\subsection{Security}

The security concerns, namely authenticity and permissions should be validated before a request is processed by the controllers. Thus, a security pipeline which validates these concerns was built based on the Spring interceptor components \cite{web:interceptors}. These components allow to implement the \verb|preHandle| method, that is executed before a request is passed to the controllers. 

\begin{figure}[!ht]
 \centering
 \includegraphics[scale=0.85]{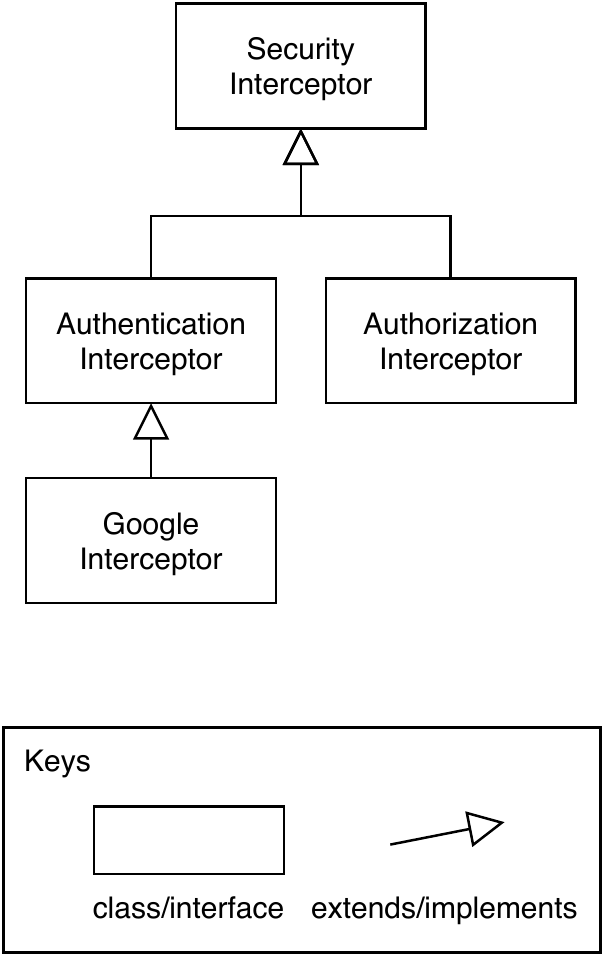}
 \caption{Security generalization architectural view.}
 \label{fig:apisecgeneralizationview}
\end{figure}

The security pipeline is composed by the \verb|AuthenticationInterceptor| and \verb|AuthorizationInterceptor| interceptors. These components implement the \verb|SecurityInterceptor| interface. This interface defines the method \verb|handle|, which shall contain the authenticity and user permissions validations. However, to ensure that these validations are executed before a request is passed to the controllers, the \verb|SecurityInterceptor| must implement the Spring \verb|HandlerInterceptor| interface and define that the \verb|handle| method is executed within the \verb|preHandle| method. This structure is represented in Figure \ref{fig:apisecgeneralizationview}.

The \verb|AuthenticationInterceptor| is the first component of the security pipeline to be executed. It supports the bearer token authentication \cite{web:bearer}. This type of authentication is based on tokens that are acquired after an authentication process with an identity provider. Thus, the requests must include the authorization header with the value \verb|Bearer| followed by a space and the authentication token. Currently, the authentication relies only on the Google Identity provider, hence the \verb|GoogleInterceptor| component extends from the \verb|AuthenticationInterceptor|. It implements the \verb|instrospect| method to define how to validate the token with Google. In case of the token being invalid, an error response with \verb|401 UNAUTHORIZED| status code is retrieved. Otherwise, the request is allowed to continue to the next security component.

The \verb|AuthorizationInterceptor| is the following component of the security pipeline. It relies on roles and annotations. Each user has a role, and each endpoint of the controller is annotated with the minimum role, which the user must hold, and with the type of operation that it represents. Hence, when this component processes a request it gathers that information and verifies if the request can proceed to the controllers. In case the needed conditions are not met, an error response with \verb|401 UNAUTHORIZED| status code is retrieved. Otherwise, the request is allowed to proceed to the controllers.

The security pipeline can be easily extended to support more authentication providers. To do so, it is only needed to create a new component that extends from \verb|AuthenticationInterceptor| and implement the \verb|instrospect| method with the specific settings to validate the token.

\subsection{Controllers}

\begin{figure}[!ht]
 \centering
 \includegraphics[scale=0.85]{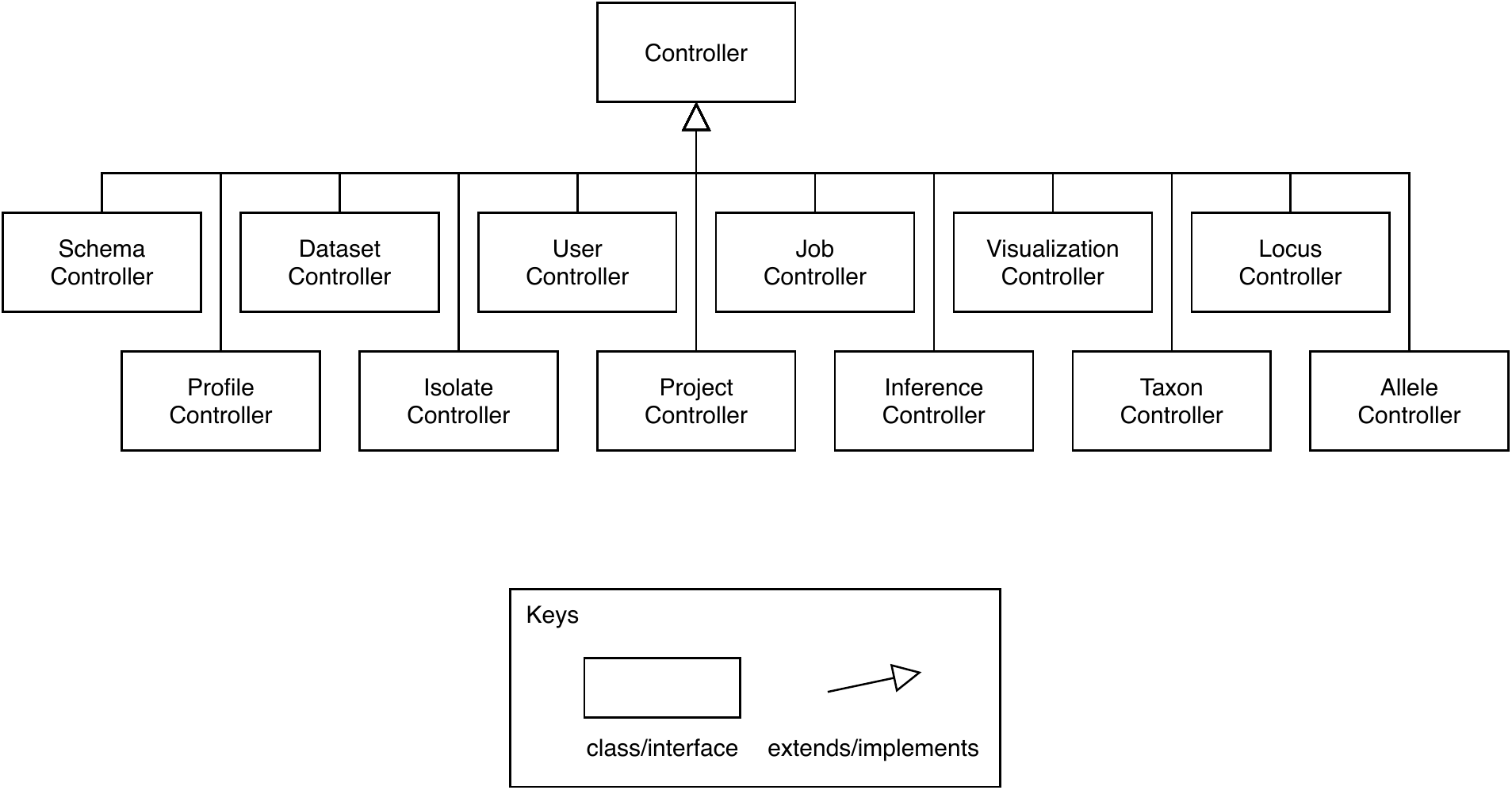}
 \caption{Controller generalization architectural view.}
 \label{fig:apigeneralazationcontroller}
\end{figure}

The controllers yield the endpoints of the \ac{API}. Thus, several instances of controllers compose the \verb|Controllers| layer. Each of these controllers are annotated with the \verb|Controller| annotation and defines different methods, however they all extend from the \verb|Controller| base class which contains utility methods that can be used by any of the extending controllers. For instance, it holds the \verb|getAllFile| method that allows to retrieve a list of domain entities from a given service and map it to a specific file format. This structure is represented in Figure \ref{fig:apigeneralazationcontroller}.

These controllers rely on input and output models to receive and retrieve data as \ac{JSON} objects, and rely on formatters to import and export large quantities of data in different file formats. Moreover, they depend on services to perform the logic associated to each endpoint. Hence, the controllers rely on the dependency injection mechanism to inject them. In our case, they are injected through the constructor of the controllers. 

A few configurations that are defined in the configuration file are used by the controllers, namely the maximum limits of the lists of domain entities that are retrieved, depending on the type of data. That is, there are different limits for the \ac{JSON} and the file formatted \verb|String| responses.

\subsection{Services}

The services provide methods to perform the business logic. Thus, several instances of them compose the \verb|Services| layer. They yield operations to manage persistent or transient domain entities. The majority of the domain entities exists independently of the time, however jobs only exist while the respective algorithm execution is not processed. Therefore, the \verb|JobService| defines operations to manage jobs, while the \verb|EntityService| define operations to manage persistent entities. Although they provide operations for different types of entities, they are annotated with the \verb|Service| annotation and extend from the \verb|Service| base class.

The \verb|EntityService| can be further decomposed depending on the need to consider versioning. That is, services which shall manage domain entities that need versioning must extend from \verb|VersionedEntityService|, while services which do not need versioning must extend from \verb|UnversionedEntityService|. This difference translates in the yielding of different methods. For example, the \verb|VersionedEntityService| provides the \verb|find| method that allows to retrieve the specified entity based on a \verb|Key| and a \verb|long|, which are the identifier and the version of the entity. However, the \verb|UnversionedEntityService| provides a similar \verb|find| method that has the same functionality but does not consider the version. That is, it retrieves the specified entity based only on the \verb|Key| that is received as argument.

\begin{figure}[!ht]
 \centering
 \includegraphics[scale=0.79]{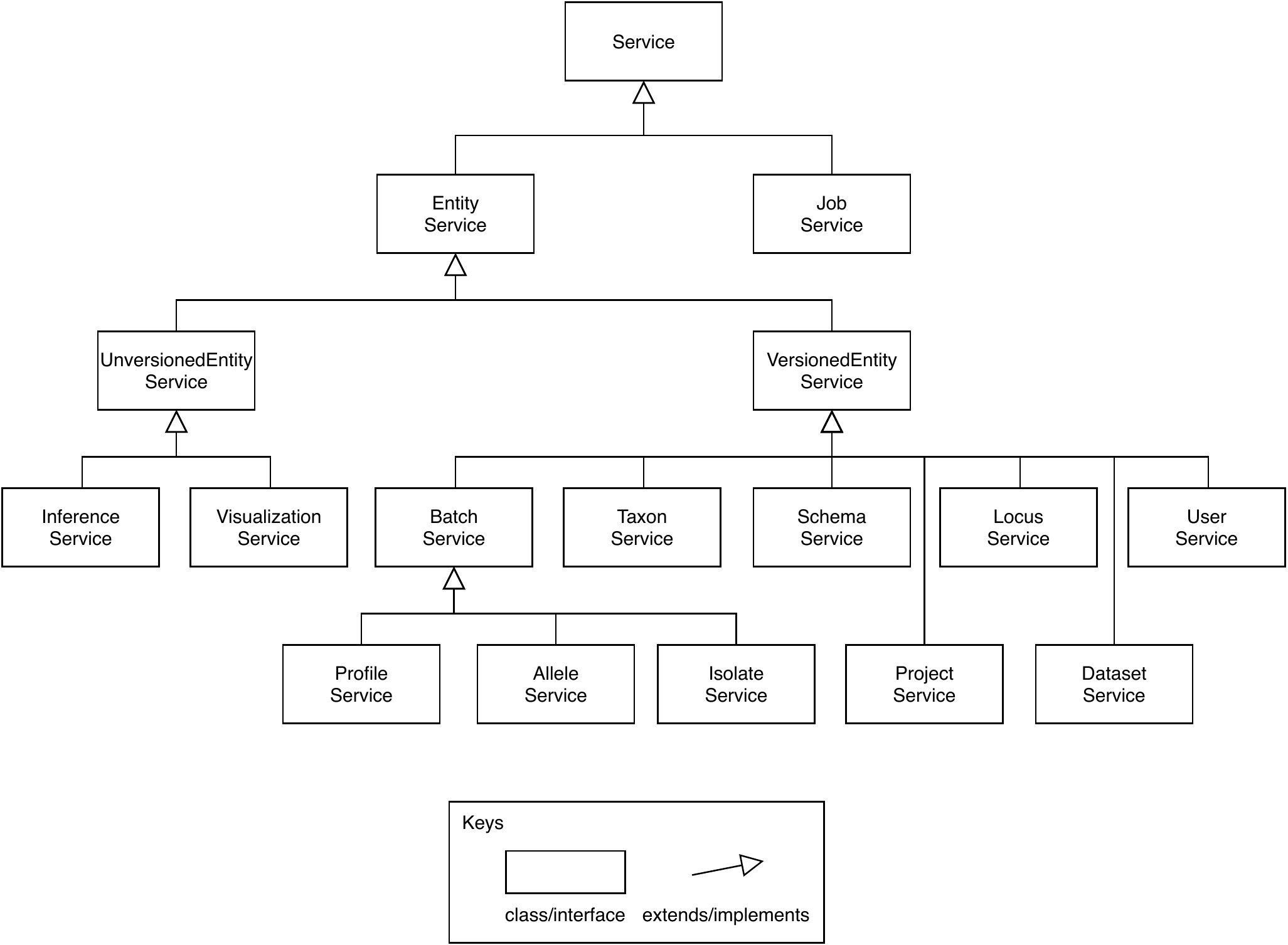}
 \caption{Service generalization architectural view.}
 \label{fig:apigeneralazationservices}
\end{figure}

The \verb|VersionedEntityService| can be further extended to support batch operations. Thus, \verb|BatchService| extends from \verb|VersionedEntityService| to provide different operations that process multiple entities. For example, it yields the \verb|saveAll| method that intends to save a list of entities that is received as arguments. This structure is represented in Figure \ref{fig:apigeneralazationservices}.

In the implementation of each of these services, it was considered that the methods must be executed inside a transaction to maintain the consistency of the data stored in the database. Hence, the \verb|Transactional| annotation was used to achieve such behaviour. This annotation is used to combine more than one write operation on a database as a single atomic operation. When a method annotated with it is called, all or none of the writes on the database are executed. 

These services normally rely on repositories that perform the interactions with the database. Hence, the services utilize a set of repositories which are also injected by the Spring mechanism of dependency injection.

The service implementation classes can either extend from \verb|BatchService|, \verb|VersionedEntityService|, \verb|UnversionedEntityService| or \verb|Service| depending on their need. For example, the \verb|ProfileService| must implement the operations inherited from \verb|BatchService|, \verb|VersionedEntityService|, and \verb|EntityService|, because it considers a persistent domain entity that needs versioning and batch operations. This structure can be easily extended to support more services. To do so, it is only needed to create the new service and extend it from one of the mentioned services depending on the needs.

\subsection{Repositories}

The repositories provide methods to perform the data access. Hence, several instances of them compose the \verb|Repositories| layer. The structure of repositories is very similar to the services one, in the sense that it also yields operations to manage persistent or transient domain entities. Therefore, the \verb|JobRepository| defines operations to manage jobs, while the \verb|EntityRepository| define operations to manage persistent entities. Although they provide operations for different types of entities, they are annotated with the \verb|Repository| annotation and extend from the \verb|Repository| base class which allows to perform queries to the database. Likewise, the \verb|EntityRepository| can also be further decomposed into \verb|VersionedEntityRepository| and \verb|UnversionedEntityRepository|, depending on the need to consider versioning.

\begin{figure}[!ht]
 \centering
 \includegraphics[scale=0.79]{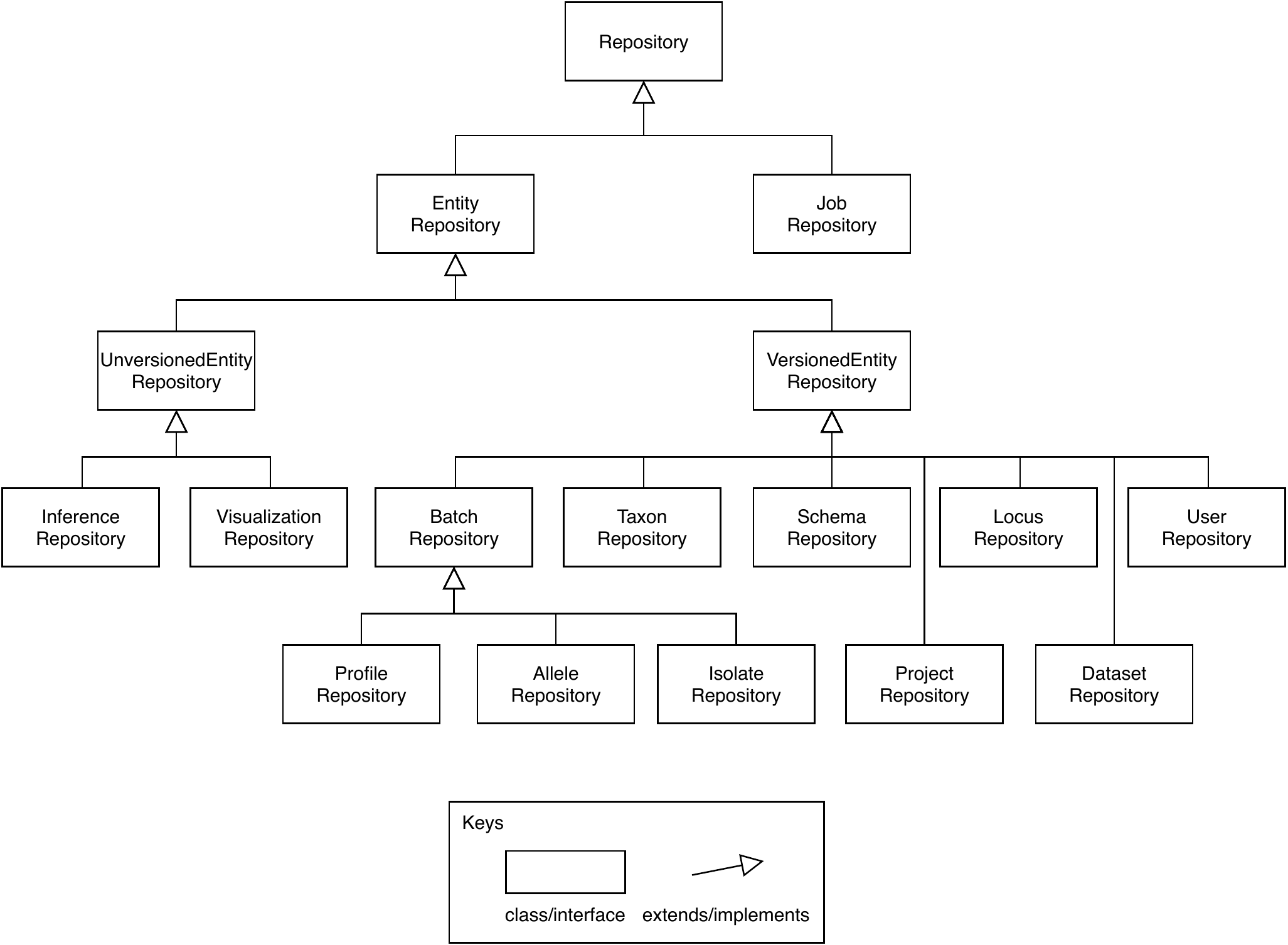}
 \caption{Repository generalization architectural view.}
 \label{fig:apigeneralazationrepositories}
\end{figure}

Moreover, the \verb|VersionedEntityRepository| can be further extended to support batch operations. Thus, \verb|BatchRepository| extends from \verb|VersionedEntityRepository| to define methods that process lists of entities. For example, it yields the \verb|saveAll| method that intends to save the list of entities that is received as arguments. This structure is represented in Figure \ref{fig:apigeneralazationrepositories}.

Several details were considered in the implementation of the methods provided by each repository, such as the use of pagination, \ac{OGM} and parameterized queries. 

An implementation consideration was to not use the usual functionalities of an \ac{OGM}. The use of these functionalities may add some overhead \cite{web:cypherfast}. Hence, it was decided that each query should be implemented from scratch to increase the performance of the data access operations. That is, the implementation of each repository method contains the respective Cypher query that it should perform, for example the \verb|findAllEntities| method implementation contains the query that allows to retrieve the resumed information of a subset of entities.

Moreover, the use of pagination in these queries was adopted because this framework is intended to handle great quantities of data. This approach allows to control the quantity of data that is dealt with in the methods that retrieve many domain entities. Thus, memory issues are less likely to happen since a maximum number of records that can be retrieved by the queries is defined. To achieve this behaviour, the respective methods must receive the page and the limit of records to retrieve. For instance, the \verb|findAllEntities| method provided by the \verb|VersionedEntityRepository| receives the page and limit values as arguments, which allows to retrieve the respective set of domain entities according to them. 

Also, each of those queries is parameterized. By using parameterized queries the performance can be increased because Neo4j can cache the query plans and reuse them in the following executions, which increases the following query speed. And, it also allows to protect from injection attacks, since parameters are never allowed to be interpreted as part of the query and have no means of escaping out of being anything other than a value of some sort \cite{web:cypherfast, web:cypherparameterized}. This can be achieved by executing a query that contains specific placeholders for each parameter and pass the values of the parameters, in the correct order, as arguments.

Another implementation characteristic is the use of indexes \cite{web:indexes}. They can increase the performance of some of the read queries. A database index is a redundant copy of some of the data in the database for the purpose of doing searches of related data more efficiently. They are useful when there are large quantities of data stored in the database, because it will allow the queries to rely on the index that points to a specific node instead of searching in the respective group of nodes for it. These indexes are created through a script that must be executed before the \ac{API} is started.

Finally, the \verb|UNWIND| feature of Neo4j was used to implement the batch operations. This feature is optimal to deal with up to fifty thousand records per operation, and allows to process a list of rows into individual rows that contain the information for each of the updates. Furthermore, queries using this feature can also be parameterized, and are constant, which allows Neo4j to use them for caching \cite{web:neo4jbatch}. Neo4j struggles to handle queries that contains many lines, therefore by using this feature small enough queries are used. 

\section{Plugin}\label{api}

The plugin of the database is based on the feature of user-defined procedures from Neo4j. A user-defined procedure is a mechanism that allows to extend Neo4j by writing custom code, which can be invoked directly from Cypher. These procedures can take arguments, perform operations on the database, and return results. Moreover, some resources can be injected into them from the database, which is similar to the dependency injection mechanism from Spring. 

This plugin intends to extend Neo4j to support inference and visualization algorithms, which shall be available as procedures. The inference algorithms procedures are executed over the profiles of a dataset, while visualization algorithms procedures are executed over the results of the inference algorithms. These procedures can be invoked directly from Cypher like any other standard procedure. For example the execution of the \ac{goeBURST} algorithm over a dataset with identifier \verb|dataset_id|, belonging to a project with identifier \verb|project_id|, only considering a maximum distance of three, and storing the result with the identifier \verb|inference_id|, can be represented as:
\begin{verbatim}
        CALL algorithms.inference.goeburst(project_id, dataset_id, 3, inference_id)
\end{verbatim}
However, these procedures are meant to be executed only by the \ac{API}, since certain validations are performed by it before executing them. For example, the existence of an inference is verified before a visualization algorithm execution is scheduled by the \ac{API}. 

\begin{figure}[!ht]
 \centering
 \includegraphics[scale=0.85]{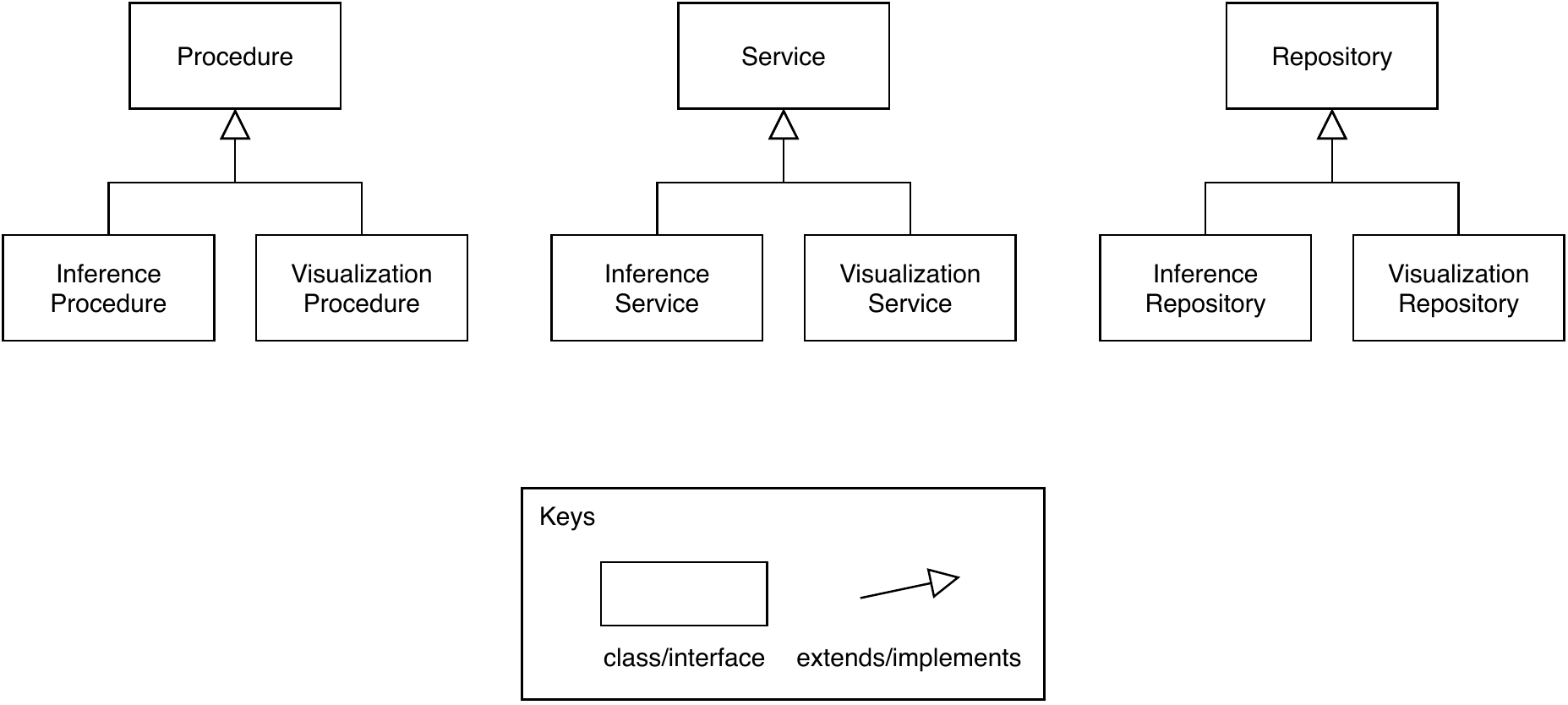}
 \caption{Plugin generalization architectural views.}
 \label{fig:plugingeneralizationview}
\end{figure}

This plugin is based on the three layers previously presented in Figure \ref{fig:pluginlayeredview}. Each of these layers was materialised into a structure of classes, that allows to easily extend themselves. These structures are presented in Figure \ref{fig:plugingeneralizationview}

\subsection{Algorithms}

Algorithms receive inputs, perform the respective computation, and produce results. Hence, inference algorithms receive distance matrices and produce graphs, while visualization algorithms receive the graphs produced by inference algorithms and generate coordinates for each of the nodes of the graph. Thus, the interface \verb|Algorithm| defines a method \verb|compute| that is extended by the \verb|InferenceAlgorithm| and \verb|VisualizationAlgorithm| interfaces to specify the respective arguments and results. For example, \verb|InferenceAlgorithm| defines that the \verb|compute| method must receive a \verb|Matrix| and retrieve an \verb|Inference|.

\begin{figure}[!ht]
 \centering
 \includegraphics[scale=0.85]{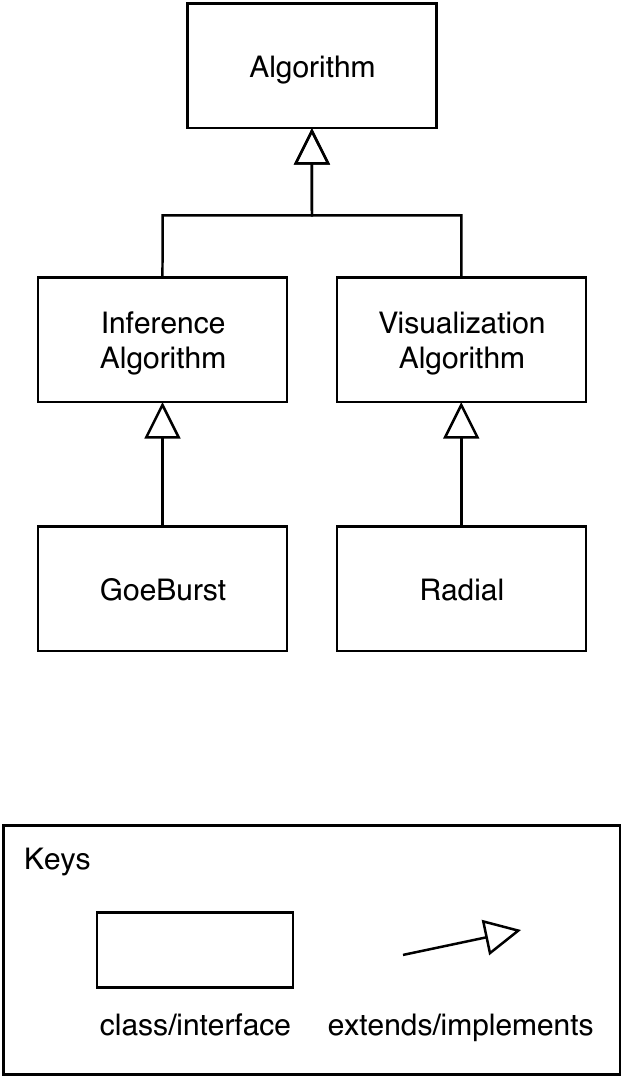}
 \caption{Algorithms generalization architectural view.}
 \label{fig:algorithmsgeneralizationview}
\end{figure}

The \ac{goeBURST} algorithm is the only inference algorithm supported, and the Radial Static Layout algorithm is the only visualization algorithm supported. Therefore, a class that implements the \verb|compute| method exists for each of these algorithms. For instance, the class \verb|GoeBurst| extends from \verb|InferenceAlgorithm| to implement the \verb|compute| method, which receives a \verb|Matrix| and retrieves an \verb|Inference|, with the logic of the \ac{goeBURST} algorithm. This structure of classes is represented in Figure \ref{fig:algorithmsgeneralizationview}. Implementations of the \ac{goeBURST} and Radial Static Layout algorithms already exist, thus our implementations are based on them \cite{web:goeburst, web:radial}.

This structure of algorithms classes can be easily extended to support new inference or visualization algorithms. It is only needed to create a class that extends from the respective base class to implement the \verb|compute| method. For instance, to support a new inference algorithm, it is needed to create a class that extends from \verb|InferenceAlgorithm| and implements the \verb|compute| method with the respective algorithm logic.

\subsection{Procedures}

The procedures yield the interface of the plugin. Thus, different instances of procedures compose the \verb|Procedures| layer. The methods declared within these procedure classes, that are annotated with the \verb|Procedure| annotation, are allowed to be executed as a standard procedure. This annotation allows to define the designation that shall be used by the users to execute these procedures. For instance, the \verb|InferenceProcedure| defines a method named \verb|goeburst|, however the annotation holds the designation \verb|algorithms.inference.goeburst|. Hence, the users need to use that designation to execute the \verb|goeburst| method. 

Each of these procedures defines different methods, however they all extend from the \verb|Procedure| base class. However, the base class only holds the \verb|GraphDatabaseService| object, which is injected by the database. This object allows to interact with the database and can only be injected at the procedures level, thus, it must be passed as an argument to the services, and then to the repositories, since the procedures rely on services to perform the respective logic. This structure is represented in Figure \ref{fig:plugingeneralizationview}. These procedures are similar to the controllers presented before, in the sense that they intend to yield the endpoints of the plugin.

\subsection{Services}

The services provide methods to perform the business logic. Thus, different instances of them compose the \verb|Services| layer. They yield methods to execute algorithms. That is, the methods defined by each service are responsible for gathering the algorithms input data from the database, computing the algorithms, and storing the results back to the database. For instance, \verb|InferenceService| provides the \verb|goeburst| method that obtains the distance matrix for the specified dataset, executes the \ac{goeBURST} algorithm, and then stores the resulting graph in the database.

Each of these services defines different methods, however they all extend from the \verb|Service| base class, as shown in Figure \ref{fig:plugingeneralizationview}. The base class holds the \verb|GraphDatabaseService| object that is passed by the \verb|Procedure|. Since each of those methods relies on the repositories to interact with the database, this object must also be passed to them. 

These services rely on repositories to perform the interactions with the database. For example, the \verb|goeburst| method from the \verb|InferenceService| uses the \verb|InferenceRepository| to obtain the distance matrix that will be used by the algorithm.

\subsection{Repositories}

The \verb|Repositories| layer is composed by different repositories that provide methods to interact with the database. The objectives of these repositories are to retrieve the input data for the algorithms from the database, and to write their results back to the database. To accomplish these objectives, they rely on the \verb|GraphDatabaseService| object, which allows them to perform operations over the database.

Also, based on those objectives, the \verb|Repository| interface is provided, which defines the \verb|read| and \verb|write| methods. These methods are meant to be implemented to retrieve and store a certain type of data respectively. Thus, the \verb|InferenceRepository| and \verb|VisualizationRepository| implement those methods to read and write the respective types of data. For example, the \verb|InferenceRepository| implements the read operation to retrieve a \verb|Matrix| and implements the write operation to store an \verb|Inference|. This structure is also presented in Figure \ref{fig:plugingeneralizationview}.

The implementation of these methods is based on the Java Core API \cite{web:javacoreapi} of Neo4j. There were other alternatives to implement the data access, such as using Cypher and using the Traversal framework. However, Cypher takes more time to execute and the Traversal framework became deprecated.

The structure of repositories classes can be easily extended to support new repositories. It is only needed to create a new class which extends from the base class, specify which type of data shall be retrieved by the read operation and received by the write operation, and implement those methods based on the Java Core API.

\section{Discussion}

This chapter describes and reasons about the implementation details of the \ac{API} and plugin components that compose the framework. The \ac{API} section includes implementation details, such as the compliance to the \ac{REST} architecture, the use of dependency injection, and how the authentication and authorization processes are accomplished. The plugin section describes details such as, how it relied on the user-defined procedures mechanism of Neo4j, and how the algorithms were implemented. Besides these details, each section explains how the respective architecture for each component was materialized and presents several exemplifications of the different concerns that were described.
\chapter{Experimental Evaluation}

This chapter explains the types of tests, the operations, the datasets, and the system settings used to evaluate the framework. Afterwards, it provides an analysis of the results obtained from these tests. 

There are three types of tests that were used to analyse the framework, namely functional, performance, and load testing. The functional tests aim to validate the functionality of the framework, that is, verify if each operation provided by the framework works as expected. Thus, unit tests were designed for almost every functionality, achieving a code coverage of more than 90\%. The performance tests objective is to check how some operations behave in terms of time and memory usage, under a particular workload. This type of tests allows to uncover what can still be improved in future work. The load tests purpose is to test how the framework behaves when handling several clients at a given moment. The framework might perform well for one client during performance testing, but it might degrade when many clients try to use it simultaneously during load testing due to lack of system resources. This type of tests also allows to uncover what can still be improved in future work.

The performance and load tests were executed over a set of read and write operations, which were picked to be analysed in terms of time and memory. These operations are composed by the \verb|Save Profiles|, \verb|Run Inference|, \verb|Run Visualization|, \verb|Save Profile|, \verb|Get Profiles|, \verb|Get Resumed Profiles|, \verb|Get| \verb|Profile|, \verb|Get Inference|, and, \verb|Get Visualization|. In these tests, the \verb|Save Profiles| saves a file containing an increasing number of profiles. The \verb|Run Inference| executes the \ac{goeBURST} algorithm over an increasing number of profiles that are stored in the database. The \verb|Run Visualization| executes the Radial Static Layout algorithm over the result of the \ac{goeBURST} algorithm execution with an increasing number of profiles and edges. It must be taken into account that these two operations not only compute the algorithm, but also include the work of gathering the data and storing the results. Note that, in the case of the inference algorithms, the calculation of the distance matrix is also included in this process. Then, the \verb|Save Profile| saves a single profile in the database that holds an increasing number of profiles. The \verb|Get Profiles| and \verb|Get Resumed Profiles| retrieve the respective information about an increasing number of profiles. The \verb|Get Profile| retrieves only a single profile independently of the quantity of profiles stored in the database. Finally, the \verb|Get Inference| and \verb|Get Visualization| retrieves the results of the algorithms executed over an increasing number of profiles and edges.

The tests relied on the \textit{Streptococcus pneumoniae} \cite{web:pubmlst} \ac{MLST} dataset, which was specifically chosen because it is part of several published studies and also because it is publicly available, which will facilitate the interpretation of the results. This dataset contains a profile length of 7 and a total of around 16000 profiles currently.

This experimental evaluation was performed on a machine with an Intel Core I7 2.40 GHz quad core processor and 8 GB of memory, where 2 GB were allocated for the database and another 4 GB were allocated for the \ac{API}.

\section{Performance Testing}

This section analyses the average running time of the operations, displaying the results in a table and a plot, as well as the average allocated memory by these operations, displayed in another table and plot. These tests were developed relying on a microbenchmarking tool for Java, namely Java Microbenchmark Harness, and using 5 warmups and 10 iterations for each operation being tested.

The average running time that each operation took to complete, over an increasing number of profiles, is represented in Table \ref{tbl:perftime}, in milliseconds. From these values, it can be observed that the \verb|Save Profiles| write operation, takes around 6000 milliseconds to process 500 profiles, while the \verb|Get Profiles| read operation takes much less time (68 milliseconds). The presented results confirm that the graph database operations that operates over a fixed amount of data are not affected by the increasing volume of data. This is confirmed by analysing the results for the \verb|Get Profile| operation, which shows that the time of execution is constant and independent of the increasing number of profiles. Additionally, it can be concluded that the presented execution times for the algorithms comply with their time complexity, which is quadratic for the \ac{goeBURST} algorithm \cite{artc:goeburst}, and linearithmic for the Radial Static Layout algorithm \cite{book:radial} since the children nodes are sorted. The presented results also reveal that relying in a graph database to handle this type of data allows to have a good performance in read and single write operations. 

\begin{table}[!ht]
    \centering
    \begin{tabular}{| c | c | c | c | c | c | c |} \hline
        \diagbox{Operation}{Profiles} & 500 & 1000 & 2000 & 5000 & 10000 & 15000 \\ \hline
        Save Profiles & 6294.861 & 12916.707 & 25434.663 & 64377.122 & 133941.541 & 212235.800\\\hline
        Run Inference & 316.292 & 1043.079 & 4199.889 & 25543.573 & 104885.648 & 244421.421\\\hline
        Run Visualization & 193.064 & 560.269 & 2055.067 & 12784.802 & 51034.175 & 119885.683\\\hline
        Save Profile & 54.371 & 58.351 & 57.265 & 50.065 & 61.265 & 55.035\\\hline
        Get Profiles & 68.595 & 89.940 & 200.015 & 550.311 & 1549.135 & 2355.374\\\hline
        Get Resumed Profiles & 6.392 & 10.320 & 19.018 & 47.900 & 96.402 & 150.546\\\hline
        Get Profile & 1.915 & 1.520 & 2.108 & 1.582 & 1.866 & 1.364\\\hline
        Get Inference & 5.954 & 10.777 & 20.097 & 48.638 & 100.737 & 152.563\\\hline
        Get Visualization & 7.625 & 13.712 & 27.242 & 81.630 & 138.806 & 214.127\\ \hline
    \end{tabular} 
    \caption{Performance testing time results in milliseconds.} 
    \label{tbl:perftime} 
\end{table}

\begin{figure}[!ht]
    \centering
    \begin{tikzpicture}
        \begin{axis}[
            xlabel={Profiles [\#]},
            ylabel={Time [ms]},
            xtick={0, 2000, 4000, 6000, 8000, 10000, 12000, 14000, 16000},
            ytick={0, 50000, 100000, 150000, 200000, 250000},
            legend pos=north west,
            cycle list name=colorful,
            ymajorgrids=true,
            xmajorgrids=true,
            grid style=dashed
        ]
            \addplot coordinates {
                (500, 6294.861)
                (1000, 12916.707)
                (2000, 25434.663)
                (5000, 64377.122)
                (10000, 133941.541)
                (15000, 212235.800)
            };
            \addplot coordinates {
                (500, 316.292)
                (1000, 1043.079)
                (2000, 4199.889)
                (5000, 25543.573)
                (10000, 104885.648)
                (15000, 244421.421)
            };
            \addplot coordinates {
                (500, 193.064)
                (1000, 560.269)
                (2000, 2055.067)
                (5000, 12784.802)
                (10000, 51034.175)
                (15000, 119885.683)
            };
            \addplot coordinates {
                (500, 54.371)
                (1000, 58.351)
                (2000, 57.265)
                (5000, 50.065)
                (10000, 61.265)
                (15000, 55.035)
            };
            \addplot coordinates {
                (500, 68.595)
                (1000, 89.940)
                (2000, 200.015)
                (5000, 550.311)
                (10000, 1549.135)
                (15000, 2355.374)
            };
            \addplot coordinates {
                (500, 6.392)
                (1000, 10.320)
                (2000, 19.018)
                (5000, 47.900)
                (10000, 96.402)
                (15000, 150.546)
            };
            \addplot coordinates {
                (500, 1.915)
                (1000, 1.520)
                (2000, 2.108)
                (5000, 1.582)
                (10000, 1.866)
                (15000, 1.364)
            };
            \addplot coordinates {
                (500, 5.954)
                (1000, 10.777)
                (2000, 20.097)
                (5000, 48.638)
                (10000, 100.737)
                (15000, 152.563)
            };
            \addplot coordinates {
                (500, 7.625)
                (1000, 13.712)
                (2000, 27.242)
                (5000, 81.630)
                (10000, 138.806)
                (15000, 214.127)
            };
            \legend{save profiles, run inference, run visualization, save profile, get profiles, get resumed profiles, get profile, get inference, get visualization}
        \end{axis}
    \end{tikzpicture}
    \caption{Performance testing time plot.}
    \label{graph:perftime}
\end{figure}
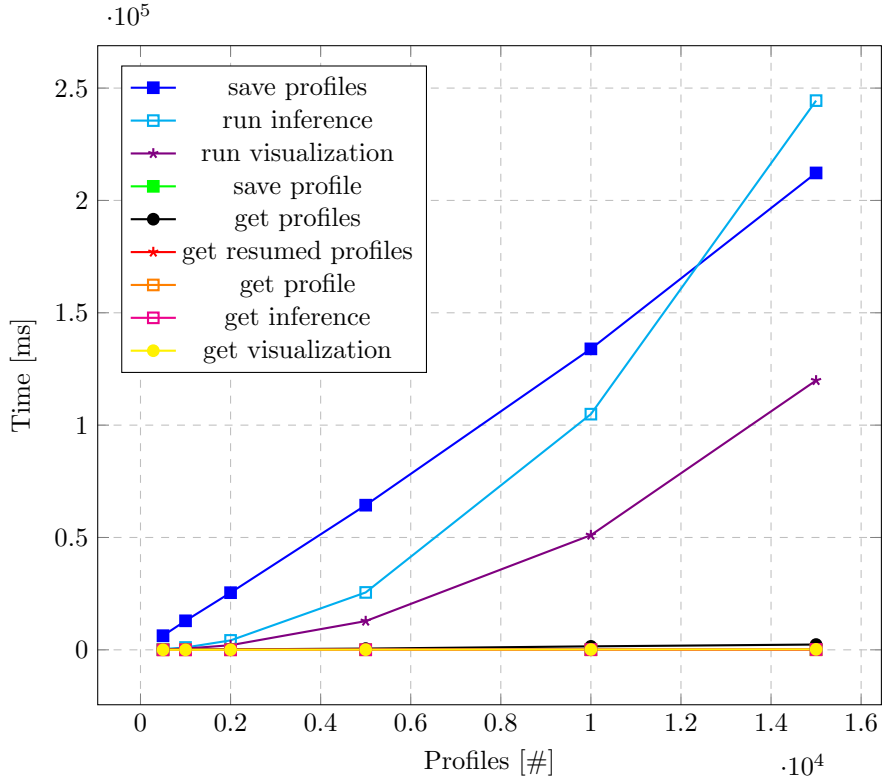

This conclusion is also backed by the plot represented in Figure \ref{graph:perftime}, which allows us to see that when the number of profiles used by a read operation is increased, it has little effect in the time taken to execute the operation. However, the time used by a batch writes operation increases almost or even linearly proportional to the number of profiles, when the latter is increased. From this plot, it can be seen that the line that corresponds to the \verb|Save Profiles| write operation grows linearly proportional to the number of profiles as it grows, while the line which corresponds to the \verb|Get Profiles}| operation is almost a flat line, and the line which corresponds to the \verb|Get Profile| operation is flat. Additionally, the lines correspondents to the algorithms are more curve than the others, that also supports that the algorithms follow their expected time complexity.

The average memory allocated that each operation consumed until completion, over an increasing number of profiles, is represented in Table \ref{tbl:perfmemory}, in megabytes. From these values, it can observed that the \verb|Save Profiles| write operation allocates around 40 megabytes to process 500 profiles, while the \verb|Get| \verb|Profiles| read operation allocates much less memory space (9 megabytes). The presented results reveal that the framework allocates the memory linearly proportional to the amount of data that it is handling, and that the read and single write operations allocate much less memory than batch writes operations. However, it must be noticed that this analysis only considers the amount of memory allocated in the \ac{API}. This is relevant because the algorithms are executed within the database, which causes them to use the database memory instead of the \ac{API} memory. Hence, the allocated memory for the algorithms executions are minimal.

\begin{table}[!ht]
    \centering
    \begin{tabular}{| c | c | c | c | c | c | c |} \hline
        \diagbox{Operation}{Profiles} & 500 & 1000 & 2000 & 5000 & 10000 & 15000 \\ \hline
        Save Profiles & 40.283 & 80.519 & 160.746 & 402.086 & 803.059 & 1203.376\\ \hline
        Run Inference & 0.018 & 0.019 & 0.020 & 0.028 & 0.053 & 0.088\\ \hline
        Run Visualization & 0.018 & 0.019 & 0.020 & 0.023 & 0.032 & 0.047\\ \hline
        Save Profile & 0.147 & 0.147 & 0.147 & 0.147 & 0.147 & 0.147\\ \hline
        Get Profiles & 9.107 & 18.170 & 36.308 & 90.620 & 181.275 & 271.595\\ \hline
        Get Resumed Profiles & 1.940 & 3.867 & 7.732 & 19.243 & 38.571 & 57.781\\ \hline
        Get Profile & 0.045 & 0.0426 & 0.0426 & 0.0426 & 0.0426 & 0.0426\\ \hline
        Get Inference & 0.820 & 1.713 & 3.548 & 9.042 & 18.195 & 27.291\\ \hline
        Get Visualization & 0.569 & 1.148 & 2.340 & 5.889 & 11.826 & 17.719\\ \hline
    \end{tabular}
    \caption{Performance testing memory results in megabytes.} 
    \label{tbl:perfmemory} 
\end{table}

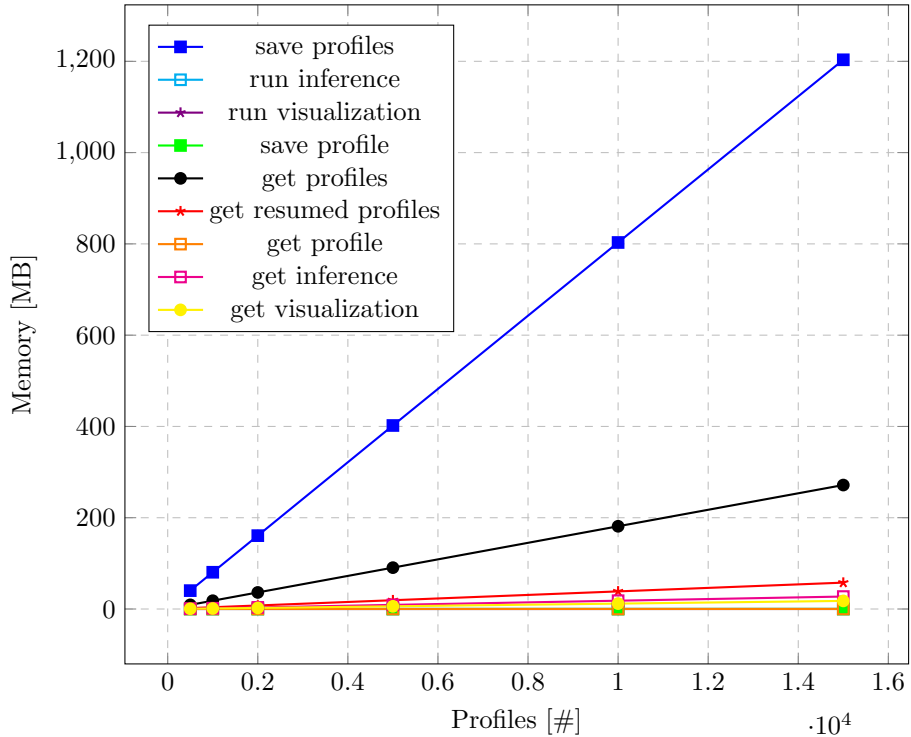
\begin{figure}[!ht]
    \centering
    \begin{tikzpicture}
        \begin{axis}[
            xlabel={Profiles [\#]},
            ylabel={Memory [MB]},
            xtick={0, 2000, 4000, 6000, 8000, 10000, 12000, 14000, 16000},
            ytick={0, 200, 400, 600, 800, 1000, 1200, 1400},
            legend pos=north west,
            cycle list name=colorful,
            ymajorgrids=true,
            xmajorgrids=true,
            grid style=dashed
        ]
            \addplot coordinates {
                (500, 40.283204)
                (1000, 80.5194224)
                (2000, 160.7456408)
                (5000, 402.0864584)
                (10000, 803.058604)
                (15000, 1203.3760864)
            };
            \addplot coordinates {
                (500, 0.0180148)
                (1000, 0.0194248)
                (2000, 0.0201936)
                (5000, 0.027648)
                (10000, 0.0530976)
                (15000, 0.0880872)
            };
            \addplot coordinates {
                (500, 0.0184726)
                (1000, 0.0193052)
                (2000, 0.0202112)
                (5000, 0.0226224)
                (10000, 0.03242)
                (15000, 0.0465392)
            };
            \addplot coordinates {
                (500, 0.146851175)
                (1000, 0.146839822)
                (2000, 0.146874072)
                (5000, 0.146820741)
                (10000, 0.146840896)
                (15000, 0.146827848)
            };
            \addplot coordinates {
                (500, 9.10662657)
                (1000, 18.169925079)
                (2000, 36.3079044)
                (5000, 90.620249867)
                (10000, 181.275324)
                (15000, 271.5950352)
            };
            \addplot coordinates {
                (500, 1.939640992)
                (1000, 3.866504774)
                (2000, 7.732114302)
                (5000, 19.243151554)
                (10000, 38.571351135)
                (15000, 57.781382048)
            };
            \addplot coordinates {
                (500, 0.043546454)
                (1000, 0.042598075)
                (2000, 0.04259576)
                (5000, 0.042596664)
                (10000, 0.042593599)
                (15000, 0.042603509)
            };
            \addplot coordinates {
                (500, 0.820017538)
                (1000, 1.713321539)
                (2000, 3.547518575)
                (5000, 9.042025816)
                (10000, 18.194895658)
                (15000, 27.2913184)
            };
            \addplot coordinates {
                (500, 0.569281774)
                (1000, 1.147687405)
                (2000, 2.33975448)
                (5000, 5.888136939)
                (10000, 11.826413171)
                (15000, 17.7193704)
            };
            \legend{save profiles, run inference, run visualization, save profile, get profiles, get resumed profiles, get profile, get inference, get visualization}
        \end{axis}
    \end{tikzpicture}
    \caption{Performance testing memory plot.}
    \label{graph:perfmemory}
\end{figure}

This conclusion is also backed by the plot represented in Figure \ref{graph:perfmemory}, which allows us to see that when the number of profiles used by any operation is increased, the amount of memory allocated grows almost linearly proportional to the number of profiles. For instance, the line that corresponds to \verb|Save Profiles| write operation grows linearly proportional to the number of profiles as it increases. 

\section{Load Testing}

This section analyses the average running times of the operations that are presented in a table and a plot. These tests are executed over a database that contains around 500 profiles, and the Save Profile operation also involves 500 profiles. These tests were developed relying on a load testing tool, namely Apache JMeter, and using 10 iterations for each operation being tested.

The average time, in milliseconds, that each operation took to complete, over an increasing number of clients, is represented in Table \ref{tbl:loadtime}. From these values it can be observed that the \verb|Save Profiles| write operation, takes around 12000 milliseconds to process 10 different requests at the same moment, while the \verb|Get Profiles| read operation takes much less time (101 milliseconds). The presented results also reveal that if the number of clients in a given moment increases, then it is still possible to maintain a good performance for the read operations. Lastly, these results also show that if there is nearly 250 clients performing the Save Profiles operation in a given moment, then the framework cannot handle all of those requests. However, this is not very problematic since the most common use cases for this framework for saving data rely on the use of the \verb|Save Profile| operation, which performs efficiently. 

\begin{table}[!ht]
    \centering
    \begin{tabular}{| c | c | c | c | c | c | c | c |} \hline
        \diagbox{Operation}{Clients} & 1 & 10 & 50 & 100 & 250 & 500 & 1000 \\ \hline
        Save Profiles & 3005 & 12348 & 62511 & 112772 & - & - & -\\ \hline
        Save Profile & 65 & 176 & 1305 & 3118 & 9164 & 18259 & 39030\\ \hline
        Get resumed Profiles & 6 & 6 & 6 & 5 & 7 & 320 & 1055\\ \hline
        Get Profiles & 97 & 101 & 975 & 1469 & 4019 & 8154 & 17095\\ \hline
        Get Profile & 2 & 2 & 2 & 2 & 2 & 2 & 319\\ \hline
        Get Inference & 33 & 21 & 26 & 102 & 493 & 1338 & 3267\\ \hline
        Get Visualization & 30 & 16 & 14 & 15 & 312 & 915 & 1701\\ \hline
    \end{tabular} 
    \caption{Load testing, simulating several parallel users, time results in milliseconds.} 
    \label{tbl:loadtime} 
\end{table}

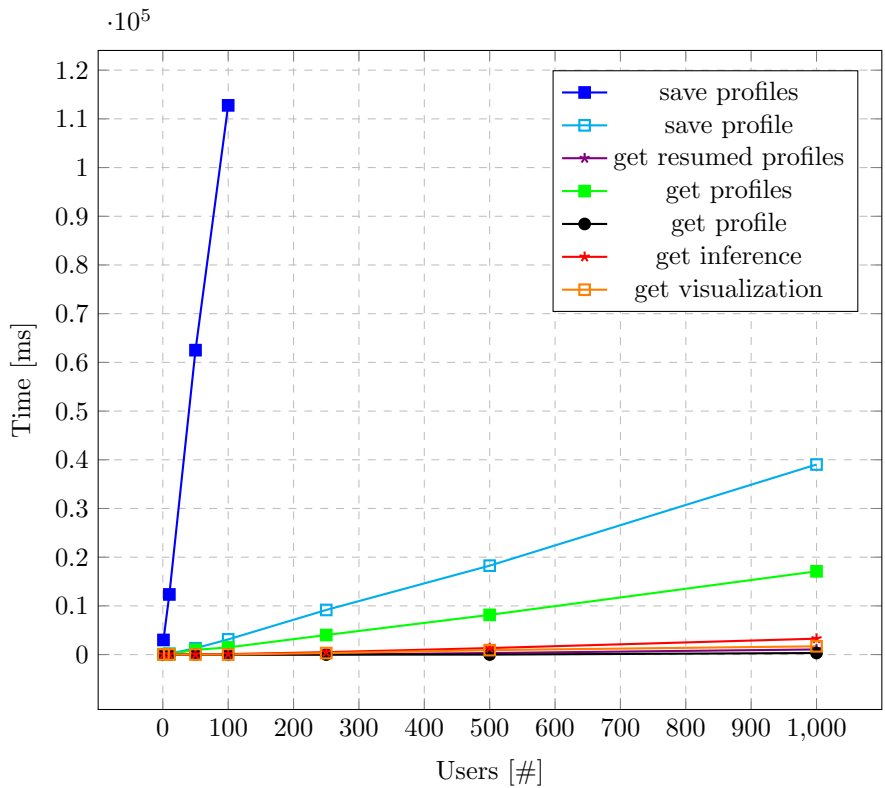
\begin{figure}[!ht]
    \centering
    \begin{tikzpicture}
        \begin{axis}[
            xlabel={Users [\#]},
            ylabel={Time [ms]},
            xtick={0, 100, 200, 300, 400, 500, 600, 700, 800, 900, 1000},
            ytick={0, 10000, 20000, 30000, 40000, 50000, 60000, 70000, 80000, 90000, 100000, 110000, 120000},
            legend pos=north east,
            cycle list name=colorful,
            ymajorgrids=true,
            xmajorgrids=true,
            grid style=dashed
        ]
            \addplot coordinates {
                (1, 3005)
                (10, 12348)
                (50, 62511)
                (100, 112772)
            };
            \addplot coordinates {
                (1, 65)
                (10, 176)
                (50, 1305)
                (100, 3118)
                (250, 9164)
                (500, 18259)
                (1000, 39030)
            };
            \addplot coordinates {
                (1, 6)
                (10, 6)
                (50, 6)
                (100, 5)
                (250, 7)
                (500, 320)
                (1000, 1055)
            };
            \addplot coordinates {
                (1, 97)
                (10, 101)
                (50, 975)
                (100, 1469)
                (250, 4019)
                (500, 8154)
                (1000, 17095)
            };
            \addplot coordinates {
                (1, 2)
                (10, 2)
                (50, 2)
                (100, 2)
                (250, 2)
                (500, 2)
                (1000, 319)
            };
            \addplot coordinates {
                (1, 33)
                (10, 21)
                (50, 26)
                (100, 102)
                (250, 493)
                (500, 1338)
                (1000, 3267)
            };
            \addplot coordinates {
                (1, 30)
                (10, 16)
                (50, 14)
                (100, 15)
                (250, 312)
                (500, 915)
                (1000, 1701)
            };
            \legend{save profiles, save profile, get resumed profiles, get profiles, get profile, get inference, get visualization}
        \end{axis}
    \end{tikzpicture}
    \caption{Load testing time plot, simulating several parallel users.}
    \label{graph:loadtime}
\end{figure}

This conclusion is also backed by the plot represented in Figure \ref{graph:loadtime}, which allows us to see that when the number of clients performing read operations is increased, it has a small effect in the time taken to execute. However, the time line of a batch writes operation grows almost vertically and grows almost linearly proportional, in the single write operations, to the number of clients when it is increased.

\section{Discussion}

In this chapter it was discussed the implications of our solution in terms of time and memory requirements in two different types of analyses, namely performance and load testing. The most important results found are related with the read operations computational cost, when comparing to write operations, since the first represents a notable difference over the latter. Overall, we consider our implementation efficient in terms of read and single write operations, and recognize that the response time of the batch writes operations can still be improved. However, the most common use cases of the framework will not be affected, since the saving of a single profile or a relatively small quantity of profiles are more typical than saving large quantities of profiles. The latter may happen in the early uses of the framework, which shall be in an offline environment and by a specific user.

Nonetheless, a possible improvement can be accomplished by using an alternative architecture for these operations, for example these operations could be handled with the same strategy applied for the algorithms, that is, the use of a queue mechanism. Using a queue mechanism would give us the advantages of providing a faster response time, as the writes are processed in the background, and of not having the constraint of only being able to process, in parallel, less than 250 clients. This alternative is proposed and explained in Chapter \ref{cpt:finalRemarks}.
\chapter{Final Remarks} \label{cpt:finalRemarks}

Epidemics have become an issue of increasing importance due to the growing exchanges of people and merchandise between countries. Hence, phylogenetic analyses are continuously generating huge volumes of typing and ancillary data. And there is no doubt about the importance of such data, and phylogenetic studies, for the surveillance of infectious diseases and the understanding of pathogen population genetics and evolution. The traditional way of performing phylogenetic analysis is not feasible anymore as a result of the amount of data generated. This analysis often produces graphs and trees, which have many relationships. Although graph oriented databases can be of much help in this setting, as far as it is known there is no solution relying on these technologies to address large scale phylogenetic analysis challenges.

\section{Conclusions}

This document starts by analysing what is the current problem with phylogenetic analysis when huge amounts of data are considered. Then, although graph oriented databases can be useful in this setting, we conclude that there is no solution based on these technologies.

Afterwards, it provides an explanation of what are graph databases, their use cases, their most important capabilities, and comparisons between them and a relational database. Then, we conclude that Neo4j offers the most interesting set of capabilities and features, beyond the general capabilities of graph databases, and also has an active community. This set of interesting capabilities and features are composed by the \ac{APOC} library, the trigger and scheduler mechanism, and the user-defined procedures.

The following topic which the document addresses is the explanation of the phylogenetic analysis process. This process starts with a scientific workflow that takes biological samples as input to obtain the genetic sequences in the form of (short) reads through a \ac{NGS} process. Then, alignment tools and/or assemblers are executed to align/assembly the genomes, which are further processed for identifying alleles at specific \textit{loci} of interest. These genomes are then typed with a given typing methodology based on their \textit{loci}, leading to several allelic profiles, and which are complemented by ancillary data. These allelic profiles can then be used as input for inferring evolutionary patterns, in the form of trees or graphs, for instance based on distance matrix methods. These patterns provide evolutionary hypotheses for the organisms being analysed and are usually presented to the analyst though different visualizations.

Then, the document provides an overview, based on the phylogenetic analysis process, of the main use cases and requirements for phylogenetic analysis. And it proposes a framework which complies with such use cases and considers all of the requirements. This framework is composed of a Spring Framework \ac{API}, a Neo4j database and a plugin for the latter, which relies on the custom procedures capabilities of Neo4j. It allows to authenticate with Google, manage all types of data that were analysed throughout the document, import and export dataset files of different formats, and execute inference and visualization algorithms. The framework shall contribute with some advantages for the phylogenetic analysis such as, by storing the results of algorithms will avoid having to compute them again, and using multilayer networks will make the comparison between them more efficient and scalable. Then, the architecture of this framework is presented with the use of different architectural views, and implementation characteristics are presented and described.

Finally, this document presents and explains how the experimental evaluation of the implemented framework was conducted, and also describes the respective results. From these results, we can conclude that the most important facts found are related with the read operations computational cost, when comparing to write operations, since the former represents a notable difference over the latter. We can also conclude that by using a graph database the operations executed over a fixed amount of data are not affected by an increasing volume of data. Furthermore, we observed that the execution times for the algorithms comply with their time complexity, which is quadratic for the \ac{goeBURST} algorithm, and linearithmic for the Radial Static Layout algorithm. Overall, we consider our implementation efficient in terms of read and single write operations. However, with the presented results and their analysis, we understand that the framework can still be improved.

\section{Future Work}

There are several possible continuations of the work done throughout this thesis. One could be the extension of our solution to provide more algorithms, and make use of parallelization to improve their performance. The already provided algorithms by Neo4j make use of parallelization, hence we could improve our algorithms execution time by parallelizing their computations. Other potential development could be on how to use the background triggers functionality to achieve the dynamic computation of inference algorithms. Another possibility could be a study of how to perform the batch writes processing based on a queue mechanism. Or even to study how to use the Neo4j functionality of integrating with big data processing engines, such as Apache Spark. With this changing the batch writes operations of the \ac{API} would have a faster response time. Finally, the \ac{API} interface could be improved to be based on hypermedia. This would allow the \ac{API} to be explorable and decouple the client applications from it.

\renewcommand{\bibname}{References}
\addcontentsline{toc}{chapter}{\bibname}
\bibliographystyle{plainurl}
\bibliography{basics/references.bib}

\begin{thebibliography}{10}

\bibitem{artc:10.1007/978-3-642-28062-7_3}
Jo{\~a}o Almeida, Jo{\~a}o Tiple, M{\'a}rio Ramirez, Jos{\'e} Melo-Cristino,
  C{\'a}tia Vaz, Alexandre P.~Francisco, and Jo{\~a}o A.~Carri{\c{c}}o.
\newblock An ontology and a rest api for sequence based microbial typing data.
\newblock In Ana~T. Freitas and Arcadi Navarro, editors, {\em Bioinformatics
  for Personalized Medicine}, pages 21--28, Berlin, Heidelberg, 2012. Springer
  Berlin Heidelberg.

\bibitem{web:titan}
Aurelius.
\newblock Titan: Distributed graph database., 2015.
\newblock Last accessed 28 December 2020.
\newblock URL: \url{http://titandb.io}.

\bibitem{book:radial}
Christian Bachmaier, Ulrik Brandes, and Falk Schreiber.
\newblock {\em Biological networks}.
\newblock 2014.

\bibitem{artc:NGS}
J~A Carriço, A~J Sabat, A~W Friedrich, M~Ramirez, and Collective on~behalf of
  the ESCMID Study Group~for Epidemiological Markers~(ESGEM).
\newblock Bioinformatics in bacterial molecular epidemiology and public health:
  databases, tools and the next-generation sequencing revolution.
\newblock {\em Eurosurveillance}, 18(4), 2013.
\newblock URL:
  \url{https://www.eurosurveillance.org/content/10.2807/ese.18.04.20382-en},
  \href {https://doi.org/https://doi.org/10.2807/ese.18.04.20382-en}
  {\path{doi:https://doi.org/10.2807/ese.18.04.20382-en}}.

\bibitem{book:architecture}
Paul Clements, David Garlan, Len Bass, Judith Stafford, Robert Nord, James
  Ivers, and Reed Little.
\newblock {\em Documenting software architectures: views and beyond}.
\newblock Pearson Education, 2002.

\bibitem{bkchp:datamodel}
Paul Clements, David Garlan, Len Bass, Judith Stafford, Robert Nord, James
  Ivers, and Reed Little.
\newblock A tour of some module style.
\newblock In {\em Documenting software architectures: views and beyonds\/}
  \cite{book:architecture}, pages 109--120.

\bibitem{web:neovis}
Neo4j Contrib.
\newblock Neovis.js.
\newblock Last accessed 28 December 2020.
\newblock URL: \url{https://github.com/neo4j-contrib/neovis.js}.

\bibitem{artc:snp}
Nicholas~J Croucher, Simon~R Harris, Christophe Fraser, Michael~A Quail, John
  Burton, Mark van~der Linden, Lesley McGee, Anne von Gottberg, Jae~Hoon Song,
  Kwan~Soo Ko, et~al.
\newblock Rapid pneumococcal evolution in response to clinical interventions.
\newblock {\em science}, 331(6016):430--434, 2011.

\bibitem{web:dgraph}
Dgraph.
\newblock Dgraph: A distributed, fast graph database., 2016.
\newblock Last accessed 28 December 2020.
\newblock URL: \url{https://dgraph.io/}.

\bibitem{book:hbase}
Nick Dimiduk and Amandeep Khurana.
\newblock {\em HBase in Actions}.
\newblock O'Reilly Media, Inc., 2012.

\bibitem{artc:eburst}
Edward~J Feil, Bao~C Li, David~M Aanensen, William~P Hanage, and Brian~G
  Spratt.
\newblock eburst: inferring patterns of evolutionary descent among clusters of
  related bacterial genotypes from multilocus sequence typing data.
\newblock {\em Journal of bacteriology}, 186(5):1518--1530, 2004.

\bibitem{web:flowcraft}
FlowCraft.
\newblock Flowcraft, 2016.
\newblock Last accessed 28 December 2020.
\newblock URL: \url{https://github.com/assemblerflow/flowcraft}.

\bibitem{web:janusgraph}
The~Linux Foundation.
\newblock Janusgraph: Distributed graph database., 2017.
\newblock Last accessed 28 December 2020.
\newblock URL: \url{http://janusgraph.org/}.

\bibitem{artc:spring}
Spring Framework.
\newblock Spring framework.
\newblock {\em Available on:< https://spring. io/>. Access in}, 3, 2018.

\bibitem{artc:cypher}
Nadime Francis, Alastair Green, Paolo Guagliardo, Leonid Libkin, Tobias
  Lindaaker, Victor Marsault, Stefan Plantikow, Mats Rydberg, Petra Selmer, and
  Andr{\'e}s Taylor.
\newblock Cypher: An evolving query language for property graphs.
\newblock In {\em Proceedings of the 2018 International Conference on
  Management of Data}, pages 1433--1445. ACM, 2018.

\bibitem{artc:goeburst}
Alexandre~P Francisco, Miguel Bugalho, M{\'a}rio Ramirez, and Jo{\~a}o~A
  Carri{\c{c}}o.
\newblock Global optimal eburst analysis of multilocus typing data using a
  graphic matroid approach.
\newblock {\em BMC bioinformatics}, 10(1):152, 2009.

\bibitem{artc:phyloviz}
Alexandre~P Francisco, C{\'a}tia Vaz, Pedro~T Monteiro, Jos{\'e} Melo-Cristino,
  M{\'a}rio Ramirez, and Joao~A Carri{\c{c}}o.
\newblock Phyloviz: phylogenetic inference and data visualization for sequence
  based typing methods.
\newblock {\em BMC bioinformatics}, 13(1):87, 2012.

\bibitem{web:googleidp}
Google.
\newblock Openidconnect.
\newblock Last accessed 28 December 2020.
\newblock URL:
  \url{https://developers.google.com/identity/protocols/OpenIDConnect}.

\bibitem{artc:rfcoauth2.0}
Dick Hardt.
\newblock The oauth 2.0 authorization framework.
\newblock 2012.

\bibitem{artc:neo4j}
Florian Holzschuher and Ren{\'e} Peinl.
\newblock Performance of graph query languages: Comparison of cypher, gremlin
  and native access in neo4j.
\newblock In {\em Proceedings of the Joint EDBT/ICDT 2013 Workshops}, EDBT '13,
  pages 195--204, New York, NY, USA, 2013. ACM.
\newblock URL: \url{http://doi.acm.org/10.1145/2457317.2457351}, \href
  {https://doi.org/10.1145/2457317.2457351}
  {\path{doi:10.1145/2457317.2457351}}.

\bibitem{web:docker}
Docker Inc.
\newblock Docker.
\newblock Last accessed 28 December 2020.
\newblock URL: \url{https://www.docker.com/}.

\bibitem{web:allegrograph}
F.~Inc.
\newblock Allegrograph., 2004.
\newblock Last accessed 28 December 2020.
\newblock URL: \url{https://franz.com/agraph/allegrograph/}.

\bibitem{web:popoto}
NHOGS Interactive.
\newblock Popoto.js.
\newblock Last accessed 28 December 2020.
\newblock URL: \url{http://www.popotojs.com/}.

\bibitem{artc:NGS2}
Keith~A. Jolley and Martin~CJ Maiden.
\newblock Bigsdb: Scalable analysis of bacterial genome variation at the
  population level.
\newblock {\em BMC Bioinformatics}, 11(1):595, 2010.
\newblock \href {https://doi.org/10.1186/1471-2105-11-595}
  {\path{doi:10.1186/1471-2105-11-595}}.

\bibitem{artc:cassandra}
Avinash Lakshman and Prashant Malik.
\newblock Cassandra: a decentralized structured storage system.
\newblock {\em ACM SIGOPS Operating Systems Review}, 44(2):35--40, 2010.

\bibitem{web:neo4jversioning}
Ljubica Lazarevic.
\newblock Keeping track of graph changes using temporal versioning.
\newblock Last accessed 28 December 2020.
\newblock URL:
  \url{https://medium.com/neo4j/keeping-track-of-graph-changes-using-temporal-versioning-3b0f854536fa}.

\bibitem{artc:workflow}
Jeremy Leipzig.
\newblock A review of bioinformatic pipeline frameworks.
\newblock {\em Briefings in bioinformatics}, 18(3):530--536, 2017.

\bibitem{artc:mlva}
Bj{\o}rn-Arne Lindstedt.
\newblock Multiple-locus variable number tandem repeats analysis for genetic
  fingerprinting of pathogenic bacteria.
\newblock {\em Electrophoresis}, 26(13):2567--2582, 2005.

\bibitem{artc:innuendo}
Ann-Katrin Llarena, Bruno~Filipe Ribeiro-Gon{\c{c}}alves, Diogo Nuno~Silva,
  Jani Halkilahti, Miguel~Paulo Machado, Mickael~Santos Da~Silva, Anniina
  Jaakkonen, Joana Isidro, Crista H{\"a}m{\"a}l{\"a}inen, Jasmin Joenper{\"a},
  et~al.
\newblock Innuendo: A cross-sectoral platform for the integration of genomics
  in the surveillance of food-borne pathogens.
\newblock {\em EFSA Supporting Publications}, 15(11):1498E, 2018.

\bibitem{web:radial}
Leonardo~Alexandre Luana~Silva and Diogo Loureiro.
\newblock Phyloviz-electron.
\newblock Last accessed 28 December 2020.
\newblock URL: \url{https://github.com/DrLDiogo/PHYLOViZ-Electron}.

\bibitem{artc:MLST2}
Martin~CJ Maiden, Jane~A Bygraves, Edward Feil, Giovanna Morelli, Joanne~E
  Russell, Rachel Urwin, Qing Zhang, Jiaji Zhou, Kerstin Zurth, Dominique~A
  Caugant, et~al.
\newblock Multilocus sequence typing: a portable approach to the identification
  of clones within populations of pathogenic microorganisms.
\newblock {\em Proceedings of the National Academy of Sciences},
  95(6):3140--3145, 1998.

\bibitem{web:bearer}
Microsoft.
\newblock The oauth 2.0 authorization framework: Bearer token usage.
\newblock Last accessed 28 December 2020.
\newblock URL: \url{https://tools.ietf.org/html/rfc6750}.

\bibitem{artc:phyloviz2.0}
Marta Nascimento, Adriano Sousa, M{\'a}rio Ramirez, Alexandre~P Francisco,
  Jo{\~a}o~A Carri{\c{c}}o, and C{\'a}tia Vaz.
\newblock Phyloviz 2.0: providing scalable data integration and visualization
  for multiple phylogenetic inference methods.
\newblock {\em Bioinformatics}, 33(1):128--129, 2016.

\bibitem{web:bolt}
Inc. Neo~Technology.
\newblock Bolt protocol.
\newblock Last accessed 28 December 2020.
\newblock URL: \url{https://boltprotocol.org/}.

\bibitem{web:indexes}
Neo4j.
\newblock Indexes for search performance.
\newblock Last accessed 28 December 2020.
\newblock URL:
  \url{https://neo4j.com/docs/cypher-manual/current/administration/indexes-for-search-performance/}.

\bibitem{web:javacoreapi}
Neo4j.
\newblock Neo4j java driver 4.2 api.
\newblock Last accessed 28 December 2020.
\newblock URL: \url{https://neo4j.com/docs/api/java-driver/current/}.

\bibitem{manl:neo4jgraphmanual}
Inc Neo4j.
\newblock {\em The Neo4j Graph Algorithms User Guide v3.5}.
\newblock Neo4j.

\bibitem{web:oidc}
OpenID.
\newblock Openidconnect.
\newblock Last accessed 28 December 2020.
\newblock URL: \url{https://openid.net/connect/}.

\bibitem{web:spring}
Inc. Pivotal~Software.
\newblock Spring framework.
\newblock Last accessed 28 December 2020.
\newblock URL: \url{https://spring.io/}.

\bibitem{web:pubmlst}
PubMLST.
\newblock Pubmlst.
\newblock Last accessed 28 December 2020.
\newblock URL: \url{Available: http://pubmlst.org/}.

\bibitem{artc:rya}
Roshan Punnoose, Adina Crainiceanu, and David Rapp.
\newblock Rya: A scalable rdf triple store for the clouds.
\newblock In {\em Proceedings of the 1st International Workshop on Cloud
  Intelligence}, Cloud-I '12, pages 4:1--4:8, New York, NY, USA, 2012. ACM.
\newblock URL: \url{http://doi.acm.org/10.1145/2347673.2347677}, \href
  {https://doi.org/10.1145/2347673.2347677}
  {\path{doi:10.1145/2347673.2347677}}.

\bibitem{web:neo4junlimitednodes}
Philip Rathle.
\newblock Official release: 3 essentials of neo4j 3.0, from scale to
  productivity and deployment, 2016.
\newblock Last accessed 28 December 2020.
\newblock URL:
  \url{https://neo4j.com/blog/neo4j-3-0-massive-scale-developer-productivity/}.

\bibitem{bkchp:GDinternals}
Ian Robinson, Jim Webber, and Emil Eifrem.
\newblock Graph database internals.
\newblock In {\em Graph Databases\/} \cite{book:GraphDatabases}, pages
  149--170.

\bibitem{book:GraphDatabases}
Ian Robinson, Jim Webber, and Emil Eifrem.
\newblock {\em Graph Databases}.
\newblock O'Reilly Media, Inc., 2013.

\bibitem{bkchp:GDintroduction}
Ian Robinson, Jim Webber, and Emil Eifrem.
\newblock Introduction.
\newblock In {\em Graph Databases\/} \cite{book:GraphDatabases}, pages 1--10.

\bibitem{bkchp:connecteddata}
Ian Robinson, Jim Webber, and Emil Eifrem.
\newblock Options for storing connected data.
\newblock In {\em Graph Databases\/} \cite{book:GraphDatabases}, pages 11--24.

\bibitem{web:goeburst}
Luana Silva.
\newblock phylolib.
\newblock Last accessed 28 December 2020.
\newblock URL: \url{https://github.com/Luanab/phylolib}.

\bibitem{artc:MLST}
BG~Spratt.
\newblock Multilocus sequence typing: molecular typing of bacterial pathogens
  in an era of rapid dna sequencing and the internet.
\newblock {\em Current opinion in microbiology}, 2(3):312—316, June 1999.
\newblock URL: \url{https://doi.org/10.1016/S1369-5274(99)80054-X}, \href
  {https://doi.org/10.1016/s1369-5274(99)80054-x}
  {\path{doi:10.1016/s1369-5274(99)80054-x}}.

\bibitem{web:interceptors}
Spring.
\newblock Spring framework documentation.
\newblock Last accessed 28 December 2020.
\newblock URL:
  \url{https://docs.spring.io/spring-framework/docs/current/reference/html/web.html}.

\bibitem{web:cypherparameterized}
Andrew Bowman~Neo4j Staff.
\newblock Neo4j security.
\newblock Last accessed 28 December 2020.
\newblock URL: \url{https://community.neo4j.com/t/neo4j-security/16044}.

\bibitem{web:neo4jbatch}
Michael Hunger~Neo4j Staff.
\newblock 5 tips \& tricks for fast batched updates of graph structures with
  neo4j and cypher.
\newblock Last accessed 28 December 2020.
\newblock URL:
  \url{https://medium.com/neo4j/5-tips-tricks-for-fast-batched-updates-of-graph-structures-with-neo4j-and-cypher-73c7f693c8cc}.

\bibitem{artc:trees}
Andreia~Sofia Teixeira, Pedro~T Monteiro, Jo{\~a}o~A Carri{\c{c}}o, M{\'a}rio
  Ramirez, and Alexandre~P Francisco.
\newblock Not seeing the forest for the trees: size of the minimum spanning
  trees (msts) forest and branch significance in mst-based phylogenetic
  analysis.
\newblock {\em Plos one}, 10(3):e0119315, 2015.

\bibitem{artc:spark}
Andreia~Sofia Teixeira, Pedro~T Monteiro, Joao~A Carri{\c{c}}o, Francisco~C
  Santos, and Alexandre~P Francisco.
\newblock Using spark and graphx to parallelize large-scale simulations of
  bacterial populations over host contact networks.
\newblock In {\em International Conference on Algorithms and Architectures for
  Parallel Processing}, pages 591--600. Springer, 2017.

\bibitem{artc:relations}
C{\'a}tia Vaz, Alexandre~P. Francisco, Mickael Silva, Keith~A. Jolley, James~E.
  Bray, Hannes Pouseele, Joerg Rothganger, M{\'a}rio Ramirez, and Jo{\~a}o~A.
  Carri{\c{c}}o.
\newblock Typon: the microbial typing ontology.
\newblock {\em Journal of Biomedical Semantics}, 5(1):43, 2014.
\newblock \href {https://doi.org/10.1186/2041-1480-5-43}
  {\path{doi:10.1186/2041-1480-5-43}}.

\bibitem{artc:10.1093/bib/bbaa147}
Cátia Vaz, Marta Nascimento, João~A Carriço, Tatiana Rocher, and Alexandre~P
  Francisco.
\newblock {Distance-based phylogenetic inference from typing data: a unifying
  view}.
\newblock {\em Briefings in Bioinformatics}, 07 2020.
\newblock bbaa147.
\newblock \href
  {http://arxiv.org/abs/https://academic.oup.com/bib/advance-article-pdf/doi/10.1093/bib/bbaa147/33551477/bbaa147.pdf}
  {\path{arXiv:https://academic.oup.com/bib/advance-article-pdf/doi/10.1093/bib/bbaa147/33551477/bbaa147.pdf}},
  \href {https://doi.org/10.1093/bib/bbaa147} {\path{doi:10.1093/bib/bbaa147}}.

\bibitem{web:cypherfast}
Christophe Willemsen.
\newblock Cypher: Write fast and furious.
\newblock Last accessed 28 December 2020.
\newblock URL: \url{https://neo4j.com/blog/cypher-write-fast-furious/}.

\bibitem{artc:grapetree}
Zhemin Zhou, Nabil-Fareed Alikhan, Martin~J Sergeant, Nina Luhmann, C{\'a}tia
  Vaz, Alexandre~P Francisco, Jo{\~a}o~Andr{\'e} Carri{\c{c}}o, and Mark
  Achtman.
\newblock Grapetree: visualization of core genomic relationships among 100,000
  bacterial pathogens.
\newblock {\em Genome research}, 28(9):1395--1404, 2018.

\end{thebibliography}

\begin{appendices}

\section*{Example Workflow}

\begin{figure}[!ht]
 \centering
 \includegraphics[scale=0.1,width=0.9\textwidth]{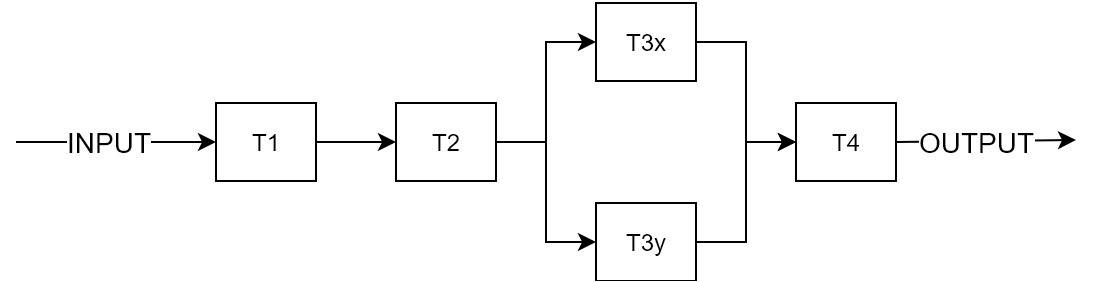}
 \caption{Example of a workflow system.}
 \label{fig:workflow}
\end{figure}

In this example of a workflow system shown in Figure \ref{fig:workflow}, can be interpreted that, an input is provided to the system, which then applies the transformations T1, and T2 sequentially. With the result of the transformation T2, it is applied the transformations T3x, and T3y parallelly. Afterwards, each result produced by the latter transformations is merged, and the transformation T4 is applied producing the final output.

\section*{Neo4j Algorithms}
\begin{table}[!ht]
        \begin{tabular}{ l l }  \hline
            Algorithm & Objective \\  \hline
            \multicolumn{1}{p{0.2\textwidth}}{Louvain} & \multicolumn{1}{p{0.7\textwidth}}{Is an algorithm for detecting communities in networks. It maximizes a modularity score for each community, where the modularity quantifies the quality of an assignment of nodes to communities by evaluating how much more densely connected the nodes within a community are, compared to how connected they would be in a random network.}\\
            \multicolumn{1}{p{0.2\textwidth}}{Label Propagation} & \multicolumn{1}{p{0.7\textwidth}}{Is a fast algorithm for finding communities in a graph. It detects these communities using network structure alone as its guide, and doesn’t require a pre-defined objective function or prior information about the communities.}\\
            \multicolumn{1}{p{0.2\textwidth}}{Connected Components} & \multicolumn{1}{p{0.7\textwidth}}{Finds sets of connected nodes in an undirected graph where each node is reachable from any other node in the same set. It differs from the Strongly Connected Components algorithm (SCC) because it only needs a path to exist between pairs of nodes in one direction, whereas SCC needs a path to exist in both directions.} \\
            \multicolumn{1}{p{0.2\textwidth}}{Strongly Connected Components} & \multicolumn{1}{p{0.7\textwidth}}{Finds sets of connected nodes in a directed graph where each node is reachable in both directions from any other node in the same set.}\\
            \multicolumn{1}{p{0.2\textwidth}}{Triangle Counting / Clustering Coefficient} & \multicolumn{1}{p{0.7\textwidth}}{Is used to determine the number of triangles passing through each node in the graph. A triangle is a set of three nodes, where each node has a relationship to all other nodes.}\\
            \multicolumn{1}{p{0.2\textwidth}}{Balanced Triads} & \multicolumn{1}{p{0.7\textwidth}}{Is used to evaluate structural balance of the graph. Balance theory differentiates between positive and negative relationships. Certain structures between individuals and objects are perceived as balanced whereas others are not.}\\  \hline
        \end{tabular} 
    \caption{Community Detection Algorithms.} 
    \label{tbl:communitydetectionalgorithms} 
\end{table}

\begin{table}[!ht]
        \begin{tabular}{ l l }  \hline
            Algorithm & Objective \\  \hline
            \multicolumn{1}{p{0.2\textwidth}}{PageRank} & \multicolumn{1}{p{0.7\textwidth}}{Measures the connectivity of nodes. It counts the number, of links to a page which determines an estimation of how important the page is. The underlying assumption is that pages of importance are more likely to receive a higher volume of links from other pages.}\\
            \multicolumn{1}{p{0.2\textwidth}}{ArticleRank} & \multicolumn{1}{p{0.7\textwidth}}{Is a variant of the PageRank Where ArticleRank differs to PageRank is that PageRank assumes that relationships from nodes that have a low out-degree are more important than relationships from nodes with a higher out-degree. ArticleRank weakens this assumption.}\\
            \multicolumn{1}{p{0.2\textwidth}}{Betweenness Centrality} & \multicolumn{1}{p{0.7\textwidth}}{Is a way of detecting the amount of influence a node has over the flow of information in a graph. It calculates the shortest (weighted) path between every pair of nodes in a connected graph, using the breadth-first search algorithm. Each node receives a score. Nodes that most frequently lie on these shortest paths will have a higher betweenness centrality score.} \\
            \multicolumn{1}{p{0.2\textwidth}}{Closeness Centrality} & \multicolumn{1}{p{0.7\textwidth}}{Is a way of detecting nodes that are able to spread information very efficiently through a graph. For each node, the Closeness Centrality algorithm calculates the sum of its distances to all other nodes, based on calculating the shortest paths between all pairs of nodes. The resulting sum is then inverted to determine the closeness centrality score for that node.}\\
            \multicolumn{1}{p{0.2\textwidth}}{Harmonic Centrality} & \multicolumn{1}{p{0.7\textwidth}}{Is a variant of closeness centrality, that was invented to solve the problem the original formula had when dealing with unconnected graphs. Rather than summing the distances of a node to all other nodes, the harmonic centrality algorithm sums the inverse of those distances.}\\
            \multicolumn{1}{p{0.2\textwidth}}{Eigenvector Centrality} & \multicolumn{1}{p{0.7\textwidth}}{Measures the connectivity of nodes. Relationships to high-scoring nodes contribute more to the score of a node than connections to low-scoring nodes. A high score means that a node is connected to other nodes that have high scores.}\\
            \multicolumn{1}{p{0.2\textwidth}}{Degree Centrality} & \multicolumn{1}{p{0.7\textwidth}}{Measures the number of incoming and outgoing relationships from a node. It can help to find popular nodes in a graph, and can be used to find the popularity of individual nodes, but it is often used as part of a global analysis where we calculate the minimum degree, maximum degree, mean degree, and standard deviation across the whole graph.}\\  \hline
        \end{tabular} 
    \caption{Centrality Algorithms.} 
    \label{tbl:centralityalgorithms} 
\end{table}

\begin{table}[!ht]
        \begin{tabular}{ l l }  \hline
            Algorithm & Objective \\  \hline
            \multicolumn{1}{p{0.2\textwidth}}{Minimum Weight Spanning Tree} & \multicolumn{1}{p{0.7\textwidth}}{Starts from a given node, and finds all its reachable nodes and the set of relationships that connect the nodes together with the minimum possible weight. Prim’s algorithm is one of the simplest and best-known minimum spanning tree algorithms.}\\
            \multicolumn{1}{p{0.2\textwidth}}{Shortest Path} & \multicolumn{1}{p{0.7\textwidth}}{Calculates the shortest (weighted) path between a pair of nodes. In this category, Dijkstra’s algorithm is the most well known.}\\
            \multicolumn{1}{p{0.2\textwidth}}{Single Source Shortest Path} & \multicolumn{1}{p{0.7\textwidth}}{Calculates the shortest (weighted) path from a node to all other nodes in the graph.} \\
            \multicolumn{1}{p{0.2\textwidth}}{All Pairs Shortest Path} & \multicolumn{1}{p{0.7\textwidth}}{Calculates the shortest (weighted) path between all pairs of nodes. This algorithm has optimisations that make it quicker than calling the Single Source Shortest Path algorithm for every pair of nodes in the graph}\\
            \multicolumn{1}{p{0.2\textwidth}}{A*} & \multicolumn{1}{p{0.7\textwidth}}{Improves on the classic Dijkstra algorithm. It is based upon the observation that some searches are informed, and that by being informed we can make better choices over which paths to take through the graph.}\\
            \multicolumn{1}{p{0.2\textwidth}}{Yen’s K-shortest paths} & \multicolumn{1}{p{0.7\textwidth}}{Computes single-source K-shortest loopless paths for a graph with non-negative relationship weights.}\\
            \multicolumn{1}{p{0.2\textwidth}}{Random Walk} & \multicolumn{1}{p{0.7\textwidth}}{Provides random paths in a graph. We start at one node, choose a neighbor to navigate to at random or based on a provided probability distribution, and then do the same from that node, keeping the resulting path in a list.}\\  \hline
        \end{tabular} 
    \caption{Path finding Algorithms.} 
    \label{tbl:pathfindingalgorithms} 
\end{table}

\begin{table}[!ht]
        \begin{tabular}{ l l }  \hline
            Algorithm & Objective \\  \hline
            \multicolumn{1}{p{0.2\textwidth}}{Jaccard Similarity} & \multicolumn{1}{p{0.7\textwidth}}{Measures similarities between sets. It is defined as the size of the intersection divided by the size of the union of two sets.}\\
            \multicolumn{1}{p{0.2\textwidth}}{Cosine Similarity} & \multicolumn{1}{p{0.7\textwidth}}{Is the cosine of the angle between two n-dimensional vectors in an n-dimensional space. It is the dot product of the two vectors divided by the product of the two vectors' lengths (or magnitudes).}\\
            \multicolumn{1}{p{0.2\textwidth}}{Pearson Similarity} & \multicolumn{1}{p{0.7\textwidth}}{Is the covariance of the two n-dimensional vectors divided by the product of their standard deviations.} \\
            \multicolumn{1}{p{0.2\textwidth}}{Euclidean Distance} & \multicolumn{1}{p{0.7\textwidth}}{Measures the straight line distance between two points in n-dimensional space.}\\
            \multicolumn{1}{p{0.2\textwidth}}{Overlap Similarity} & \multicolumn{1}{p{0.7\textwidth}}{Measures overlap between two sets. It is defined as the size of the intersection of two sets, divided by the size of the smaller of the two sets.}\\
            \multicolumn{1}{p{0.2\textwidth}}{Approximate Nearest Neighbors} & \multicolumn{1}{p{0.7\textwidth}}{Constructs a k-Nearest Neighbors Graph for a set of objects based on a provided similarity algorithm. The similarity of items is computed based on Jaccard Similarity, Cosine Similarity, Euclidean Distance, or Pearson Similarity.}\\  \hline
        \end{tabular} 
    \caption{Similarity Algorithms.} 
    \label{tbl:similarityalgorithms} 
\end{table}
 
 \begin{table}[!ht]
        \begin{tabular}{ l l }  \hline
            Algorithm & Objective \\  \hline
            \multicolumn{1}{p{0.2\textwidth}}{Adamic Adar} & \multicolumn{1}{p{0.7\textwidth}}{Is a measure used to compute the closeness of nodes, based on their shared neighbors.}\\
            \multicolumn{1}{p{0.2\textwidth}}{Common Neighbors} & \multicolumn{1}{p{0.7\textwidth}}{Captures the idea that two strangers who have a friend in common are more likely to be introduced than those who don’t have any friends in common.}\\
            \multicolumn{1}{p{0.2\textwidth}}{Preferential Attachment} & \multicolumn{1}{p{0.7\textwidth}}{Is a measure used to compute the closeness of nodes, based on their shared neighbors.} \\
            \multicolumn{1}{p{0.2\textwidth}}{Resource Allocation} & \multicolumn{1}{p{0.7\textwidth}}{Is a measure used to compute the closeness of nodes, based on their shared neighbors.}\\
            \multicolumn{1}{p{0.2\textwidth}}{Same Community} & \multicolumn{1}{p{0.7\textwidth}}{Is a way of determining whether two nodes belong to the same community.}\\
            \multicolumn{1}{p{0.2\textwidth}}{Total Neighbors} & \multicolumn{1}{p{0.7\textwidth}}{Computes the closeness of nodes, based on the number of unique neighbors that they have. It is based on the idea that the more connected a node is, the more likely it is to receive new links.}\\  \hline
        \end{tabular} 
    \caption{Link Prediction Algorithms.} 
    \label{tbl:linkpredictionalgorithms} 
\end{table}

\end{appendices}

\end{document}